# GW Notes

April to August 2010

Notes & News for GW science
Editors:
P. Amaro-Seoane and B. F. Schutz

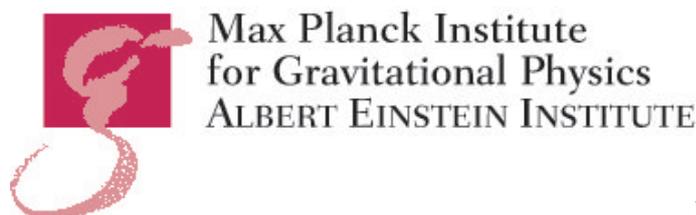

**Max Planck Institute
for Gravitational Physics
ALBERT EINSTEIN INSTITUTE**

GW Notes was born from the need for a journal where the distinct communities involved in gravitational wave research might gather. While these three communities - Astrophysics, General Relativity and Data Analysis - have made significant collaborative progress over recent years, we believe that it is indispensable to future advancement that they draw closer, and that they speak a common idiom.

---

### *Editorial*

### *Fundamental physics with GWs*

---

One of the principal goals of the LISA mission is to study the signals emitted when stellar-mass black holes spiral into massive black holes, the so-called Extreme-Mass-Ratio Inspiral (EMRI) systems. Observed systems will emit $10^5$ or more cycles of radiation in the LISA band, in a waveform whose complexity measures the fine details of the strong gravitational field near a Kerr black hole. But what if the observed signal does not conform to the family of signals expected from orbits around a Kerr hole? What if, in other words, general relativity is not quite right? To go beyond falsifying GR, to say something positive about the new theory, would be challenging. Do the observed departures in the waveform reflect a modified geometry or a modified emission mechanism, or both? We have asked Carlos F. Sopuerta to write an essay on his thoughts on how we should prepare ourselves to answer the question: If not GR, what then?

Pau Amaro-Seoane & Bernard F. Schutz, editors

> *GW Notes highlight article*
>
> *Fundamental physics and EMRIs*

A ROADMAP TO FUNDAMENTAL PHYSICS FROM LISA EMRI OBSERVATIONS

Carlos F. Sopuerta
Institut de Ciències de l'Espai (CSIC-IEEC)
Facultat de Ciències, Campus UAB, Torre C5 parells
Bellaterra, 08193 Barcelona, Spain
e-mail: sopuerta@ice.cat

### Abstract

The Laser Interferometer Space Antenna is a future space-based gravitational-wave observatory (a joint mission between the European Space Agency and the US National Aeronautics and Space Administration) that is expected to be launched during the next decade. It will operate in the low-frequency gravitational-wave band, probably the richest part of the gravitational-wave spectrum in terms of science potential, where we find: massive black hole mergers as the outcome of galaxy collisions; many galactic compact binaries; the capture and subsequent inspiral of a stellar compact object into a massive black hole; and gravitational-wave signatures from early universe physical processes connected to high-energy physics and physics not yet fully understood. In this article we focus on the third type of source, the so-called extreme-mass-ratio inspirals, a high precision tool for gravitational wave astronomy that can be used, among other things, to advance in our understanding of fundamental physics questions like the nature and structure of black holes and the details of the gravitational interaction in regimes not yet proven by other experiments/observatories. Here, we give an account of some of the progress made in the development of tools to exploit the future LISA EMRI observations, we discuss what scientific questions we can try to answer from this information and, finally, we discuss the main theoretical challenges that we face in order to develop all the necessary tools to maximize the scientific outcome and some avenues that can be followed to make progress in the near future.

# Contents

Notes & News for GW science

# 1 Introduction

Fundamental Physics is a term that is nowadays widely used and it normally refers to the area of Physics that tries to push our knowledge of the basic physical interactions, covering from the studies of their relationships to the efforts to unify them in more and more *fundamental* theories. This does not necessarily mean that every single physical phenomena will find a simple explanation from these fundamental theories, as we certainly need to understand the basic rules that govern the complex structures around us, but they are certainly crucial in understanding basic questions like the structure of the space-time and the building blocks of matter.

To pursue these goals there are several large-scale laboratories and observatories that conduct experiments to tests the limits of what we know in Fundamental Physics and also to go beyond them, towards regimes not yet explored in Physics. Thinking about examples of ground-based research facilities one cannot avoid nowadays to refer to the Large Hadron Collider (**LHC**) at the European Organization for Nuclear Research (CERN) where, in a tunnel underground 27 kilometers (17 miles) in circumference, protons moving at 99.999999% of the speed of light are smashed into each other. These collisions will create physical conditions not yet reached before that will enable researchers to test various predictions of High-Energy Physics, including the existence of the hypothesized Higgs boson and of the large family of new particles suggested by supersymmetry. Other predictions that may be tested are the possible existence of extra dimensions, as predicted by various models inspired by string theory, and the nature of the dark matter that shapes the large-scale structure of the Universe.

Regarding facilities in space, they have increased in recent times and cover a wide range of wavelengths and (astro)physical phenomena. One can mention here observatories like the **Hubble Space Telescope** and **WMAP** (and presently we have **Planck** starting scientific operations) which have made observations that have improved our knowledge of Cosmology, with strong implications in Fundamental Physics, including the discovery and confirmation of the acceleration of the expansion of the Universe (Riess et al., 1998, Perlmutter et al., 1999 and Spergel et al., 2003), which constitutes an enormous challenge for Fundamental Physics. Moreover, there are several observatories (for instance, **Chandra** and **XMM-Newton** in the X-ray band and **Fermi** in the Gamma-ray band) that look at the high-energy end of the electromagnetic spectrum for radiation coming from the most powerful events in the Universe: supernovae, Gamma-ray bursts (GRBs), neutron stars (NSs), etc. Some of the physical mechanisms behind these energetic events are still unknown, like in the case of GRBs, and also constitute a challenge for Astrophysics and Fundamental Physics.

These high-energy electromagnetic observations, together with observations of cosmic rays and the promising developments in neutrino astronomy, constitute the core of what has been called Astroparticle Physics, which can be defined as the branch of Particle Physics that studies elementary particles of astronomical origin, and their relation to Astrophysics and Cosmology. Cosmic rays are particles that originated in outer space and that reach Earth's atmosphere. The energy spectrum is very broad and extends up to very high energies, much higher than those that will be achieved at the LHC. There are measurements of particles with energies up to ∼ 100 EeV = $10^{20}$ eV, the so-called ultra-high-energy or extreme-energy cosmic rays. Although active galactic nuclei, presumably powered by supermassive black holes (BHs), are believed to be the engine that produce them, the details of the production mechanisms are still unclear. There are observatories, like the **Pierre Auger Observatory**, that have started operations to unveil the mysteries of cosmic rays. On the other hand, neutrinos are expected to be produced in significant amounts from the high-energy scenarios that have been discussed. Although neutrino astronomy is still in its infancy (the only extraterrestrial neutrinos detected so far come from the Sun and the well-known supernova SN1987A), detectors like **IceCube** are expected, in a 5 year timescale, to do science beyond the detection of neutrinos associated with cosmic rays, including the study of dark matter and the study of neutrino physics at energies far beyond the reach of accelerators.

As it is obvious from this discussion, Astroparticle Physics is an area that also has a very important intersection with Fundamental Physics, with a high potential for discovery in areas like dark matter, AGN physics, etc. An important fact about Astroparticle Physics is that it includes *messengers* of astronomical information different from photons: Protons, neutrinos, etc., and even gravitational waves, whose direct detection is around the corner (advanced detectors like **Advanced LIGO** and **Advanced VIRGO** are expected to make the first discoveries within the present decade), and are also considered part of Astroparticle Physics. This opens the door for what has been called *Multi-messenger* Astronomy, which essentially consists in combining observations of astronomical sources from different messengers, which can produce the necessary synergy to supply new insights on the physics of the most violent/energetic events in the Universe. There are many examples of this, for instance, the detection of gravitational waves and neutrinos from supernovae, or multi-messenger emissions associated with Gamma-ray bursts (Stamatikos et al., 2009).

The focus of this article is on Gravitational Wave Astronomy (see (Sathyaprakash and Schutz, 2009) for a recent review), another emergent area in Astronomy based on the detection and analysis of the gravitational radiation produced by astrophysical and cosmological sources, like the merger of binary BHs or NSs. In order to understand the process of generation, propagation, and detection of gravitational waves (GWs) we have to take into account the fact that gravity, as we perceive it from our four-dimensional spacetime perspective, appears as the weakest of all physical interactions known to date. In spite of this fact, it is the force that governs the large-scale structure of the universe. In the context of Einstein's theory of relativity, gravity also determines the spacetime geometry and hence the relations between the events that take place on it. This means that GWs, as an astronomical messenger, are a double-edged sword: On the one hand, their detection is quite a difficult problem that requires very advanced technology. On the other hand, they are an ideal tool to test *strong gravity* since GWs carry almost uncorrupted information from their sources.

Here, there is the question of what strong gravity means? It is a concept that appears in many places but not always refers to the same physical phenomena. We know that Newtonian gravity is sufficient to describe a wide range of gravitational phenomena. Relativistic effects become relevant when we make very precise observations of astronomical systems, and also of systems that involve either very compact objects or fast motions. However, due to the weakness of gravity, these effects are difficult to measure and, as a consequence, only certain regimes of relativistic gravity have been tested so far. These tests confirm, to their level of precision, the validity of general relativity (GR) (see (Will, 2006) for a review). These experimental tests include observations of the motion of different objects in the solar system and observations of millisecond binary pulsars. To assess how strong is the gravitational field in different cases we can use a dimensionless measure of the Newtonian gravitational potential, namely $\phi_N \equiv c^{-2} \Phi_{\text{Newtonian}}$. In the case of solar system dynamics it yields $\phi_N \sim (GM_\odot)/(c^2 \, 1\text{AU}) \sim 10^{-8}$.[1] In the case of pulsars, if we take the well-known Hulse and Taylor binary pulsar (PRS B1913+16) (Weisberg and Taylor, 2005), the one that provided the first strong evidence for the existence of GWs (Hulse and Taylor, 1975), the same dimensionless measure yields $\phi_N \sim (GM_\odot)/(c^2 \, r_{\text{Hulse}-\text{Taylor}}^{\text{Periastron}}) \sim 10^{-6}$. This quantity is a couple of orders bigger than in the case of the Solar System and therefore these binary systems allow us to test the post-Newtonian regime. Actually, millisecond pulsars have already being used to test alternative theories of gravity, like scalar-tensor theories (see, e.g. (Damour, 2007b, 2007a and Stairs, 2003)). However, this has to be put in contrast with the estimate we get for the case of a binary BH merger, for which we have that $\phi_N \sim (GM_{\text{BH}})/(c^2 \, [\text{a few } r_{\text{Horizon}}]) \sim 10^{-1} - 1$, which sets what we will call throughout this article the *strong gravitational field regime*. Here, one can also argue that the study of millisecond pulsars, in particular of the recently discovered *double pulsar* (J0737-3039A,B) (Lyne et al., 2004) already allows tests in this regime because of the large self-gravitation of neutron stars (since $M_{\text{NS}} \sim 1.4 M_\odot$ and $R_{\text{NS}} \sim 10$ km we have $\phi_N \sim (GM_{\text{NS}})/(c^2 \, R_{\text{NS}}) \sim 0.2$, i.e. strong gravitational field) in this binary system does not influence their orbital motion and hence there is no violation of the strong equivalence principle. That is, for systems like binary pulsars we have two measures to consider, the one associated with the self-gravity of each component of the binary, and the one associated with the binary binding gravitational potential. While the first measure is relevant for tests of the strong equivalence principle, when we want to distinguish between theories that satisfy it (at least within observational limits) only the second measure is relevant, and this is going to be the case in the main subjects discussed in this article.

---

[1] Here, $G$ denotes the gravitational Newton constant, $c$ the speed of light, and $M_\odot$ the mass of the Sun.

To access the gravitational regime not yet tested one can try to resort to electromagnetic observations (see (Psaltis, 2008) for a review) as new observatories in the high-energy end of the spectrum have good potential for such a goal. Nevertheless, due to reasons mentioned above, Gravitational Wave Astronomy has a unique potential to explore the strong regime (which can be enhanced when used in a future multi-messenger Astronomy setup in combination, e.g. with electromagnetic observations (Bloom et al., 2009 and Phinney, 2009)). It is an emergent area that promises to bring revolutionary discoveries to the areas of Astrophysics, Cosmology, and Fundamental Physics. There are a number of ground laser interferometer detectors (**Advanced LIGO**, **Advanced VIRGO**, **LCGT**, etc.) that will detect GWs, in the high frequency range ($f \sim 10-10^3$ Hz) during the next decade. There are also ongoing developments for future detectors in space, like the Laser Interferometer Space Antenna (LISA) (**ESA site**, **NASA site**) or **DECIGO**. In particular, LISA will operate in the low frequency band ($f \sim 10^{-4} - 1$ Hz), a band not accessible from the ground due to seismic noise, and probably the richest band in terms of interesting astrophysical and cosmological sources. Apart from these detectors, there is work in progress to use networks of millisecond pulsars to detect GWs in the ultra-low frequency range ($f \sim 10^{-9} - 10^{-8}$ Hz). These pulsars are remarkable stable rotators, and as such, they require only simple models to describe their spin-down and times-of-arrival (TOAs) with a precision of $< 1$ $\mu$s over many years of observations. GWs are not included in the analysis, so their existence will induce differences between the measured and theoretical TOAs, the so-called *timing residuals*. To determine the exact origin of the timing residuals, that is, effects different from GWs (calibration effects, errors in planetary ephemeris, irregularities in the pulsar spin-down, etc.) it is necessary to correlate the timing residuals of multiple pulsars. It has been estimated that with a timely progress in technology, a successful detection of GWs should happen within a decade (Verbiest et al., 2010) (or alternatively the experiments will rule out current predictions for GW sources in this frequency band).

For information on tests of the different gravitational theories see (Will, 1987, 1992, 1993a, 1993b, 1996, 1998, 2006, 2010, Damour, 1995, Shapiro, 1999 and Turyshev, 2008). Throughout this article, and otherwise stated, we use geometry units, in which $G = c = 1$.

## 2   LISA and Extreme-Mass-Ratio Inspirals

In this article we will focus on LISA, a joint mission between the European Space Agency (ESA) and the US National Aeronautics and Space Administration (NASA). LISA has been designed to detect and analyze the gravitational radiation coming from astrophysical and cosmological sources in the low-frequency band (Prince et al., 2009) (i.e. oscillation periods in the range 10 s - 10 hours). LISA consists of three identical spacecrafts that fly in a triangular constellation, with equal arms of $5 \cdot 10^6$ km each. The LISA constellation is in orbit around the Sun, at a plane inclined by 60 degrees to the ecliptic. The triangle appears to rotate once around its center in the course of a year's revolution around the Sun. The center of the LISA triangle traces an Earth-like orbit in the ecliptic plane, trailing Earth by 20 degrees. The free-fall orbits of the three spacecraft around the Sun maintain this triangular formation, with the triangles appearing to rotate about its center once per year (see the LISA International Science Team **LIST** website and the LISA International Science Community **LISC** for more detailed information and illustrations of the LISA observatory). As GWs from distant sources reach LISA, they warp the space-time geometry stretching and compressing the triangle. The sensitivity of LISA as a GW observatory is described by the strength of its response to impinging GWs as a function of frequency. At low frequencies it is limited by acceleration noise; at mid frequencies by laser shot noise and optical-path measurement errors; and at high frequencies by the fact that the GW wavelength becomes shorter than the LISA arm length, reducing the efficiency of the interferometric measurement. Monitoring the separation between the spacecrafts to a high degree of precision (at the picometer level), we can measure the GWs and from there infer the physical parameters that characterize the systems that emitted them. An essential tool for the detection of GWs with LISA is the so-called Time-Delay Interferometry (Dhurandhar and Tinto, 2005), which allows phase measurements many orders of magnitude below the intrinsic phase stability of the lasers despite delays due to the fact that the arms are not exactly equal. Different models of the LISA sensitivity and response functions have been constructed over the years (Larson et al., 2000, Prince et al., 2002, Cornish and Rubbo, 2003, Rubbo et al., 2003 and Petiteau et al., 2008).

Within the LISA frequency band we find several important sources of GWs: (i) Massive BH mergers. Mergers of BHs sitting at galactic centers that form binaries after their host galaxies collide (which seem to happen several times in the life of most galaxies). At some point, mainly because of dynamical friction, the BHs are dragged towards each other and become gravitationally bounded. From here, gas or passing stars extracting angular momentum from the binary can shrink the orbit until the binary gets to the point in which it is solely driven by GW emission (see Berti, 2006 and Madau et al., 2009) and references therein for details). (ii) Galactic binaries. Millions of these are emitting GWs inside the LISA band. They are almost monochromatic sources and form a GW foreground from which only a fraction of them can be individually resolved (Nelemans, 2009). A few of them are known nearby galactic binary systems, the so-called *verification binaries*, are guaranteed sources for LISA (Stroeer and Vecchio, 2006 and Nelemans, 2006). (iii) EMRIs. The capture and subsequent inspiral of a stellar-mass compact objects (white dwarfs, or neutron stars, or stellar BHs) into a massive BH sitting at a galactic center. This source, commonly referred to as an Extreme-Mass-Ratio Inspirals (EMRIs), is the one we will focus on throughout this article as a high-precision tool for fundamental physics tests. (iv) Stochastic GW backgrounds. There are also prospects of detecting stochastic GW backgrounds produced by high-energy processes in the early universe (inflation, superstrings, topological defects, etc.). For recent reviews of the potential of LISA for Fundamental Physics see (Hogan, 2007, Schutz, 2009, Schutz et al., 2009 and Hogan and Binetruy, 2009).

In this article we focus on EMRIs and the possibilities of doing Fundamental Physics by means of observations of these systems by LISA. Standard EMRIs are binaries composed of a stellar-mass compact object (SCO) that inspirals into a massive BH (MBH) located in a galactic center, i.e. extreme-mass-ratio binaries in the stage where the dynamics is driven by GW emission. The masses of interest for the SCO are in the range $m_\star = 1 - 10^2 \, M_\odot$, and for the MBH in the range $M_\bullet = 10^5 - 10^7 \, M_\odot$. Then, the mass-ratio for these systems is in the interval: $\mu = m_\star/M_\bullet \sim 10^{-7} - 10^{-3}$. During the inspiral phase, the system losses energy and angular momentum via the emission of GWs, producing a shrinking of the orbit, which implies a decrease in the orbital periods and therefore, an increase of the GW frequencies. Regarding the formation of these systems, there are several astrophysical mechanisms that have the potential to produce EMRIs (see (Amaro-Seoane et al., 2007) for a review on astrophysics, science applications and detection using LISA). The most likely and studied mechanism consist in the capture of SCOs from a stellar cusp or core that surrounds the MBH at a galactic center. It turns out that for a given SCO there is a small but non-negligible probability that, via gravitational interactions with the other stars, it will end up in a bounded trajectory with respect to the MBH. A number of these SCOs will evolve in such a way that interactions with the rest of stellar objects will be negligible and therefore they will follow an almost geodesic trajectory around the MBH. The evolution of the orbit is slow in the sense that it can be approximated by a sequence of geodesics dictated by the GW emission (see (Glampedakis, 2005)). The orbit decays and at some point the SCO plunges into the MBH. For a number of systems, the whole inspiral process takes a time significantly smaller than the Hubble time scale, and hence they can be considered LISA EMRI sources. The EMRIs produced by this mechanism are initially very eccentric, with eccentricities at the capture time in the range $1 - e \sim 10^{-6} - 10^{-3}$, but by the time they enter inside the LISA band the eccentricity is expected to be substantially reduced (due to GW emission), although it will still be quite significant (in the range $e \sim 0.5 - 0.9$). A remarkable fact about EMRIs is that during the last year before plunge they emit on the order of $10^5$ or more GW cycles (Finn and Thorne, 2000), with a high number of harmonics contributing to the GW emission (Barack and Cutler, 2004), which means that these waves carry a wealth of information about the MBH strong field region. Regarding the event rates for EMRIs produced via the capture mechanism, it has been estimated that LISA will be able to detect GW signals of around $10 - 10^3$ EM-RIs per year up to distances of $z \lesssim 1$ (Gair et al., 2004 and Hopman and Alexander, 2006).

The detection and analysis of EMRI signals will not be an easy task. These signals will be hidden in the LISA instrumental noise and in the GW foreground produced mainly by compact binaries in the LISA band. Thus, in order to extract the EMRI signals we need a very accurate theoretical description of the gravitational waveforms. The main difficulty is the description of the gravitational backreaction, that is, the effect of the SCO's gravitational field on its own trajectory. This backreaction mechanism is the responsible for the inspiral of the SCO into the MBH. The extreme mass ratios involved imply the existence of very different spatial scales (e.g. the size of the SCO as compared to the size of the MBH) and also of very different time scales, namely the orbital time scale versus the inspiral time scale: $T_{\text{orbital}}/T_{\text{inspiral}} \sim \mu$. The advantage of having an extreme mass ratio in the problem is that we can assume that the spacetime geometry is the one due to the MBH with deviations produced by the SCO. Then, one can use BH perturbation theory to deal with EMRIs. In this context, the inspiral can be described as the action of a local force, the so-called *self-force*, according to the equation of motion derived by Mino, Sasaki and Tanaka (Mino et al., 1997), and Quinn and Wald (Quinn and Wald, 1997), known as the *MiSaTaQuWa* equation of motion. The MiSaTaQuWa equation can be used as the foundation of a self-consistent scheme to simulate an EMRI event in an *adiabatic* way by coupling it to the equations describing the perturbations produced by the SCO, together with a scheme to identify the regular part of the perturbations responsible for the inspiral (see (Gralla and Wald, 2008 and Pound, 2010b, 2010a) for recent studies on these issues). There is an entire research program devoted to the computation of the self-force carried out by a community that annually gathers at the CAPRA Ranch Meetings on Radiation Reaction. This research is certainly improving our understanding of General Relativity and can be considered as a first contribution of EMRI research to Fundamental Physics. In the last few years, there has been a lot progress in this direction which, among other things, has lead to the first computations of the gravitational self-force for generic orbits around a non-spinning MBH (Barack and Sago, 2009, 2010). Details on the self-force research program and recent advances can be found in the reviews (Barack, 2009 and Poisson, 2004) and references therein. Given the amount of cycles required for LISA EMRIs and their computational cost we cannot expect to generate complete waveform template banks by means of full self-force calculations. Instead, the goal of these studies should be to understand all the details of the structure of the self-force so that we can formulate efficient and precise algorithms to create the waveforms needed for LISA data analysis, perhaps complementing present approximate schemes (Barack and Cutler, 2004, Babak et al., 2007, Drasco and Hughes, 2006 and Yunes et al., 2010).

LISA observations of EMRIs have the potential to make revolutionary discoveries in Astrophysics, Cosmology, and Fundamental Physics. In Astrophysics (Miller et al., 2009) we expect to understand much better the dynamics in galactic centers (mass segregation, resonant relaxation, massive perturbers, etc); to obtain information about the mass spectrum of stellar BHs in galactic nuclei in order to understand the formation of stellar BHs and their relation to their progenitors; to obtain information on the distribution of the MBH spins for masses up to a few million solar masses, which has implications for galaxy formation models; perhaps to detect the inspiral of an Intermediate-Mass BH (IMBH) into a MBH, which will give direct evidence for the existence of IMBHs; etc. Regarding Cosmology, it has been proposed (MacLeod and Hogan, 2008) that precise measurements of the Hubble parameter are possible by correlating LISA EMRI observations (which can be considered as standard *sirens*, with calibration provided by General Relativity, that provide precise measurements of the luminosity distance up to $z \sim 1$) with galaxy redshift surveys, which would provide statistical redshift information of the EMRI events (which cannot be inferred from the GW observations). Applying this idea to a simplified cosmological model, it has been estimated that using 20 or more EMRI events to $z \sim 0.5$, one could measure the Hubble constant to better than one percent precision.

In the case of Fundamental Physics, the potential of EMRIs is based on the fact that the GWs emitted by EMRIs, and in principle detectable by LISA, are quite long ($1-2$ yrs or even more) and have significantly high Signal-to-Noise Ratios (SNRs) (SNR $\gtrsim 30$). Using a typical EMRI system consisting of an SCO with mass $m_{\star} = 10 M_{\odot}$ inspiraling into a MBH with mass $M_{\bullet} = 10^6 M_{\odot}$ at SNR = 30, and based on the last year of inspiral, Barack and Cutler (Barack and Cutler, 2004) have estimated (using GW parameter estimation theory; see appendix in section 6) the following precision in the estimation of the EMRI parameters:

$$\Delta(\ln M_{\bullet}), \quad \Delta\left(\ln \frac{m_{\star}}{M_{\bullet}}\right), \quad \Delta\left(\frac{S_{\bullet}}{M_{\bullet}^2}\right) \sim 10^{-4}, \qquad (1)$$

where $S_{\bullet}$ denotes the MBH spin (for spinning Kerr BHs: $0 \leq |S_{\bullet}/M_{\bullet}^2| \leq 1$), and

$$\Delta e_o \sim 10^{-4}, \quad \Delta\Omega_S \sim 10^{-3}, \quad \Delta\Omega_K \sim 5 \cdot 10^{-2}, \qquad (2)$$

where $e_o$ is the initial eccentricity, $\Delta\Omega_S$ is the error in the angular position of the source in the sky (solid angle), and $\Omega_K$ is the ellipse error in the determination of the MBH spin direction.

Taking into account that most of the EMRI GW emission comes from orbital motion close to the MBH horizon, in the region $r \lesssim 10\, r_{\text{Horizon}}$ (where $r_{\text{Horizon}}$ is the gravitational or horizon radius of the MBH. For spinning BHs, it is given by $r_{\text{Horizon}} = M_{\bullet} + \sqrt{M_{\bullet}^2 - a_{\bullet}^2}$, with $a_{\bullet} = S_{\bullet}/M_{\bullet}$), that is, the orbital motion of the SCO gets fully combed the strong gravitational field region of the MBH. Then, the GWs carry a detailed *map* of the MBH spacetime, i.e., of the MBH multipole moments. In this sense, it is expected (see (Ryan, 1995b, 1997a and Barack and Cutler, 2007)) that LISA will be able to measure $3 - 5$ MBH multipole moments with good accuracy.

Therefore, it is always said that LISA EMRI observations provide a unique opportunity to test the no-hair theorem and also to perform tests of alternative theories of gravity. This is going to be the main subject of the remainder of this contribution.

Although this article focuses on LISA EMRI observations, the discussion that follows can also be relevant in part for ground-based observatories, in particular for advanced (second generation) and third-generation detectors. These observatories are expected to be sensitive to GWs from intermediate-mass-ratio inspirals (IMRIs), where the central massive object can have masses up to $M_\bullet \sim 10^3 M_\odot$ (Brown et al., 2007 and Mandel et al., 2008), which corresponds to the IMBHs. The formation channels for these IMBHs are different from the ones of stellar BHs and also from the ones of MBHs and supermassive BHs (with masses $M_\bullet \gtrsim 10^8 M_\odot$). It is not yet completely certain that IMBHs exist (Miller and Colbert, 2004), although there have been several claims of observational evidence. In the case they exist, the expected mass range is $M_\bullet \sim 10^2 - 10^4 M_\odot$. In the same way as IMRIs, in which an SCO inspirals into an IMBH, would be detectable by future advanced ground observatories, the inspiral of an IMBH into an MBH would be detectable by LISA. The modeling of these systems is difficult in the sense that the approximations made for EMRI modeling are not going to be accurate enough in order to obtain precise IMRI waveform templates for data analysis purposes. In particular, the non-linear gravitational dynamics is going to play an important role and feedback from Numerical Relativity and post-Newtonian theory is going to be required.

## 3   What hypothesis we can try to test with EMRIs?

Let us now discuss what we can actually try to do with LISA EMRI observations and what are the theoretical challenges that we face. First of all, we are going to summarize what is the theoretical status of EMRIs. Until here we have been assuming that EMRIs are composed by a MBH and a SCO (that can also be a BH, or a NS, or a white dwarf, or even a main sequence star). However, although astronomical observations are providing accumulating evidence of the existence of *dark* compact objects at almost all galactic centers (including our own galaxy, the Milky Way), we have no definitive arguments to conclude that these objects must be MBHs as we understand them in theory: Objects whose spacetime geometry is asymptotically flat and contain a region of no escape, the BH region, in the sense that neither particles nor light rays can escape from this region towards null infinity (see (Chandrasekhar, 1992 and Frolov and Novikov, 1997) for exhaustive accounts on different aspects of BH physics).

There are different ways in which MBHs can form: Direct collapse of a massive star; collapse of the core of a star cluster; collapse of overdensities in the early Universe; etc. In General Relativity, it is known that under certain conditions on the matter distribution, the end point of spherical collapse must be a spherical BH described by the Schwarzschild solution (see, e.g. (Wald, 1984)). The Schwarzschild solution has a very special status in General Relativity as it is the only possible vacuum spherically symmetric solution of Einstein's field equations, and hence it describes the exterior of all spherical matter distributions. Regarding nonspherical collapse, the situation is by no means as clear as in the spherical case. In relation to this Thorne (Thorne, 1972) proposed the so-called *hoop* conjecture that says that BHs with horizons form when and only when a mass $M$ gets compacted into a region whose circumference $C$ in every direction is bounded by $C < 4\pi M$.

On the other hand, the final outcome of nonspherical collapse, appealing to the singularity theorems (see (Hawking and Ellis, 1973)), must be a singularity. But the question now is whether this singularity is covered by a horizon and hence the end point of the collapse is a BH. This is what is called the *cosmic censorship* conjecture and a satisfactory formulation and proof of it remains as the open key issue in gravitational collapse (see (Wald, 1984) for a detailed discussion of these issues). In this sense, it is worth mentioning that the hoop conjecture is compatible with the appearance of singularities without horizons, known as *naked* singularities.

Assuming that the cosmic censorship conjecture is correct, we should expect that the final state of the collapse is a stationary BH. It turns out that in General Relativity there is a theorem (actually a series of theorems; see (Hawking and Ellis, 1973 and Misner et al., 1973) for details) that asserts that the only vacuum stationary BH solutions are those described by the Kerr family (Kerr, 1963) of BHs.

One important characteristic of the Kerr spacetime geometry is that is it fully determined by two numbers, the BH mass $M_\bullet$ and the BH intrinsic angular momentum (spin) $S_\bullet$ (also the charge $Q_\bullet$ in the case of charged BHs described by the Kerr-Newman spacetime geometry (Newman et al., 1965), although we expect those BHs to discharge very quickly in astrophysical situations).

This leads to the so-called *no-hair* theorem (although it should be called conjecture since it has not been proven), associated with Wheeler (Ruffini and Wheeler, 1971) as well as the BH denomination, which postulates that all BH solutions of the Einstein (Einstein-Maxwell) equations, in General Relativity, can be completely characterized by only two (three) externally observable parameters: Mass $M$ and spin $S$ (and electric charge $Q$). All other information (to what the word *hair* refers to) about the matter fields that evolved towards the formation of a BH are erased (they either disappear behind the horizon or are radiated away). It can be said that $M$ and $S$ (and $Q$) are unique conserved quantities associated with the external gravitational field of the BH.

From this discussion, there is an obvious hypothesis that we want to test by means of LISA EMRI observations:

**H1:** *Kerr Hypothesis*: The dark, compact and very massive objects sitting at the galactic centers are stationary BHs described by the Kerr solution of the General Theory of Relativity.

This hypothesis, which assumes the validity of General Relativity, is equivalent to the no-hair theorem in the sense that it is saying that the collapsed objects at the galactic centers are BHs, and the geometry of these BHs is described well by the Kerr metric. Now, the question is how to test this hypothesis in practice. It is clear that we must have a method to discriminate between the Kerr geometry and other possible geometries for collapsed objects from the EMRI signals present in the future LISA data stream (see appendix in section 6 for a review on the theory of parameter estimation in Gravitational Wave Astronomy). To that end, we are going to assume that whatever the geometry of the dense objects at the galactic centers is, it is going to be stationary, or in other words, independently of how these objects were formed they have settled down to a final state whose *spatial* geometry is time independent to a high degree of precision. We also assume that the exterior geometry of these objects can be considered to be a vacuum and asymptotically flat geometry. The first of these two assumptions basically means that although they are surrounded by dark matter and they are embedded in an expanding universe whose accelerated expansion may be due to some sort of dark energy, the local density of these energy components produces a spacetime curvature that is negligible when we compare with the situation in which the BH is in a truly vacuum spacetime. The second assumption means that for practical purposes the objects are isolated enough from other massive objects.

Then, we are in a situation in which the goal is to distinguish between different stationary and asymptotically-flat geometries. In Newtonian theory, the gravitational potential of an isolated body can be characterized by an infinite set of multipole moments. It turns out that in General Relativity there is a similar result for vacuum, stationary, and asymptotically-flat spacetimes that can be used for our purposes: It has been shown that any stationary asymptotically-flat vacuum spacetime geometry is fully determined by an infinite set of multipole moments (actually two in comparison with the Newtonian case). These multipole moments were introduced first by Geroch (Geroch, 1970) for the case of static spacetimes, and later by Hansen (Hansen, 1974) for stationary spacetimes. Relevant results regarding these multipole moments are: Xanthopoulos (Xanthopoulos, 1979) proved that a vacuum, stationary, and asymptotically-flat spacetime is static if and only if all its angular moments vanish; moreover, he proved that a vacuum, static, and asymptotically-flat spacetime is flat if and only if all multipole moments vanish; Beig and Simon (Beig and Simon, 1980) proved for the static case that two solutions with the same multipole moments are identical at least in some neighborhood of infinity. Kundu (Kundu, 1981) and Beig and Simon (Beig and Simon, 1981) also proved that the multipole moments uniquely determine the local structure of a stationary, asymptotically flat, vacuum metric. Gürsel (Gürsel, 1983) showed that a stationary spacetime is axisymmetric if and only if all its multipole moments are axisymmetric.

On the other hand, in the context of gravitational-wave theory, Thorne (Thorne, 1980) introduced a multipolar expansion of a stationary, and asymptotically-flat spacetime metric. This expansion is a special case of the multipolar expansion of Thorne for the gravitational radiation emission of slow-motion sources of GWs. From this expansion one can read off the multipole moments that characterize the spacetime geometry. The equivalence between the multipole moments of Geroch and Hansen and those of Thorne was established by Gürsel (Gürsel, 1983).

While the Geroch and Hansen formalism is a geometrical one, it is based on the analysis of the three-dimensional space associated with the stationarity of the spacetime and the conformal structure of the induced 3-metric and has the elegance of being *general covariant* (i.e. it does not rely on any special choice of coordinate system), the Thorne one is based on expansions which are valid in spacetime regions whose geometry is *close* to the Minkowski flat geometry and is expressed in a particular type of coordinates, namely asymptotically Cartesian and mass centered (ACMC) coordinates (de Donder/harmonic coordinates are ACMC coordinates). Moreover, when the compact body is not a BH, and its interior can be covered by de Donder coordinates, the Thorne multipole moments are related to integrals over the body of an *effective* stress-energy tensor (Thorne, 1980) (see also (Landau and Lifshitz, 1971 and Misner et al., 1973)).

Since the approach of Thorne is more intuitive and more closely related to gravitational-wave theory, we use it here to show the way in which the multipole moments can be introduced. For any stationary and asymptotically-flat spacetime, the form of the metric tensor in ACMC coordinates (which also defines this type of coordinate systems), $\{t, x^i\}$ ($i = 1, 2, 3$), can be written as[2]:

$$g_{tt} = -1 + \frac{2M}{r} + \sum_{\ell=2}^{\infty} \frac{1}{r^{\ell+1}} \left[ \frac{2(2\ell-1)!!}{\ell!} \mathcal{M}_{A_\ell} N_{A_\ell} + \mathcal{R}_{\ell-1} \right] ,$$

$$g_{ti} = \sum_{\ell=1}^{\infty} \frac{1}{r^{\ell+1}} \left[ -\frac{4\ell(2\ell-1)!!}{(\ell+1)!} \epsilon_{ija_\ell} \mathcal{J}_{jA_{\ell-1}} N_{A_\ell} + \mathcal{R}_{i;\ell-1} \right] , \qquad (3)$$

$$g_{ij} = \delta_{ij} + \sum_{\ell=0}^{\infty} \frac{1}{r^{\ell+1}} \mathcal{R}_{ij;\ell-1} ,$$

where we are using the multi-index notation of Thorne (Thorne, 1980): $A_\ell = a_1 a_2 \cdots a_\ell$, and with this $Q_{A_\ell} = Q_{a_1 a_2 \cdots a_\ell}$. There is summation over repeated (multi-)indices and $N_{A_\ell} \equiv n_{a_1} n_{a_2} \cdots n_{a_\ell}$ with $n_i \equiv x^i/r$ ($r^2 \equiv \delta_{ij} x^i x^j$). In these expressions, $\epsilon_{ijk}$ is the usual flat-space Levi-Civita antisymmetric tensor. There are two sets of multipole moments, the *mass* multipole moments, $\mathcal{M}_{A_\ell} = \mathcal{M}_{a_1 a_2 \cdots a_\ell}$, and the *current* multipole moments, $\mathcal{J}_{A_\ell} = \mathcal{J}_{a_1 a_2 \cdots a_\ell}$. The multipole moments are symmetric and trace-free quantities that are invariant under coordinate transformation from an ACMC coordinate system to another one. The terms $\mathcal{R}_\ell$, $\mathcal{R}_{i;\ell}$, and $\mathcal{R}_{ij;\ell}$ are quantities that only depend on angles (the two angles that appear in the standard transformation from Cartesian to spherical coordinates), and this dependence can be expressed in terms of spherical harmonics of order $\leq \ell$. The multipole moments of Geroch and Hansen, which we denote by $M_{A_\ell} = M_{a_1 a_2 \cdots a_\ell}$ and $J_{A_\ell} = J_{a_1 a_2 \cdots a_\ell}$, are related to the Thorne ones by the following simple relations (Gürsel, 1983):

$$\mathcal{M}_{a_1 a_2 \cdots a_\ell} = \frac{1}{(2\ell - 1)!!} M_{a_1 a_2 \cdots a_\ell},$$

$$\mathcal{J}_{a_1 a_2 \cdots a_\ell} = \frac{\ell + 1}{2\,\ell\,(2\ell - 1)!!} J_{a_1 a_2 \cdots a_\ell}.$$

---

[2] We use the notation: $n!! \equiv n(n-2)(n-4) \cdots (2 \text{ or } 1)$.

Up to now we have only assumed that the spacetime is stationary and asymptotically flat (and that the exterior of the central object can be considered to be vacuum). Things simplify significantly when we assume that in addition the spacetime is axisymmetric, which means there is an extra symmetry which, as the stationarity is described by a Killing vector field. This is the case of the Kerr spacetime and is also expected to be the case for most compact massive astrophysical objects. The axial symmetry has associated an axis of symmetry (the rotation axis in the case of a compact object), which has a direction associated with it. The multipole moments, like other geometrical quantities, are invariant under the axial symmetry. Since the multipole moments are invariant under this symmetry (see (Hansen, 1974) for details), they must be multiples of the symmetric, trace-free products of the axis vector with itself. Therefore, the multipole moments introduced above can be fully determined by two sets of numbers, $M_\ell$ and $J_\ell$, defined by (in terms of the Geroch and Hansen multipole moments[3]):

$$M_\ell = -\frac{1}{\ell!} M_{A_\ell} Z_{A_\ell},$$
$$J_\ell = \frac{1}{\ell!} J_{A_\ell} Z_{A_\ell}, \tag{4}$$

where $z^i$ is the (unit norm) axis vector and $Z_{A_\ell} = z_{a_1} z_{a_2} \cdots z_{a_\ell}$. Axisymmetric configurations provide, in addition to the axial symmetry that we have already exploited, a reflection symmetry. This reflection symmetry reverses the sign of the axis vector, the mass multipole moments are invariant, and the current multipole moments are also reversed in sign. As a consequence, the non-vanishing multipole moments are the $\ell$-even mass moments and the $\ell$-odd current moments. There is an algorithm, given an stationary axisymmetric and asymptotically-flat vacuum metric, for computing the multipole moments that was introduced in (Fodor et al., 1989). This method can also be used in principle to reconstruct a spacetime from the knowledge of the set of multipole moments. Finally, Ashtekar and collaborators (Ashtekar et al., 2004) have introduced mass and multipole moments for axisymmetric *isolated horizons*, which provide a diffeomorphism invariant characterization of the horizon geometry. These moments have been interpreted as the *source multipoles* of BHs in equilibrium and hence, they are potentially interesting for the purposes described in this article.

---

[3] We have introduced a minus sign for the $M_\ell$'s in order to get a positive monopole [see Eq. (5)].

In the particular case of a Kerr BH, it turns out that the two sets of multipole moments of Eqs. (4) satisfy the following elegant relation:

$$M_\ell + iJ_\ell = M_\bullet \ (ia_\bullet)^\ell \ , \tag{5}$$

that is, $M_{2\ell} = (-1)^\ell M_\bullet \ a_\bullet^{2\ell}$ and $J_{2\ell+1} = (-1)^\ell M_\bullet \ a_\bullet^{2\ell+1}$, and the rest vanish identically as expected from the reflection symmetry. Note that in the general case, the number of non-vanishing multipoles is infinite, whereas in the non-spinning case (Schwarzschild static spacetime, $a_\bullet = 0$) the only non-vanishing multipole is the first one, namely the mass monopole $M_0 = M_\bullet$. The remarkable feature (although expected from our previous discussion of the no-hair theorem) is that all the multipole moments depend only on the BH mass and spin. That is, only $M_0$ and $J_1 = M_\bullet a_\bullet$ are independent moments, the rest is a combination of them. For instance, the next mass and current multipoles can be related to these two by the following simple relations:

$$-M_\bullet a_\bullet^2 = M_2 = -\frac{J_1^2}{M_0} \ ,$$
$$-M_\bullet a_\bullet^3 = J_3 = \frac{M_2 J_1}{M_0} \ . \tag{6}$$

Then, after all this discussion on multipole moments, we can change our hypothesis **H1** by another similar one that is more amenable to be tested by LISA EMRI observations:

**H1'**: *Kerr Hypothesis*: The exterior gravitational field of the dark, compact and very massive objects sitting at the galactic centers can be well described by the vacuum, stationary, and axisymmetric solutions of the General Theory of Relativity whose multipole moments $\{M_\ell, J_\ell\}_{\ell=0,...,\infty}$ satisfy the Kerr relations of Eq. (5).

The question now is whether we can do this in practice and how. Here, we have to mention previous work by Poisson (Poisson, 1996), which can be seen as a first attempt at attacking the hypothesis **H1**, where using a simplified model (the inspiral is a sequence of circular orbits in the equatorial plane of the MBH) estimates of the accuracy with which the main EMRI parameters (mass ratio and MBH mass and spin) can be measured were presented (based on the techniques summarized in the appendix, in section 6). Poisson's work has a certain independence from the theory or gravity which allowed some discussion on how to use the techniques of this paper to even test the theory of gravity.

The first attempt to look at the hypothesis **H1'** was done by Ryan in a series of papers (Ryan, 1995a, 1995b, 1996, 1997a, 1997b). In this work we can distinguish two parts. In the first one, Ryan (Ryan, 1995b) dealt with the question of to what extend the GW emission of EMRIs depend on the values of the different multipole moments of the central massive object. In principle, we expect that EMRI GWs have a strong dependence on them due to the simple fact that the SCO spends a large number of cycles in the strong field region of the MBH. However, apart from this intuition there was no other argument that supports this expectation before Ryan's work. In his study, Ryan makes certain simplifications of the EMRI scenario. He considers very nearly circular and very nearly equatorial orbits and the inspiral takes place in a slowly and adiabatic manner (avoiding analyzing the GW emission in detail, a very complex task), in which the orbit can be seen as a series of geodesics, and the central object is assumed not to be affected by the orbital evolution. Then, three quantities that depend on the GW emission are analyzed: (i) The GW energy spectrum, $\delta E(f)$, i.e. the energy emitted per unit logarithmic frequency interval ($\delta E(f) \equiv f \, dE_{GW}/df$); (ii) the frequency of orbital precession (due to the small eccentricity), $\Omega_p$; and (iii) the frequency of precession of the instantaneous *orbital plane* with respect to the equatorial plane (due to the small inclination), $\Omega_z$. By doing a power series expansion of these three quantities in the variable[4] $v = (\pi M f)^{1/3}$, where $f$ is the dominant GW frequency (twice the orbital frequency), it was found that the coefficients of this power series, for which an explicit computational algorithm was derived, can be expressed entirely in terms of the multipole moments $\{M_\ell, J_\ell\}$ of the central object. The GW phase $\Phi(t)$, a key observable GW quantity (in contrast to the three previous quantities) that can be measure with high precision by most GW observatories, was also analyzed. However, things are much more involved for this quantity as it is obvious that it depends on the details of the GW generation mechanism, and hence it is quite difficult to find clear relations between the multipole moments and the GW phase. Nevertheless, for the purely circular and equatorial case it is shown that the number of cycles that the dominant GW components spend in a logarithmic frequency interval, $\delta N(f) \equiv f^2/(df/dt) = f \, \delta E(f)/(dE_{GW}/df)$, which contains equivalent information to the GW phase $\Phi(t)$, contains full information of the whole set of multipole moments. Moreover, it is shown how to extract at least the first three moments, $(M_0, J_1, M_2)$, which is enough in principle to perform a partial test of our hypothesis **H1'** [essentially, a check of Eq. (6)]. However, as it has already been advanced, to construct an algorithm to read off the multipole moments from the GW phase evolution would require to elaborate the full complexities of the GW generation theory which, as we have discussed above, is an ongoing research program that has still several milestones to be reached.

---

[4] In the Newtonian limit $v$ is the norm of the SCO's velocity.

The second part of Ryan's study (Ryan, 1997a) is a more pragmatical work in which the goal is to estimate the accuracy of LIGO and LISA in determining the multipole moments from the GWs emitted by an IMRI (LIGO) or an EMRI (LISA). His works uses the parameter estimation theory summarized in the appendix, in section **6**. Then, the basic idea is to take an EMRI waveform template model that includes a certain number of multipole moments in the set of source parameters $\vec{\theta}$. In order to look at this in more detail, let us call $\vec{\theta}_K$ to the parameters of EMRIs consisting of a Kerr MBH as the central massive body. For generic EMRI configurations we need 14 parameters that can be taken to be (Barack and Cutler, 2004): (1) A fiducial time $t_o$; (2) $\ln m_\star$; (3) $\ln M_\bullet$; (4) $S_\bullet/M_\bullet^2$; (5) $e_o = e(t_o)$, the eccentricity at the fiducial time; (6) $\tilde{\gamma}_o$, where $\tilde{\gamma}(t)$ is the pericenter angle in the orbital plane; (7) $\Phi_o$, where $\Phi(t)$ is the mean anomaly; (8) and (9) $\theta_S$ and $\phi_S$, sky location of the source; (10) $\hat{\mathbf{L}} \cdot \hat{\mathbf{S}}_\bullet$ (=const.), where $\hat{\mathbf{L}}$ and $\hat{\mathbf{S}}_\bullet$ are unit vectors in the direction of the orbital angular momentum and MBH spin respectively; (11) $\alpha_o$, where $\alpha(t)$ is the azimuthal direction of $\hat{\mathbf{L}}$; (12) and (13) $\theta_K$ and $\phi_K$, polar and azimuthal angles of the MBH spin; (14) $\ln(m_\star/D)$, where $D$ is the distance to the source. Then, we need to add extra parameters describing a number of multipole moments (excluding obviously $M_0$ and $J_1$), and we call these parameters $\vec{\theta}_M$. In this way, the modeled signals in the LISA detector are going to be time series of the form: $\mathbf{h}(t; \vec{\theta})$, where the parameters $\vec{\theta} = (\theta^I)$ will be: $\theta^I = \theta_K^I$ for $I = 1, \ldots, 14$, and $\theta^I = \theta_M^I$ for $I = 15, \ldots, 14 + N_{mm}$, where $N_{mm}$ is the number of multipole moments included in the EMRI modeling. The waveform model of Ryan is essentially the circular and equatorial model used in his description of the GW phase of the first part of the work here described (this reduces the 14 parameters of a generic EMRI system to only 7). Then, only the leading order effects of each multipole in the waveforms are considered. Another important ingredient are the priors for the multipole moments considered in the analysis, which try to favor models appropriate for compact bodies of a characteristic size, and which have an impact in the results for the estimated accuracies. In order to compute the Fisher and variance-covariance matrices (see appendix in section **6**) Ryan adopts a simplified model for the LISA noise that, for instance, does not include the confusion noise generated by the foreground of millions of galactic compact binaries. The conclusions of Ryan's computations are optimistic for LISA and open the door to the possibility of testing the hypothesis introduced above. In order to illustrate his results, let us mention that for the particular case of an SCO with mass $m_\star = 10 M_\odot$ inspiraling into a central massive object of mass $M = 10^5 M_\odot$, during 2 years before plunge and with an SNR of 10, it is found that LISA could measure $M$ with a fractional error of $\Delta M/M \sim 10^{-3}$, $m_\star$ with fractional error $\Delta m_\star/m_\star \sim 10^{-3}$, for the spin Ryan finds $\Delta(S/M^2) \sim 10^{-3}$, and finally for the mass quadrupole moment the accuracy is $\Delta(M_2/M^3) \sim 0.5$. These errors change with the number of multipole moments considered in the analysis, although a tendency to a stabilization of the errors with increasing $N_{mm}$ is observed.

Recently, Barack and Cutler (Barack and Cutler, 2007) conducted a study in similar lines as the second part of Ryan's work. There are several differences, most of them improvements. First of all, the waveform model of Barack and Cutler is the same as the one (Barack and Cutler, 2004) used for the estimations of standard Kerr EMRIs given in Eqs. (**1**) and (**2**). In this model, the EMRI system follows at any instant a Newtonian orbit emitting lowest-order quadrupolar GWs (essentially the Peters-Matthews waveforms (Peters and Mathews, 1963)), but contains PN modifications to secularly evolve the parameters of the orbit in such a way that the model contains all the features that we expect from generic EMRIs. These waveforms are currently known as Analytic *Klugde* (AK) waveforms, referring to the fact that they are semi-analytic models that capture all the ingredients of EMRIs but are not good enough for data analysis purposes, and also to distinguish them from other kludge waveform models that have appear more recently in the literature. Regarding the LISA model, Barack and Cutler also used a more detailed treatment as compared with Ryan's work, including the main features of the LISA constellation motion. The LISA noise model is also more complete, including the confusion noise from galactic and extragalactic binaries. However, regarding the inclusion of multipole moments, they only include the effects of the mass quadrupole $M_2$. With all these ingredients, the parameter estimation analysis of Barack and Cutler (Barack and Cutler, 2007) yields the following result for the errors on the mass quadrupole [same type of system as in Eqs. (**1**) and (**2**) but now with SNR = 100]:

$$\Delta\left(\frac{M_2}{M_0^3}\right) \sim 10^{-4} - 10^{-2},\qquad(7)$$

which is a considerably better error estimate than Ryan's estimate. This is mainly due to the full complexity of the EMRI dynamics encoded in the AK waveform model of Barack and Cutler (Barack and Cutler, 2004).

The program initiated by Ryan, although it is far from being complete, has been continued by other authors. Apart from the work of Barack and Cutler (Barack and Cutler, 2007) that we have just discussed, Collins and Hughes (Collins and Hughes, 2004) initiated a program to construct exact General Relativistic solutions that are almost BHs (they have been named *bumpy* BHs), with the difference that they have some multipole moments with the *wrong* value. Apart from this, these bumpy BHs have an exterior that is stationary, axisymmetric, asymptotically flat and vacuum. In the work of Collins and Hughes (Collins and Hughes, 2004) they build the geometry of bumpy BHs with no angular momentum (bumpy Schwarzschild BHs) by perturbing the hole with a pair of negative masses at the poles and also by a ring of positive mass around its equator. An important feature of these geometries is that they are valid into the strong field of the object. Then, they show that the hole's bumpiness is encoded in the strong-field periapsis precession. These bumpy BHs may be used for performing *null* experiments, i.e. by comparing their properties with measurements of astrophysical sources. The drawback of these solutions is that the changes introduced are not smooth and leads to a somehow pathological strong-field structure. Nevertheless, this work has been extended recently by Vigeland and Hughes (Vigeland and Hughes, 2010) in two ways. First, to fix the mentioned problems of the Collins and Hughes construction and, second, to introduce angular momentum (bumpy Kerr BHs). More recently, a method to compute the multipole moments of these bumpy BHs has been presented (Vigeland, 2010). In a related work, Glampedakis and Babak (Glampedakis and Babak, 2006), using the exterior Hartle-Thorne rotating geometry (Hartle, 1967 and Hartle and Thorne, 1968), have constructed non-Kerr geometries and have analyzed their geodesic orbits. They concluded that the deviations from Kerr induced from the changes in the multipole moments may be difficult to detect and also pointed out that there may be a confusion problem between waveforms of the specific non-Kerr metric and a Kerr metric with different orbital parameters. However, their analysis was based only on geodesics, without including radiation reaction effects. Thus, in order to reach definite conclusions an extended analysis including those effects would be required. Generalization of the work of Ryan to include possible electromagnetic fields of the source was done in (Sotiriou and Apostolatos, 2005).

After all this discussion on the multipolar structure associated with the geometry of stationary, axisymmetric, asymptotically flat and vacuum solutions that describe the exterior of compact objects, its importance for EMRI dynamics, and its use via LISA observations for testing the geometry of the massive objects sitting at galactic centers, we can conclude that it seems a realistic task with good prospects in terms of the precision that LISA is expected to achieve. Then, it seems also reasonable, for the same price that we can test the hypothesis **H1′**, to perform tests of the following more general hypothesis:

**H2:** *X-Body Hypothesis*: The exterior gravitational field of the dark, compact and very massive objects sitting at the galactic centers can be well described by the vacuum, stationary, axisymmetric, and asymptotically-flat solutions of the General Theory of Relativity whose multipole moments $\{M_\ell, J_\ell\}_{\ell=0,...,\infty}$ satisfy the following set of relations: $\mathcal{R}_a[\{M_\ell, J_\ell\}] = 0 \quad (a = 1, \dots)$.

A justification for this relies on the fact that despite of the well-known theorems in General Relativity regarding the outcome of the gravitational collapse and the formation of singularities covered by BH horizons, there is still room to propose models for the gravitational field different from the Kerr BH solutions of General Relativity. These models circumvent these theorems by not satisfying one or several of their premises. Normally, these non-BH models do not satisfy the energy-momentum conditions assumed in these theorems, which means that they are formed by some form of *exotic* matter (supported by some exotic physics). It is also possible to circumvent the topological/geometrical assumptions. In any case, independently of the possibly weak grounds in which some of these models may be based and the skepticism they may originate in the community, it is still interesting to study them from the point of view of this article in other to rule them out or to confirm their existence. Among the candidates that have been proposed in the literature we mention here a few: Boson stars (see (Jetzer, 1992 and Schunck and Mielke, 2003) for reviews and (Amaro-Seoane et al., 2010)), fermion *balls* (Munyaneza and Viollier, 2002), *gravastars* (Mazur and Mottola, 2001, 2004), naked singularities (see, e.g. (Nakamura et al., 1988, Shapiro and Teukolsky, 1991 and Christodoulou, 1994)), *dark energy* stars (Chapline et al., 2001), different quantum modifications of the spacetime geometry in the strong field region (near of what would be the classical BH horizon) that would lead to objects different from BHs (Chapline et al., 2003 and Barcelo et al., 2008), etc. Some of these models have been compared with data from measurements of the candidate for MBH at our galactic center, Sgr A*, which puts stringent constraints on these models, to the point that some of them are practically ruled out or close to (see (Broderick and Narayan, 2006) for a discussion on the nature of Sgr A*).

There have been other proposals, not based on the use of the multipole moments of the central massive object, to try to distinguish a Kerr geometry from a different one, or at least to characterize trends that can be found in the LISA EMRI GWs and that can either prove or disprove the Kerr hypothesis. Drasco (Drasco, 2009) analyzed the GWs for about $\sim 10^4$ EMRI generic orbital configurations using Teukolsky's BH perturbation formalism (Teukolsky, 1972, 1973). Each binary radiates power exclusively in modes with frequencies that are integer-linear-combinations of the orbit's three fundamental frequencies (radial, polar, and azimuthal; see (Schmidt, 2002 and Drasco and Hughes, 2004) for the definitions of these frequencies and how they can be used). General spectral properties are found: 99% of the radiated power is carried by a few hundred modes; the dominant frequencies can be grouped into a small number of families defined by fixing two of the three integer frequency multipliers; and the specifics of these trends can be qualitatively inferred from the geometry of the orbit under consideration. Kesden, Gair, and Kamionkowski (Kesden et al., 2005) investigated the possibility of distinguishing between a central MBH and a boson star from signatures in the waveforms produced by the inspiral of an SCO, given that for a BH the waveform will end at the plunge whereas for the boson star it will continue until the SCO reaches its center. In (Barausse et al., 2007), a more astrophysically oriented non-Kerr geometries were constructed by adding to the BH a self-gravitating and homogeneous compact torus with comparable mass and spin. It was found that for most configurations the EMRI waveforms produced in these systems are indistinguishable from pure-Kerr waveforms, indicating a possible confusion problem. In (Barausse and Rezzolla, 2008) the dissipative effect of the hydrodynamic drag exerted by the torus on the SCO has been found to be much smaller than the corresponding one due to radiation reaction, although there may be configurations in which these two effects are of similar magnitude. In another study by Gair, Li, and Mandel (Gair et al., 2008), which also goes along the lines of using bumpy BHs, a particular exact solution of the Einstein equations was studied, namely the Manko-Novikov solution (Manko and Novikov, 1992), whose multipoles differ from those of the Kerr solution starting already at the mass quadrupole. One of the outcomes of this study is that, due to the lost of integrability of the geodesic equations that come from certain configurations of this class of spacetime geometries, certain regions of the phase space associated with the orbital motion exhibit chaotic behavior. This chaotic behavior may induce observable signatures in the EMRI waveforms. This work was extended by Apostolatos, Lukes-Gerakopoulos and Contopoulos (Apostolatos et al., 2009 and Lukes-Gerakopoulos et al., 2010), where it was shown that in the Manko-Novikov solutions, seen as perturbations of the Kerr solution, the phase space tori (corresponding to the Kerr integrable motion) whose associated fundamental frequencies (two of them or the three) are commensurable (their ratio is a rational number) disintegrate so as to form a chain of the so-called *Birkhoff islands*, where all orbits share the same ratio of frequencies. This constancy of the ratio of frequencies during the evolution of the GW signal can be seen as a fingerprint of a non-Kerr geometry that may be distinguished in the future LISA EMRI observations. Other questions related to the integrability of the geodesic orbits and the impact to the EMRI problem have been discussed recently by Brink in a series of papers (Brink, 2008b, 2010a, 2010b). Finally, tests of the no-hair theorem have been proposed in the electromagnetic spectrum (Johannsen and Psaltis, 2010a, 2010b).

Nevertheless, and despite the good prospects that come from all the discussion of the work that has been done until now, there is still an important question that we have to face and solve: How are we going to carry out this program once we get the real LISA data? This is a practical question and essentially is the question of how are we going to produce the necessary waveform models to perform the estimation of the physical parameters (see appendix in section 6) necessary to test the different hypothesis we have listed before. This is a crucial point since once we have the LISA data stream, containing GW signals from a number of different sources, the EMRI signals are expected to be, at least, about an order of magnitude below LISA's instrumental noise. This means that to extract these signals and estimate the physical parameters is going to be a difficult task and it will require very precise GW waveform templates. It has been estimated that the error in the GW phase of these templates with respect to the true signals should be smaller than 1 radian over 1 year. Given that such a waveform can contain of the order of $10^5$ cycles or more, this supposes a considerable theoretical challenge for the modeling of EMRI waveforms. We have already commented above the basics and present status of EMRI modeling. Here, we just want to recall what are the main ingredients of the modeling in order to discuss the challenges that we face in order to be in a position to test the hypothesis **H1'** and **H2**. In the perturbative general-relativistic framework in which EMRIs are modeled, and given the smallness of the mass ratios of relevance, it is believed that the corrections to the geodesic orbital motion should be locally small. Nevertheless, these corrections build up in time and at some point the SCO has deviated significantly from the original geodesic orbit that approximated its motion. At that point, it should be close to another geodesic, which should give a better approximation to the motion for another interval of time. Then, we can try to patch all the solutions corresponding to all these intervals of time for which different fiducial geodesic orbits have been used in order to construct a worldline that describes the orbital motion for a longer time. The worldline that would result from shrinking the duration of these time intervals to zero should be governed by a *self-consistent* differential equation that approximates well the motion as long as it remains locally close to geodesic motion. The notion of self-consistency here has been discussed in detail by Gralla and Wald in (Gralla and Wald, 2008), where they suggest that the MiSaTaQuWa equation may be such an equation, in which case it they should provide an accurate and global (in time) solution to the EMRI modeling problem. Nevertheless, it is still uncertain to what extend the accuracy that one would achieve is good enough for the purposes of doing LISA science, and whether we still should care for second-order effects (or even higher) in our perturbative schemes. In any case, and for the purposes of the present discussion, let us assume that the MiSaTaQuWa formulation (or a similar one) of the EMRI dynamics is good enough for the purposes of doing science with LISA EMRI observations. This formulation has two main ingredients: (i) The computation of the gravitational metric perturbations generated by SCO in the *background* of the MBH, and (ii) The solution of the equation of motion, which has the form of the geodesic equation of motion but with an added force term (the gravitational self-force). Obviously, the most complicated part in computational terms is point (i), since point (ii) can be reduced to the solution of ordinary differential equations. We have already mentioned that in the General Relativistic case and for the case where the central massive body is a MBH, the non-spinning (Schwarzschild) case is presently under control, and progress is being made towards the solution of the spinning (Kerr) case. Since the perturbative equations are partial differential equations which need to be solved using complex numerical methods, the solution of them takes a substantial amount of computer time. As a consequence, it is not feasible to produce full template banks of EMRI gravitational waveforms (that cover the whole parameter space $\vec{\theta}$ of the EMRI system, 14-dimensional in the case of MBHs in General Relativity) using these methods. Instead, one can use these studies to understand the details of the structure and properties of the self-force in such a way that we can use this information to design and construct efficient and precise algorithms to create the waveforms needed for LISA EMRI data analysis. There are already several schemes (Barack and Cutler, 2004, Babak et al., 2007, Drasco and Hughes, 2006, Yunes et al., 2010 and Sopuerta and Yunes, 2010), among them the kludge waveform models, that can be used as possible seeds to build the desired algorithms.

All these ingredients for EMRI modeling that we have just described refer to the standard case, in which the central compact object is a MBH described by the Kerr metric. However, to test the different hypothesis that we have considered, we need to extend the modeling in two ways: (i) To consider objects different from Kerr BHs.

(ii) To extend the parameter space to include multipole moments (or equivalent information about the central compact object). Point (i) implies a number of changes since different geometries lead to different perturbative equations and also to different equations for the geodesic orbital motion. In this respect, we have to take into account the special character of the Kerr geometry in the sense that it has certain symmetries that lead to simplifications of these equations. Indeed, as we have already mentioned, the geodesic equations of motion are completely integrable as the associated Hamilton-Jacobi equation can be completely separated (Carter, 1968 and Carter, 1968), which is associated with the existence of a 2-rank Killing tensor symmetry. This is very likely to change for other geometries as it has already been found by several authors and has been mentioned above. Moreover, this extra symmetry, together with the special structure of the Kerr curvature tensor (it is Petrov type D; see (Stephani et al., 2003) for an exhaustive summary of the Petrov classification of the Weyl curvature tensor) makes possible, as it was found first by Teukolsky (Teukolsky, 1972, 1973), to decouple the perturbative equations in terms of a single complex equation for a single complex variable, $\Psi_4$ (a component of the Weyl tensor in a complex basis), that describes all the gauge-invariant information about the perturbation of a Kerr BH. Again, this decoupling of the perturbative equations is likely not to happen in other geometries, certainly we know it is not going to happen in geometries that are not in the Petrov type D class.

From this discussion we can see that extending, within General Relativity, the EMRI modeling to non-Kerr spacetimes may be a quite difficult task. Nevertheless, we do not want to advocate here a pessimistic view of the problem. There are some avenues that one can follow. For instance, it would be very useful to get more insight into the possible outcomes of the gravitational collapse in General Relativity (even for cases where exotic matter or pathologies in the spacetime topology/geometry are allowed), specially it would be very interesting to learn about the possible geometries that would result of it and their structure. This can provide a very valuable guidance for the study of bumpy BHs, since it is not clear at all that the particular features that some of the cases studied in the literature exhibit are going to be present in physically realistic cases. In other words, perhaps not all possible combinations of the multipole moments $\{M_\ell, J_\ell\}$ are acceptable as the moments of collapsed objects, instead, it may be possible that only certain restrictive classes of moments are physically realistic. Another possibility is to advance further in the development of kludge waveforms. There are two types of benefits from this line of research. First, as we produce better EMRI waveforms we can obtain better parameter error estimates, and in this way we can confirm and improve the estimates already given by Barack and Cutler (Barack and Cutler, 2004, 2007). In this sense, the computational algorithms to produce these waveforms are less complex and sometimes they are related to approximations that can be connected in a relatively simple way with the multipole moments (here the use of the ACMC coordinate systems can help (Sopuerta and Yunes, 2010)). Therefore, they are a convenient tool to continue the work on the study of the error estimation for multipoles moments higher than the mass quadrupole $M_2$, and try to confirm and improve the results obtained by Ryan (Ryan, 1997a). Second, as we have already mentioned, improving the present kludge waveform algorithms may be the way to create fast algorithms for EMRI waveform production for data analysis purposes. In this respect, these algorithms should provide simple mechanisms to incorporate relevant information that can be obtained from other tools in General Relativity, like Numerical Relativity and/or post-Newtonian theory (recently, the calculations of the perturbative approach have been successfully compared with the post-Newtonian predictions (Blanchet et al., 2010a, 2010b), and also with the Effective-One-Body formalism (Barack et al., 2010)). In any case, the design of good data analysis strategies for the search of EMRI signals (see (Cornish, 2008) for a discussion) will also have an impact on the theoretical side. For instance, some strategies may only require precise waveforms for a restricted time interval, making the modeling easier than it would be otherwise. Finally, another feasible avenue is to consider solutions that in a certain sense do not deviate much from Kerr, in such a way that we can treat those deviations as perturbations. This may allow us to maintain the same line of computations as in the Kerr case with modifications that hopefully only add computational difficulties but not conceptual ones.

Up to this point, our discussion has taken place in the framework of General Relativity and the possibility to test in this context a series of hypothesis on the geometry of the collapsed objects that may describe the massive compact objects at the galactic centers. From now on, we discuss what are the prospects of testing also the theory of gravity using LISA EMRI observations. This is obviously a much more complex goal and requires a series of reflections. First of all, the range of alternative theories of gravity available nowadays is so overwhelming that just trying to classify them is a difficult task, and the amount of literature on them is huge. There are different aspects of the theory of gravity that we can change: we have modifications at the Newtonian level, changing the equations of motion; we can consider theories of gravity based only on the metric tensor but with different Lagrangian densities, a very popular exercise these days in the search for an explanation of the acceleration of the Universe; we can change the type of gravitational fields by considering the whole spectrum of additional fields (with all possible associated spin) that can complement the metric, and we can even consider non-metric theories; we can consider theories that are Lorentz invariant and theories that are not; we can also find theories with different spacetime dimensionality; we can find quantum theories like string theory or loop quantum gravity (which add new ingredients like branes, dualities, etc.), or simplified versions of them; there are non-gravitational physical theories that try to explain gravity as an emergent phenomenon; etc.

Many of the alternative theories of gravity available in the literature have been proposed as the solution to a particular observational or theoretical problem. Therefore, they are tailored to certain physical situations and it can happen that they can-not even accommodate in a consistent way objects like BHs. Hence, we need to have a set of criteria to decide which theories of gravity we can test with GW ob-servations of EMRI systems. There are many ways to establish such criteria. One possibility that has been proposed in the context of GW observations in (Yunes and Pretorius, 2009a) is to consider theories that satisfy the following: (i) It must be a metric theory of gravity, i.e. the gravitational field is described by a symmetric metric tensor that satisfies the weak-equivalence principle (see, e.g. (Misner et al., 1973)). (ii) It must be consistent in the weak-field regime: It must reduce to Gen-eral Relativity for weak gravitational fields and for small velocities, in such a way that experimental and observational tests are passed. This implies the existence and stability of physical solutions, such as the Newtonian limit that describes to a high accuracy the dynamics of the Solar system. (iii) It must be *inconsistent* in the strong-field regime: It must deviate from General Relativity in the dynamical sector of strong gravitational fields, where gravity is strong and speeds are comparable to the speed of light. Apart from these criteria we can impose additional ones, such as well-posedness of the initial-value formulation, the existence of a well-defined and complete relativistic action, stability of the solutions, links to Fundamental Physics theories, etc. These additional criteria, however, might be too stringent, in part be-cause only General Relativity has been sufficiently studied to determine whether they are satisfied, and also because we can eliminate theories that are valid in the physical regimes in which we are interested but not in other regimes where they are not supposed to work, as for instance can happen with *effective* theories that derive from a more fundamental theory in a certain limit.

After this discussion, we are going to assume that we have a theory of gravity in which an EMRI system can be properly defined and studied. Then, we have to take into account that there are several aspects of EMRIs and their dynamics that in the context of General Relativity we give for granted, but that when we change the the-ory of gravity we should put into question. In this sense, a very important aspect is the question of the final outcome of gravitational collapse. As we have seen, in Gen-eral Relativity there are very strong indications that once a self-gravitating system has gone through all stages of the collapse it should form a Kerr BH. In alternative theories of gravity, the Kerr metric may not even be a solution. There may be, how-ever, a different solution that plays a similar role (existence of uniqueness theorem, etc.). But in that case, it may be that it is not characterized by just two numbers (mass and spin) but by more (or less, just the mass). This may be the case of theo-ries with extra fields, and in which these fields can confer extra hair (e.g. charges associated with these fields) to the BHs (or whatever objects are produced) after col-lapse. In that case, it may not be possible to make an straightforward translation of hypothesis **H1** to the alternative theory of gravity. As an alternative, it may be possible to make tests of an hypothesis in the lines of **H1′** and **H2**. For this, it is required that multipole moments can be consistently defined, but it may be that the multipole moments, as they are in General Relativity, are not enough to determine uniquely the geometry of the massive collapsed objects, specially in the cases where extra fields may contribute to their properties. Nevertheless, since multipolar expansions, which are essentially expansions in powers of $1/r$ [as we can see from the expansions of Thorne (Thorne, 1980) in Eqs. (3)], are likely to exist with possibly extra sets of numbers associated with extra fields/degrees of freedom, say $\{\mathcal{M}_\ell^A\}$ (where $A$ labels the possible different sets), in such a way that they describe univocally the final states of the collapsed objects. In such a case, we can try to use LISA EMRI observations to test hypothesis of the following kind:

**H3:** *X-Body in Y-Gravity Theory Hypothesis*: The exterior gravitational field of the dark, compact and very massive objects sitting at the galactic centers can be well described by the stationary, axisymmetric, and asymptotically-flat solutions of the Y-Gravity Theory whose multipole moments $\{M_\ell, J_\ell, \mathcal{M}_\ell^A\}_{\ell=0,\ldots,\infty}^{A=1,\ldots}$ satisfy the following set of relations: $\mathcal{R}_a[\{M_\ell, J_\ell, \mathcal{M}_\ell^A\}] = 0 \quad (a = 1, \ldots)$.

This discussion illustrates the variety of cases that we may face when we open the door to other theories of gravity. What makes things difficult is that we want to deal with systems like EMRIs that probe the strong field sector of the gravitational theory under consideration. Then, modifications to the theory that in the weak-field regime look tiny can make a difference in the strong-field regime. And the behaviour in this regime cannot be described, in principle, by perturbative schemes that make expansions around the General Relativistic case. Another difficulty for the data analysis and also from the point of view of estimating errors in the parameters, is that the waveform models that we have to consider are going to be more complex. They can be more complex in two ways. First, the waveforms may contain more polarization states as compared with General Relativity (see (Eardley et al., 1973b, 1973a) for a classification of gravitational theories with respect to the possible polarization states). Second, they will contain extra parameters, so we will facing waveforms models of form $\mathbf{h}(t; \vec{\theta})$, where the parameters $\vec{\theta} = (\theta^I)$ will be: $\theta^I = \theta_K^I$ for $I = 1, \ldots, 14$, $\theta^I = \theta_M^I$ for $I = 15, \ldots, 14 + N_{mm}$, and $\theta^I = \theta_{XY}^I$ for $I = 15 + N_{mm}, \ldots, 14 + N_{mm} + N_{XY}$, where $N_{XY}$ is the number of extra parameter required, which contains extra parameters that are necessary to describe the orbital motion and the properties of the central massive object, also extra types of multipoles, and finally, extra parameters related to the theory of gravity.

The next thing we can think of is of trying to test the theory of gravity itself by using LISA EMRI observations. However, this is a already implicit in the previous discussion, where we have also taken into account parameters associated with the specific theory of gravity under consideration. These parameters may be *coupling constants* of different nature that are in principle unknown (or even if we have a prediction for them, we can always consider them as unknown for the purposes of testing the theory). Following what we have just said, this can be reduced to estimate the parameters that characterize the alternative theory of gravity. However, this may become a quite difficult task from the practical point of view if the total number of parameters necessary to characterize an EMRI system in an alternative theory of gravity (including the ones associated with the theory) is too high. One may find problems with possible degeneracies among parameters, or confusion problems in the identification of the physical parameters.

A clear conclusion that one can extract from all this is that the spectrum of possible cases is too vast and we know too little in order to draw a clear plan for doing LISA EMRI science in the context of hypothesis **H3**. Then, continuing this line of argument may end up in a too vague and also too speculative discussion. To avoid reaching that point and with the aim of illustrating part of what has been said, in the remainder of this article we present an example of an study of EMRIs in a particular alternative theory of gravity.

We refer to this particular alternative theory of gravity as Dynamical Chern-Simons Modified Gravity (DCSMG; see (Alexander and Yunes, 2009) for a recent review). This type of modifications of General Relativity was initiated by Jackiw and Pi (Jackiw and Pi, 2003). Their main modification consists in the addition of a new term to the Einstein-Hilbert Lagrangian that is a generalization of the well-known three-dimensional Chern-Simons term. This new term represents a parity violating interaction, and its inclusion is motivated by several quantum gravity approaches. It appears in four-dimensional compactifications of perturbative string theory due to the Green-Schwarz anomaly-canceling mechanism (Polchinski, 1998), but it also can appear in the non-perturbative sector (Alexander and Gates, 2006). This term also arises naturally in loop quantum gravity when the Barbero-Immirzi parameter is promoted to a scalar field coupled to the Nieh-Yan invariant (Taveras and Yunes, 2008, Calcagni and Mercuri, 2009 and Mercuri and Taveras, 2009). More generically, the Pontryagin correction is also motivated from an effective field theory standpoint, through the inclusion of high-curvature terms to the action (see (Weinberg, 2008) for an application of this approach to inflationary cosmology). In this 4D theory, the action can be written as the following sum of terms:

$$\mathcal{S} = \mathcal{S}_{\text{EH}} + \mathcal{S}_{\text{CS}} + \mathcal{S}_{\phi} + \mathcal{S}_{\text{matter}} , \qquad (8)$$

where $\mathcal{S}_{\text{EH}}$ is the Einstein-Hilbert action

$$S_{\text{EH}} = \kappa \int d^4x \sqrt{-g}\, R , \qquad \kappa = \frac{1}{16\pi G} , \qquad (9)$$

where $R$ is the scalar curvature. The second term, $\mathcal{S}_{\text{CS}}$, is the modification that contains the Pontryagin density $\,^*\!R\,R$:

$$\mathcal{S}_{\text{CS}} = \frac{\alpha}{4} \int d^4x \sqrt{-g}\, \phi \,\,^*\!R\,R \qquad (10)$$

where ${}^{\ast}R\,R = R_{\mu\nu\rho\sigma}\,{}^{\ast}R^{\mu\nu\rho\sigma} = \frac{1}{2}\epsilon^{\mu\nu\rho\sigma}R_{\mu\nu\gamma\delta}R^{\gamma\delta}{}_{\rho\sigma}$.

Here $\alpha$ is the coupling constant associated with the Pontryagin density term, and $\epsilon_{\mu\nu\rho\sigma}$ and $R_{\mu\nu\rho\sigma}$ are the spacetime Levi-Civita antisymmetric and Riemann curvature tensors respectively. It is important to mention that the Pontryagin density, a topological invariant in a four-dimensional spacetime, is multiplied by a scalar field, $\phi$, to produce a modification of the General Relativistic field equations. Then, we also have the contribution to the action of the scalar field:

$$\mathcal{S}_\phi = -\beta \int d^4 x \, \sqrt{-g} \, \left[\frac{1}{2}g^{\mu\nu}\left(\nabla_\mu\phi\right)\left(\nabla_\nu\phi\right) + V(\phi)\right],\tag{11}$$

where $\beta$ is another coupling constant. In the original version of the theory, the scalar field $\phi$ was forced to be a fixed function (without dynamics and hence with no contribution to the action). It turns out that this leads to an additional constraint, the vanishing of the Pontryagin density, which is too restrictive. In particular, it disallows the possibility of spinning BH solutions for scalar fields whose gradient is time-like (Grumiller and Yunes, 2008) and forbids perturbations of non-spinning BHs (Yunes and Sopuerta, 2008). Hence, this version of the theory is an example of theory where we cannot study EMRIs. Finally, $\mathcal{S}_{\text{matter}}$ is the action of any additional matter fields.

We now summarize the main results obtained so far on the study of EMRIs in DC-SMG (Sopuerta and Yunes, 2009 and Yunes and Sopuerta, 2010). The first point is the question of the geometry of spinning MBHs in this theory. They are no longer described by the Kerr metric (although non-spinning MBHs are still described by the Schwarzschild metric). Using small-coupling [$\xi/M_\bullet^4 \ll 1$, where $\xi \equiv \alpha^2/(\beta\kappa)$; see Eqs. (10) and (11)] and slow-rotation ($a_\bullet/M_\bullet \ll 1$) approximations, the exterior, stationary and axisymmetric gravitational field of a rotating MBH in dynamical DC-SMG modified gravity, in Boyer-Lindquist-type coordinates $\{t, r, \theta, \phi\}$ (see (Misner et al., 1973, Chandrasekhar, 1992 and Frolov and Novikov, 1997) for details on this coordinate system), is given by (see (Yunes and Pretorius, 2009b) for the derivation):

$$g_{\mu\nu}dx^\mu dx^\nu = ds^2 = ds^2_{\text{Kerr}} + \frac{5\,\xi\,a_\bullet}{4r^4}\left[1 + \frac{12M_\bullet}{7r} + \frac{27M_\bullet^2}{10r^2}\right]\sin^2\theta\,dt\,d\varphi\,,\tag{12}$$

where $ds^2_{\text{Kerr}}$ is the line element for the Kerr metric. The multipolar structure of the modified metric, as in the case of the Kerr metric, remains completely determined by only two numbers: the mass $M_\bullet$ and spin parameter $a_\bullet$. Therefore, we can try to establish a no-hair theorem for this metric in this theory, as with the Kerr metric in General Relativity. However, the multipole moments do not satisfy the relation given in Eq. (5) for all $\ell$. It can be seen that the multipole moments satisfy different relations for $\ell \geq 4$, which involve the parameter $\xi$. On the other hand, the solution for the DCSMG scalar field $\phi$ is:

$$\phi = \frac{5}{8}\frac{\alpha}{\beta}\frac{a_\bullet}{M_\bullet}\frac{\cos(\theta)}{r^2}\left(1 + \frac{2M}{r} + \frac{18M_\bullet^2}{5r^2}\right),\tag{13}$$

which is axisymmetric and fully determined by the MBH geometry (Yunes and Pretorius, 2009b). At this point, we want to remark that the MBH solution of Eq. (12) is an approximate one. There is no yet an exact solution for spinning MBHs in this theory. Thus, there is no yet studies on the role of such a metric in DCSMG, like whether it is unique under some conditions, etc. We expect that future studies can shed some light on these questions.

At this point we can think about EMRIs in DCSMG by assuming that the MBH geometry is well described by Eq. (12). Then, we have to look at the equations of motion of the SCO around the MBH. It has been shown (Sopuerta and Yunes, 2009) that point-particles follow geodesics orbits in this theory, as in General Relativity. Moreover, it turns out that the metric given in Eq. (12), up to the same level of approximation, has the same Killing symmetries as the Kerr metric (i.e. it is also stationary and axisymmetric), and also has a 2-rank Killing tensor. As a consequence, the geodesic equations are fully integrable as in the case of Kerr. Once the equations are separated, the difference with respect to the equations in the Kerr geometry can be encoded in a single function (Sopuerta and Yunes, 2009). Analyzing the modified geodesic equations, one can see that the location of the innermost-stable circular orbit is shifted with respect to the Kerr ISCO as follows (Yunes and Pretorius, 2009b):

$$R_{\text{ISCO}} = \underbrace{6M \mp \frac{4\sqrt{6}\,a_\bullet}{3} - \frac{7\,a_\bullet^2}{18\,M_\bullet}}_{\text{Kerr ISCO}} \pm \underbrace{\frac{77\sqrt{6}\,a_\bullet}{5184}\,\frac{\alpha^2}{\beta\kappa M_\bullet^4}}_{\text{DCSMG Modification}}, \qquad (14)$$

where the upper (lower) signs correspond to co- and counter-rotating geodesics (with respect to the MBH spin rotation). Notice that the DCSMG correction acts *against* the spin effects. One can also check that the three fundamental frequencies of motion change with respect to the values they take for geodesic orbital motion in Kerr.

The next important question to address is how GW emission and propagation is affected in DCSMG. First of all, it has been shown (Sopuerta and Yunes, 2009) that observers far away from the GW sources can only perceive the two same polarizations as in General Relativity, although there is an additional mode, a breathing mode, that has an impact in the strong-field dynamics but decays too fast with distance to be observable in the radiation zone (see (Garfinkle et al., 2010) for a related study on the propagation of GWs in DCSMG). Then, the analysis of the dynamics of EMRIs in DCSMG done in (Sopuerta and Yunes, 2009) was carried out in the spirit of the so-called *semi-relativistic* approximation (Ruffini and Sasaki, 1981), where the orbital motion is assumed to be geodesic and GWs emitted are assumed to propagate in a flat spacetime background geometry. In contrast with (Ruffini and Sasaki, 1981), where the waveforms are computed using the gravitational Lienard-Wiechert potentials, in (Sopuerta and Yunes, 2009) the waveforms are computed using the GW multipolar expansion (Thorne, 1980). This method of computing EMRI waveforms is similar to the one used in (Babak et al., 2007) for Kerr to produce the so-called *numerical kludge* waveforms. The DCSMG EMRI waveforms obtained in this way do not contain radiation reaction effects and the dephasing between these waveforms and the ones computed in the framework of General Relativity is due only to the differences in the MBH geometry. These differences imply that a given set of fundamental frequencies corresponds to different sets of orbital parameters in General Relativity and DCSMG, and vice versa. This can be seen in a different way: Given a General Relativistic waveform with its associated fundamental frequencies and corresponding orbital parameters, we can always find orbital parameters in DCSMG that can match the General Relativistic waveforms. In order to break this degeneracy radiation reaction effects need to be included.

In view of the importance of radiation reaction effects, a first examination of this question in DCSMG was carried out in (Sopuerta and Yunes, 2009) by using the *short-wave* approximation (Isaacson, 1968a, 1968b) (see also (Misner et al., 1973)). It was found that to leading order the GW emission formulae are unchanged with respect to General Relativity. More specifically, in the short-wave approximation we can introduce an effective energy-momentum tensor for the GWs (the Isaacson tensor (Isaacson, 1968b)), and it turns out that the one found in DCSMG has exactly the same form (dependence on the metric waveform) as the one obtained in General Relativity. There are subdominant contributions to the radiation reaction mechanism due to the presence of the DCSMG scalar field $\phi$, which is also radiated away in the EMRI dynamics.

In spite of not including radiation reaction effects in the analysis, a rough estimate (Sopuerta and Yunes, 2009) of the accuracy to which DCSMG gravity could be constrained via LISA EMRI observations was given by comparing waveforms computed in GR with waveforms computed in DCSMG (assuming the same orbital parameters and also that the degeneracy mentioned above can be broken). This estimate can be expressed as: $\xi^{1/4} \lesssim 5 \cdot 10^5 \, \Delta\xi \, (M_\bullet/M_{MW}) \, \text{km}$, where $\Delta\xi$ is a measure of the accuracy to which $\xi$ can be measured, which depends on the integration time, the SNR, the type of orbit, etc. Moreover, $M_{MW}$ is an approximation to the mass of the presumable BH at Sgr A* in our Milky Way, taken to be $4.5 \cdot 10^6 M_\odot$. This estimate is to be compared with the constraint that can be obtained from observations of the double pulsar (Yunes and Pretorius, 2009b): $\xi^{1/4} \lesssim 10^4$ km. Notice that IMRIs (with total masses in the range $10^3 - 10^4 M_\odot$) are favored over EMRIs, in the sense that IMRI observations can produce better constraints, and these constraints are likely to be more stringent than the one from the double pulsar.

We have seen that in DCSMG there are some similarities with General Relativity in what refers to the dynamics of EMRIs, at least at the level of approximation in which the DCSMG case has been studied. The MBH metric has essentially the same geometric structure as the Kerr metric but different multipolar expansion. Assuming that all these similarities will survive in deeper studies of this theory, it is the type of theory where hypothesis of the type **H3** can be meaningful. In that case, we would only have an extra parameter, namely $\xi$, as the multipole moments of the approximate metric of Eq. (12) only depend on the mass, spin, and $\xi$. The scalar field does not introduce any new hair. Then, in a situation where testing an alternative theory of gravity only means to add one extra parameter in the waveform models, it seems feasible to perform reliable tests, but more work in this direction is required to confirm these prospects. In the same way, in order to understand better this theory and the dynamics of EMRIs within it we need to get a deeper knowledge of several subjects. In this sense, it is specially important to analyze the role of radiation reaction effects and to perform parameter estimation analysis to see how well LISA can distinguish between General Relativity and DCSMG.

## 4 Remarks, Conclusions, and Prospects for the Future

In this article a personal view of the possibilities that future observations of EMRIs by LISA offer for Fundamental Physics has been presented. First of all, we have tried to put in perspective the emergent area of Gravitational Wave Astronomy, and within it, the importance of the observations in the low-frequency band that LISA is expected to carry out in the future. The focus has been in a particular source of GWs for LISA, the EMRIs. We have tried to support the idea that LISA will make very precise measurements of these systems, in such a way that this will allow us to produce diverse scientific discoveries that will impact, in particular, the area of Fundamental Physics. Here, there are two types of questions that LISA can answer: First, whether or not the geometry of the very massive compact objects in the galactic centers is well described by the Kerr solution of General Relativity. And in the negative case, it may be possible to test other types of geometries. Second, whether or not the strong field regime of the gravitational field, as determined by the gravitational field of compact objects like neutron stars or BHs, is well described by the General Theory of Relativity.

For the first type of question, we have followed the line of thought initiated by Ryan and others, who used the fact that the geometry of self-gravitating compact objects can be determined univocally by their multipole moments. Different sets of moments corresponds to different objects, and what is more important, the BHs of General Relativity have multipole moments that are fully determined in terms of the mass and the spin, and they satisfy a very particular set of relations. Then, we have discussed the type of hypothesis that can be tested with LISA EMRI observations, the theoretical challenges that we face to develop the necessary tools, and the present status. As we have seen, although a lot of work has been done, there is also a lot of research to be carried out in the future. In this sense, we want to remark the fact that the relation between the multipole moments and observables like the GW phase seems to be quite intricate. A possible reason that has been pointed out (for instance by Collins and Hughes (Collins and Hughes, 2004)) is that the multipole moments are numbers that appears in the coefficients of an expansion in the inverse of the distance to the source, an expansion that works well in the weak-field regime and even in the moderately strong-field regime but not in the strong one, whereas the GW phase is more affected precisely by the orbital motion in the strong field region. Then, it may be a good idea that in the process of learning about the deep connection between multipole moments and the features of the GW signals, we try to find alternatives to the multipolar description. These alternatives should provide equivalent information to the multipole moments but should be more directly and clearly related to the waveform models.

Another important question that has arisen is the question of integrability of the geodesic orbits in non-Kerr spacetimes, and the possibility of ergodic/chaotic motion. In many cases, it has been found that chaotic behavior can occur only near the innermost stable circular orbit and in the strong-field region of the central massive body. This is precisely the region where the radiation-reaction effects are stronger and operate in a time-scale significantly smaller than in the initial (observable) parts of the inspiral. An interesting question is then to know how important these radiation reaction effects can be and whether they can change qualitatively the chaotic picture. Another important question is to know how realistic or illustrative are some of the examples of non-Kerr objects that have been studied in the literature and to understand whether or not the new behavior that some of them exhibit is what we can expect to see in the Universe.

Finally, regarding tests of alternative theories of gravity, we have seen that given the wide spectrum of theories available, it is quite complicated to predict in general what kind of physical phenomena we can expect to test by means of LISA EMRI observations. This situation is partly due to the fact that few theories of gravity have been studied in the same great detail as General Relativity. Therefore, we know very little a priori of the dynamics of EMRIs in other theories. In this regard, it is worth mentioning that, in general, the problem of testing the geometry of massive collapsed objects and the theory of gravity with EMRIs is a coupled problem. Nevertheless, we can always fix the theory of gravity and study the objects with the strongest gravitational field there. Given this complex situation and our state of knowledge, we would like to advocate here the view that studying particular cases, as the one described here about Chern-Simons modifications of General Relativity, is a good way of starting to get some insight into what we can expect from different classes of theories. It is also a way of making progress in the more general area of Gravitational Physics.

## 5 Acknowledgments

We would like to thank P. Amaro-Seoane, S. Babak, L. Barack, C. Barceló, E. Berti, P. Canizares, S. Drasco, J. Gair, J. L. Jaramillo, L. Lehner, E. Poisson, B. Schutz, U. Sperhake, and N. Yunes for discussions related to the topic of this article. CFS acknowledges support from the Ramón y Cajal Program of the Spanish Ministry of Education and Science (MEC) and by a Marie Curie International Reintegration

Grant (MIRG-CT-2007-205005/PHY) within the 7th European Community Framework Programme. Financial support from the contracts ESP2007-61712 (MEC) and FIS2008-06078-C03-01 (Ministry of Science and Innovation of Spain) is gratefully acknowledged. The research of CFS uses the resources of the Centre de Supercomputació de Catalunya (CESCA) and the Centro de Supercomputación de Galicia (CESGA; project number ICTS-2009-40).

## 6 Appendix: Basics of the Theory of Parameter Estimation

The theory of detection and measurement of GW signals was given a solid foundation in the works of Finn (Finn, 1992), Cutler and Flanagan (Cutler and Flanagan, 1994), and others (see (Finn and Chernoff, 1993 and Poisson and Will, 1995) and (Weinstein and Zubakov, 1962 and Helstrom, 1968) for general aspects of the theory of signal extraction). Here, we summarize some of the aspects of the Parameter Estimation theory that are relevant for the discussion of this article. This material has been presented by many other authors in different places and the presentation that we provide here is quite standard.

Let us start by considering the output of a set of $D$ detectors (which is appropriate for LISA as it has several different channels), a set of time series that can be represented by the vector $s^\alpha(t)$ ($\alpha = 1, 2, \ldots, D$). In GW theory it is convenient to work with the Fourier transform of the signals, which we denote with a tilde and use the following convention for its definition

$$\tilde{s}^\alpha(f) \equiv \int_{-\infty}^{+\infty} dt \, s^\alpha(t) \, e^{2\pi i f t} \,. \tag{15}$$

The detector output can be split into the instrumental noise, $n^\alpha(t)$, and the different GW signals that have produced a response, $h^\alpha(t)$, i.e. $s^\alpha(t) = n^\alpha(t) + h^\alpha(t)$. For our purposes, we assume that the instrumental noise is *stationary*, which means that the different Fourier components $\tilde{n}^\alpha(f)$ are uncorrelated, that is (we denote complex conjugation by an asterisk)

$$< \tilde{n}^\alpha(f) \, \tilde{n}^\beta(f')^* > = \frac{1}{2} \delta(f - f') \, S_n^{\alpha\beta}(f) \,, \tag{16}$$

where $< \cdots >$ means to take the average over all possible realizations of the instrumental noise and $S_h^{\alpha\beta}(f)$ is the so-called one-sided spectral density of the noise (which is the Fourier transform of the noise autocorrelation function: $C_n^{\alpha\beta}(\tau) \equiv < n^\alpha(t) \, n^\beta(t + \tau) >_t$, where $< \cdots >_t$ means time average). We also assume that the noise is Gaussian, which means that the probability of a given Fourier component of the noise, $\tilde{n}^\alpha(f)$, follows a Gaussian probability distribution function. Then, the probability of a particular noise realization, $n_o^\alpha$, can be written as (Finn, 1992) (we use the vector notation $\mathbf{n} = (n^\alpha)$)

$$p(\mathbf{n} = \mathbf{n_o}) \propto e^{-\frac{1}{2}(\mathbf{n_o}|\mathbf{n_o})} \,, \tag{17}$$

where $(\cdot|\cdot)$ is a natural inner product on the vector space of signals which, given two signals $\mathbf{p}$ and $\mathbf{q}$, is defined as follows (there is summation over repeated indices)

$$(\mathbf{p}|\mathbf{q}) = 4 \, \mathrm{Re} \int_0^\infty df \, \left[\mathbf{S_n^{-1}}(f)\right]_{\alpha\beta} \tilde{p}^\alpha(f)^* \, \tilde{q}^\beta(f) \,. \tag{18}$$

Thus, if the actual GW signal that passed through the detectors is $\mathbf{h}$, the probability of measuring the signals $\mathbf{s}$ is proportional to $\exp\left\{-\frac{1}{2}(\mathbf{s}-\mathbf{h}|\mathbf{s}-\mathbf{h})\right\}$. Let us now assume that the GWs signals in the detector can be described by a set of functions that depend on a number $N$ of source parameters: $\mathbf{h}(t;\vec{\theta})$ with $\vec{\theta} = (\theta^I)$ and $I = 1, \ldots, N$. Let us also assume that we have some criterion to decide that such a signal is present in the detectors data stream. Then, the probability of a certain value of the parameters given the detector data is

$$p(\vec{\theta}|\mathbf{s}) = p(\mathbf{s}|\vec{\theta}) \, \exp\left\{-\frac{1}{2}\left(\mathbf{s}-\mathbf{h}(\vec{\theta})|\mathbf{s}-\mathbf{h}(\vec{\theta})\right)\right\} \,, \tag{19}$$

where $p(\mathbf{s}|\vec{\theta})$ is the *a priori* probability that the signal is characterized by the parameters $\vec{\theta}$, the so-called *priors*. Then, the true value of the source parameters, $\vec{\theta}_{\mathrm{true}}$, can be estimated by finding the values of the parameters at which this probability distribution function has a maximum, which we call the *best-fit* parameters $\vec{\theta}_{\mathrm{bf}}$. It can then be seen that the *matched filtering* SNR of the signal $\mathbf{h}$ is simply given by

$$\mathrm{SNR} = \sqrt{(\mathbf{h}|\mathbf{h})} \,, \tag{20}$$

evaluated at $\vec{\theta} = \vec{\theta}_{\mathrm{bf}}$. Different realizations of the noise will give rise to different best-fit parameters. For a given realization of the noise, say $\mathbf{n}$, we denote the difference between the best-fit and true parameters by $\delta\vec{\theta}(\mathbf{n}) \equiv \vec{\theta}_{\mathrm{bf}}(\mathbf{n}) - \vec{\theta}_{\mathrm{true}}$. For large SNR, the best-fit parameters will assume a Gaussian distribution centered on the correct values and given by

$$p(\delta\theta^I) = \sqrt{\det\left(\frac{\Gamma}{2\pi}\right)} \, \exp\left\{-\frac{1}{2}\,\Gamma_{IJ}\,\delta\theta^I\delta\theta^J\right\} \,, \tag{21}$$

where $\Gamma_{IJ}$ denotes the Fisher information matrix, which is defined by

$$\Gamma_{IJ} \equiv \left(\frac{\partial\mathbf{h}}{\partial\theta^I}\,\bigg|\,\frac{\partial\mathbf{h}}{\partial\theta^J}\right) \,, \tag{22}$$

evaluated at $\vec{\theta} = \vec{\theta}_{\mathrm{bf}}$. Then, the variance-covariance matrix of the errors $\delta\theta^I$, $\Sigma^{IJ}$, is given by

$$< \delta\theta^I\,\delta\theta^J > \equiv \Sigma^{IJ} + O\left(\mathrm{SNR}^{-1}\right) = (\Gamma^{-1})^{IJ} + O\left(\mathrm{SNR}^{-1}\right) \,. \tag{23}$$

We define the measurement error in the parameter $\theta^I$ by (there is no summation over repeated indices in the next two equations)

$$\Delta\theta^I \equiv \sqrt{\Sigma^{II}} , \qquad (24)$$

and the correlation coefficient between the parameters $\theta^I$ and $\theta^J$ as

$$C^{IJ}(\vec{\theta}) \equiv \frac{\Sigma^{IJ}}{\sqrt{\Sigma^{II}\,\Sigma^{JJ}}} , \qquad (25)$$

which, by definition, are in the range: $-1 \leq C^{IJ}(\vec{\theta}) \leq 1$. It is important to remarks that these quantities are meaningful in the large SNR limit. See (Vallisneri, 2008) for a detailed discussion of the use of the Fisher matrix analysis in the context of GW parameter estimation.

# References in the highlight article

Alexander, S. and Yunes, N. (2009). Chern-Simons Modified General Relativity. *Physics Reports*, 480:1-55.

Alexander, S. H. S. and Gates, S. J. (2006). Can the string scale be related to the cosmic baryon asymmetry?. *Journal of Cosmology and Astroparticle Physics*, 0606:018.

Amaro-Seoane, P., Barranco, J., Bernal, A. and Rezzolla, L. (2010). Constraining scalar fields with stellar kinematics and collisional dark matter. *ArXiv e-prints, arXiv:1009.0019*.

Amaro-Seoane, P. and others (2007). Astrophysics, detection and science applications of intermediate- and extreme mass-ratio inspirals. *Classical and Quantum Gravity*, 24:R113-R169.

Apostolatos, T. A., Lukes-Gerakopoulos, G. and Contopoulos, G. (2009). How to Observe a Non-Kerr Spacetime Using Gravitational Waves. *Physical Review Letters*, 103:111101.

Ashtekar, A., Engle, J., Pawlowski, T. and Van Den Broeck, C. (2004). Multipole moments of isolated horizons. *Classical and Quantum Gravity*, 21:2549-2570.

Babak, S., Fang, H., Gair, J. R., Glampedakis, K. and Hughes, S. A. (2007). 'Kludge' gravitational waveforms for a test-body orbiting a Kerr black hole. *Physical Review D*, 75:024005.

Barack, L. (2009). Gravitational self force in extreme mass-ratio inspirals. *Classical and Quantum Gravity*, 26:213001.

Barack, L. and Cutler, C. (2004). LISA capture sources: Approximate waveforms, signal-to- noise ratios, and parameter estimation accuracy. *Physical Review D*, 69:082005.

Barack, L. and Cutler, C. (2007). Using LISA EMRI sources to test off-Kerr deviations in the geometry of massive black holes. *Physical Review D*, 75:042003.

Barack, L., Damour, T. and Sago, N. (2010). Precession effect of the gravitational self-force in a Schwarzschild spacetime and the effective one-body formalism. .

Barack, L. and Sago, N. (2009). Gravitational self-force correction to the innermost stable circular orbit of a Schwarzschild black hole. *Physical Review Letters*, 102:191101.

Barack, L. and Sago, N. (2010). Gravitational self-force on a particle in eccentric orbit around a Schwarzschild black hole. *Physical Review D*, 81:084021.

Barausse, E. and Rezzolla, L. (2008). The influence of the hydrodynamic drag from an accretion torus on extreme mass-ratio inspirals. *Physical Review D*, 77:104027.

Barausse, E., Rezzolla, L., Petroff, D. and Ansorg, M. (2007). Gravitational waves from extreme mass ratio inspirals in non-pure kerr spacetimes. *Physical Review D*, 75:064026.

Barcelo, C., Liberati, S., Sonego, S. and Visser, M. (2008). Fate of gravitational collapse in semiclassical gravity. *Physical Review D*, 77:044032.

Beig, R. and Simon, W. (1980). Proof of a multipole conjecture due to geroch. *Communications on Mathematical Physics*, 78:75–82.

Beig, R. and Simon, W. (1981). On the multipole expansion for stationary space-times. *Proceedings of the Royal Society of London. Series A, Mathematical and Physical Sciences*, 376:333-341.

Berti, E. (2006). LISA observations of massive black hole mergers: event rates and issues in waveform modelling. *Classical and Quantum Gravity*, 23:S785-S798.

Blanchet, L., Detweiler, S. L., Le Tiec, A. and Whiting, B. F. (2010b). High-Order Post-Newtonian Fit of the Gravitational Self- Force for Circular Orbits in the Schwarzschild Geometry. *Physical Review D*, 81:084033.

Blanchet, L., Detweiler, S. L., Le Tiec, A. and Whiting, B. F. (2010a). Post-Newtonian and Numerical Calculations of the Gravitational Self-Force for Circular Orbits in the Schwarzschild Geometry. *Physical Review D*, 81:064004.

Bloom, J. S. and others (2009). Astro2010 Decadal Survey Whitepaper: Coordinated Science in the Gravitational and Electromagnetic Skies. .

Brink, J. (2008b). Spacetime Encodings II - Pictures of Integrability. *Physical Review D*, 78:102002.

Brink, J. (2010a). Spacetime Encodings III - Second Order Killing Tensors. *Physical Review D*, 81:022001.

Brink, J. (2010b). Spacetime Encodings IV - The Relationship between Weyl Curvature and Killing Tensors in Stationary Axisymmetric Vacuum Spacetimes. *Physical Review D*, 81:022002.

Broderick, A. E. and Narayan, R. (2006). On The Nature of the Compact Dark Mass at the Galactic Center. *The Astrophysical Journal Letters*, 638:L21-L24.

Brown, D. A. and others (2007). Gravitational waves from intermediate-mass-ratio inspirals for ground-based detectors. *Physical Review Letters*, 99:201102.

Calcagni, G. and Mercuri, S. (2009). The Barbero-Immirzi field in canonical formalism of pure gravity. *Physical Review D*, 79:084004.

Carter, B. (1968). Global Structure of the Kerr Family of Gravitational Fields. *Physical Review*, 174:1559-1571.

Carter, B. (1968). Hamilton-Jacobi and Schrodinger separable solutions of Einstein's equations. *Communications on Mathematical Physics*, 10:280.

Chandrasekhar, S. (1992). *The Mathematical Theory of Black Holes*. Oxford University Press, New York.

Chapline, G. (2005). Dark energy stars. In Chen, P., Bloom, E., Madejski, G. and Patrosian, V., editors, *Proceedings of 22nd Texas Symposium on Relativistic Astrophysics (SLAC-R-752)*, pages 0205, San Francisco. SLAC.

Chapline, G., Hohlfeld, E., Laughlin, R. B. and Santiago, D. (2001). . *Philosophical Magazine B*, 81:235.

Chapline, G., Hohlfeld, E., Laughlin, R. B. and Santiago, D. I. (2003). Quantum phase transitions and the breakdown of classical general relativity. *International Journal of Modern Physics A*, 18:3587-3590.

Christodoulou, D. (1994). Examples of naked singularity formation in the gravitational collapse of a scalar field. *Annals of Mathematics*, 140:607-653.

Collins, N. A. and Hughes, S. A. (2004). Towards a formalism for mapping the spacetimes of massive compact objects: Bumpy black holes and their orbits. *Physical Review D*, 69:124022.

Cornish, N. J. (2008). Detection Strategies for Extreme Mass Ratio Inspirals. .

Cornish, N. J. and Rubbo, L. J. (2003). The LISA response function. *Physical Review D*, 67:022001.

Cutler, C. and Flanagan, E. E. (1994). Gravitational waves from merging compact binaries: How accurately can one extract the binary's parameters from the inspiral wave form?. *Physical Review D*, 49:2658-2697.

Damour, T. (1995). Gravitation, experiment and cosmology. In Gazis, E. N., Koutsoumbas, G., Tracas, N. D. and Zoupanos, G., editors, *Proceedings of the 5th Hellenic School and Workshops on Elementary Particle Physics*, pages 332–368. Corfu Summer Institute.

Damour, T. (2007b). Binary Systems as Test-beds of Gravity Theories. In Colpi, M. and Casella, P. and Gorini V. and Moschella, U. and Possenti, A., editor, *Physics of Relativistic Objects in Compact Binaries: from Birth to Coalescence*. Springer.

Damour, T. (2007a). Black hole and neutron star binaries: Theoretical challenges. In Gursky, H. and Ruffini, R., editors, *Revised Edition of: Neutron Stars, Black Holes and Binary X-Ray Sources*.

Dhurandhar, S. V. and Tinto, M. (2005). Time-delay interferometry. *Living Reviews in Relativity*, 8(4).

Drasco, S. (2009). Verifying black hole orbits with gravitational spectroscopy. *Physical Review D*, 79:104016.

Drasco, S. and Hughes, S. A. (2004). Rotating black hole orbit functionals in the frequency domain. *Physical Review D*, 69:044015.

Drasco, S. and Hughes, S. A. (2006). Gravitational wave snapshots of generic extreme mass ratio inspirals. *Physical Review D*, 73:024027.

Eardley, D. M., Lee, D. L. and Lightman, A. P. (1973a). Gravitational-wave observations as a tool for testing relativistic gravity. *Physical Review D*, 8(10):3308–3321.

Eardley, D. M., Lee, D. L., Lightman, A. P., Wagoner, R. V. and Will, C. M. (1973b). Gravitational-wave observations as a tool for testing relativistic gravity. *Physical Review Letters*, 30(18):884–886.

Finn, L. S. (1992). Detection, Measurement and Gravitational Radiation. *Physical Review D*, 46:5236-5249.

Finn, L. S. and Chernoff, D. F. (1993). Observing binary inspiral in gravitational radiation: One inferferometer. *Physical Review D*, 47:2198–2219.

Finn, L. S. and Thorne, K. S. (2000). Gravitational waves from a compact star in a circular, inspiral orbit, in the equatorial plane of a massive, spinning black hole, as observed by LISA. *Physical Review D*, 62:124021.

Fodor, G., Hoenselaers, C. and Perjés, Z. (1989). Multipole moments of axisymmetric systems in relativity. *Journal of Mathematical Physics*, 30:2252–2257.

Frolov, V. P. and Novikov, I. D. (1997). *Black Hole Physics: Basic Concepts and New Developments*. Kluwer.

Gair, J. R., Barack, L., Creighton, T., Cutler, C. and Larson, S. L.et al. (2004). Event rate estimates for LISA extreme mass ratio capture sources. *Classical and Quantum Gravity*, 21:S1595-S1606.

Gair, J. R., Li, C. and Mandel, I. (2008). Observable Properties of Orbits in Exact Bumpy Spacetimes. *Physical Review D*, 77:024035.

Garfinkle, D., Pretorius, F. and Yunes, N. (2010). Linear Stability Analysis and the Speed of Gravitational Waves in Dynamical Chern-Simons Modified Gravity. .

Geroch, R. (1970). Multipole moments. ii. curved space. *Journal of Mathematical Physics*, 11:2580-2588.

Glampedakis, K. (2005). Extreme Mass Ratio Inspirals: LISA's unique probe of black hole gravity. *Classical and Quantum Gravity*, 22:S605-S659.

Glampedakis, K. and Babak, S. (2006). Mapping spacetimes with LISA: inspiral of a test-body in a 'quasi-Kerr' field. *Classical and Quantum Gravity*, 23:4167-4188.

Gralla, S. E. and Wald, R. M. (2008). A Rigorous Derivation of Gravitational Self-force. *Classical and Quantum Gravity*, 25:205009.

Grumiller, D. and Yunes, N. (2008). How do Black Holes Spin in Chern-Simons Modified Gravity?. *Physical Review D*, 77:044015.

Gürsel, Y. (1983). Multipole moments for stationary systems: The equivalence of the geroch-hansen formulation and the thorne formulation. *General Relativity and Gravitation*, 15:737-754.

Hansen, R. O. (1974). Multipole moments of stationary space-times. *Journal of Mathematical Physics*, 15:4652.

Hartle, J. B. (1967). Slowly rotating relativistic stars. 1. Equations of structure. *The Astrophysical Journal*, 150:1005-1029.

Hartle, J. B. and Thorne, K. S. (1968). Slowly Rotating Relativistic Stars. II. Models for Neutron Stars and Supermassive Stars. *The Astrophysical Journal*, 153:807.

Hawking, S. W. and Ellis, G. F. R. (1973). *The Large Scale Structure of Space-Time*. Cambridge University Press, Cambridge.

Helstrom, C. W. (1968). *Statistical Theory of Signal Detection*. Pergamon, Oxford, UK.

Hogan, C. J. (2007). The New Science of Gravitational Waves. .

Hogan, C. J. and Binetruy, P. (2009). Gravitational Waves from New Physics. Astro2010 Science White Paper.

Hopman, C. and Alexander, T. (2006). The effect of mass-segregation on gravitational wave sources near massive black holes. *The Astrophysical Journal Letters*, 645:L133-L136.

Hulse, R. A. and Taylor, J. H. (1975). Discovery of a pulsar in a binary system. *The Astrophysical Journal*, 195:L51-L53.

Isaacson, R. A. (1968a). Gravitational Radiation in the Limit of High Frequency. I. The Linear Approximation and Geometrical Optics. *Physical Review*, 166:1263-1271.

Isaacson, R. A. (1968b). Gravitational Radiation in the Limit of High Frequency. II. Nonlinear Terms and the Effective Stress Tensor. *Physical Review*, 166:1272-1279.

Jackiw, R. and Pi, S. Y. (2003). Chern-simons modification of general relativity. *Physical Review D*, 68:104012.

Jetzer, P. (1992). Boson stars. *Physics Reports*, 220:163-227.

Johannsen, T. and Psaltis, D. (2010a). Testing the No-Hair Theorem with Observations in the Electromagnetic Spectrum: I. Properties of a Quasi-Kerr Spacetime. *The Astrophysical Journal*, 716:187-197.

Johannsen, T. and Psaltis, D. (2010b). Testing the No-Hair Theorem with Observations in the Electromagnetic Spectrum: II. Black-Hole Images. *The Astrophysical Journal*, 718:446-454.

Kerr, R. P. (1963). Gravitational field of a spinning mass as an example of algebraically special metrics. *Physical Review Letters*, 11:237-238.

Kesden, M., Gair, J. and Kamionkowski, M. (2005). Gravitational-wave signature of an inspiral into a supermassive horizonless object. *Physical Review D*, 71:044015.

Kundu, P. (1981). On the analyticity of stationary gravitational fields at spatial infinity. *Journal of Mathematical Physics*, 22:2006–2011.

Landau, L. D. and Lifshitz, E. M. (1971). *The Classical Theory of Fields*. Pergamon, Oxford.

Larson, S. L., Hiscock, W. A. and Hellings, R. W. (2000). Sensitivity curves for space-borne gravitational wave interferometers. *Physical Review D*, 62:062001.

Lukes-Gerakopoulos, G., Apostolatos, T. A. and Contopoulos, G. (2010). An observable signature of a background deviating from Kerr. *Physical Review D*, 81:124005.

Lyne, A. G. and others (2004). A Double-Pulsar System - A Rare Laboratory for Relativistic Gravity and Plasma Physics. *Science*, 303:1153-1157.

MacLeod, C. L. and Hogan, C. J. (2008). Precision of Hubble constant derived using black hole binary absolute distances and statistical redshift information. *Physical Review D*, 77:043512.

Madau, P. and others (2009). Massive Black Holes Across Cosmic Time. .

Mandel, I., Brown, D. A., Gair, J. R. and Miller, M. C. (2008). Rates and Characteristics of Intermediate-Mass-Ratio Inspirals Detectable by Advanced LIGO. *The Astrophysical Journal*, 681:1431.

Manko, V. S. and Novikov, I. D. (1992). Generalizations of the Kerr and Kerr-Newman metrics possessing an arbitrary set of mass-multipole moments. *Classical and Quantum Gravity*, 9:2477-2487.

Mazur, P. O. and Mottola, E. (2001). Gravitational condensate stars. .

Mazur, P. O. and Mottola, E. (2004). Gravitational vacuum condensate stars. *Proceedings of the National Academy of Sciences of the USA*, 101:9545-9550.

Mercuri, S. and Taveras, V. (2009). Interaction of the Barbero–Immirzi Field with Matter and Pseudo-Scalar Perturbations. *Physical Review D*, 80:104007.

Miller, M. C. (2008). Intermediate-Mass Black Holes as LISA Sources. *Classical and Quantum Gravity*.

Miller, M. C. and Colbert, E. J. M. (2004). Intermediate-Mass Black Holes. *International Journal of Modern Physics D*, 13:1-64.

Miller, M. C. and others (2009). Probing Stellar Dynamics in Galactic Nuclei. .

Mino, Y., Sasaki, M. and Tanaka, T. (1997). Gravitational radiation reaction to a particle motion. *Physical Review D*, 55:3457-3476.

Misner, C. W., Thorne, K. S. and Wheeler, J. A. (1973). *Gravitation*. W. H. Freeman & Co., San Francisco.

Munyaneza, F. and Viollier, R. D. (2002). The motion of stars near the Galactic center: A comparison of the black hole and fermion ball scenarios. *The Astrophysical Journal*, 564:274-283.

Nakamura, T., Shapiro, S. L. and Teukolsky, S. A. (1988). Naked singularities and the hoop conjecture: An analytic exploration. *Physical Review D*, 38(10):2972–2978.

Nelemans, G. (2006). Astrophysics of white dwarf binaries. *AIP Conf. Proc.*, 873:397-405.

Nelemans, G. (2009). The Galactic Gravitational wave foreground. *Classical and Quantum Gravity*, 26:094030.

Newman, E. T., Couch, E., Chinnapared, A., Exton, A. and Prakash, A.et al. (1965). . *Journal of Mathematical Physics*, 6:918.

Perlmutter, S. and others (1999). Measurements of Omega and Lambda from 42 High-Redshift Supernovae. *The Astrophysical Journal*, 517:565-586.

Peters, P. (1964). Gravitational Radiation and the Motion of Two Point Masses. *Physical Review*, 136:B1224–B1232.

Peters, P. and Mathews, J. (1963). Gravitational Radiation from Point Masses in a Keplerian Orbit. *Physical Review*, 131:435–440.

Petiteau, A. and others (2008). LISACode : A scientific simulator of LISA. *Physical Review D*, 77:023002.

Phinney, E. S. (2009). Finding and Using Electromagnetic Counterparts of Gravitational Wave Sources. .

Poisson, E. (1996). Measuring black-hole parameters and testing general relativity using gravitational-wave data from space-based interferometers. *Physical Review D*, 54:5939.

Poisson, E. (2004). The motion of point particles in curved spacetime. *Living Reviews in Relativity*, 7:6.

Poisson, E. and Will, C. M. (1995). Gravitational waves from inspiraling compact binaries: Parameter estimation using second-post-Newtonian waveforms. *Physical Review D*, 52:848–855.

Polchinski, J. (1998). *String theory. Vol. 2: Superstring theory and beyond*. Cambridge University Press, Cambridge, UK.

Pound, A. (2010b). Self-consistent gravitational self-force. *Physical Review D*, 81:024023.

Pound, A. (2010a). Singular perturbation techniques in the gravitational self-force problem. *Physical Review D*, 81:124009.

Prince, T. A. and others (2009). The Promise of Low-Frequency Gravitational Wave Astronomy. .

Prince, T. A., Tinto, M., Larson, S. L. and Armstrong, J. W. (2002). The LISA optimal sensitivity. *Physical Review D*, 66:122002.

Psaltis, D. (2008). Probes and tests of strong-field gravity with observations in the electromagnetic spectrum. *Living Reviews in Relativity*, 11(9).

Quinn, T. C. and Wald, R. M. (1997). An axiomatic approach to electromagnetic and gravitational radiation reaction of particles in curved spacetime. *Physical Review D*, 56:3381-3394.

Riess, A. G. and others (1998). Observational Evidence from Supernovae for an Accelerating Universe and a Cosmological Constant. *Astronomical Journal*, 116:1009-1038.

Rubbo, L. J., Cornish, N. J. and Poujade, O. (2003). The LISA simulator. *American Institute of Physics Conference Proceedings*, 686:307-310.

Ruffini, R. and Sasaki, M. (1981). ON A SEMIRELATIVISTIC TREATMENT OF THE GRAVITATIONAL RADIATION FROM A MASS THRUSTED INTO A BLACK HOLE. *Progress of Theoretical Physics*, 66:1627-1638.

Ruffini, R. and Wheeler, J. A. (1971). Introducing the black hole. *Physics Today*, 24:30-41.

Ryan, F. D. (1995a). Effect of gravitational radiation reaction on circular orbits around a spinning black hole. *Physical Review D*, 52:3159-3162.

Ryan, F. D. (1995b). Gravitational waves from the inspiral of a compact object into a massive, axisymmetric body with arbitrary multipole moments. *Physical Review D*, 52:5707-5718.

Ryan, F. D. (1996). Effect of gravitational radiation reaction on nonequatorial orbits around a Kerr black hole. *Physical Review D*, 53:3064-3069.

Ryan, F. D. (1997a). Accuracy of estimating the multipole moments of a massive body from the gravitational waves of a binary inspiral. *Physical Review D*, 56:1845-1855.

Ryan, F. D. (1997b). Scalar waves produced by a scalar charge orbiting a massive body with arbitrary multipole moments. *Physical Review D*, 56:7732-7739.

Sathyaprakash, B. S. and Schutz, B. F. (2009). Physics, Astrophysics and Cosmology with Gravitational Waves. *Living Reviews in Relativity*, 12(2).

Schmidt, W. (2002). Celestial mechanics in Kerr spacetime. *Classical and Quantum Gravity*, 19:2743.

Schunck, F. E. and Mielke, E. W. (2003). General relativistic boson stars. *Classical and Quantum Gravity*, 20:R301-R356.

Schutz, B. F. (2009). Fundamental physics with LISA. *Classical and Quantum Gravity*, 26:094020.

Schutz, B. F., Centrella, J., Cutler, C. and Hughes, S. A. (2009). Will Einstein Have the Last Word on Gravity?. .

Shapiro, I. I. (1999). A century of relativity. *Reviews of Modern Physics*, 71:S41–S53.

Shapiro, S. L. and Teukolsky, S. A. (1991). Formation of naked singularities: The violation of cosmic censorship. *Physical Review Letters*, 66(8):994–997.

Sopuerta, C. F. and Yunes, N. (2009). Extreme- and Intermediate-Mass Ratio Inspirals in Dynamical Chern-Simons Modified Gravity. *Physical Review D*, 80:064006.

Sopuerta, C. F. and Yunes, N. (2010). Improved Approximate Waveforms for Extreme-Mass-Ratio Inspirals: The Chimera Scheme. *In preparation*.

Sotiriou, T. P. and Apostolatos, T. A. (2005). Tracing the geometry around a massive, axisymmetric body to measure, through gravitational waves, its mass moments and electromagnetic moments. *Physical Review D*, 71:044005.

Spergel, D. N. and others (2003). First Year Wilkinson Microwave Anisotropy Probe (WMAP) Observations: Determination of Cosmological Parameters. *The Astrophysical Journal Supplement*, 148:175-194.

Stairs, I. H. (2003). Testing General Relativity with Pulsar Timing. *Living Reviews in Relativity*, 6(5).

Stamatikos, M., Gehrels, N., Halzen, F., Meszaros, P. and Roming, P. W. A. (2009). Multi-Messenger Astronomy with GRBs: A White Paper for the Astro2010 Decadal Survey. .

Stephani, H., Kramer, D., MacCallum, M., Hoenselaers, C. and Herlt, E. (2003). *Exact solutions of Einstein's field equations*. Cambridge University Press, Cambridge.

Stroeer, A. and Vecchio, A. (2006). The LISA verification binaries. *Classical and Quantum Gravity*, 23:S809-S818.

Tavaras, V. and Yunes, N. (2008). The Barbero-Immirzi Parameter as a Scalar Field: K- Inflation from Loop Quantum Gravity?. *Physical Review D*, 78:064070.

Teukolsky, S. A. (1972). Rotating Black Holes: Separable Wave Equations for Gravitational and Electromagnetic Perturbations. *Physical Review Letters*, 29:1114–1118.

Teukolsky, S. A. (1973). Perturbations of a Rotating Black Hole. I. Fundamental Equations for Gravitational, Electromagnetic, and Neutrino-Field Perturbations. *The Astrophysical Journal*, 185:635–648.

Thorne, K. S. (1972). Nonspherical Gravitational Collapse–A Short Review. In Klauder, J. R., editor, *Magic Without Magic: John Archibald Wheeler*, pages 231, San Francisco. W. H. Freeman.

Thorne, K. S. (1980). Multipole expansions of gravitational radiation. *Reviews of Modern Physics*, 52:299-339.

Turyshev, S. G. (2008). Experimental Tests of General Relativity. *Annual Review of Nuclear and Particle Science*, 58:207-248.

Vallisneri, M. (2008). Use and abuse of the Fisher information matrix in the assessment of gravitational-wave parameter-estimation prospects. *Physical Review D*, 77:042001.

Verbiest, J. P. W. and others (2010). Status Update of the Parkes Pulsar Timing Array. *Classical and Quantum Gravity*, 27:084015.

Vigeland, S. J. (2010). Multipole moments of bumpy black holes. .

Vigeland, S. J. and Hughes, S. A. (2010). Spacetime and orbits of bumpy black holes. *Physical Review D*, 81:024030.

Wald, R. M. (1984). *General Relativity*. The University of Chicago Press, Chicago.

Weinberg, S. (2008). Effective Field Theory for Inflation. *Physical Review D*, 77:123541.

Weinstein, L. A. and Zubakov, V. D. (1962). *Extraction of Signals from Noise*. Prentice-Hall, Englewood Cliffs, NJ.

Weisberg, J. M. and Taylor, J. H. (2005). Relativistic binary pulsar b1913+16: Thirty years of observations and analysis. *Astronomical Society of the Pacific Conference Series*, 328:25.

Will, C. M. (1987). Experimental gravitation from newton's principia to einstein's general relativity. In Hawking, S. W. and Israel, W., editors, *Three Hundred Years of Gravitation*, pages 80–127. Cambridge University Press, Cambridge, UK.

Will, C. M. (1992). The confrontation between general relativity and experiment: A 1992 update. *International Journal of Modern Physics D*, 1:13–68.

Will, C. M. (1993b). *Theory and Experiment in Gravitational Physics*. Cambridge University Press, Cambridge, UK, 2nd edition.

Will, C. M. (1993a). *Was Einstein Right?: Putting General Relativity to the Test*. Basic Books, New York, USA, 2nd edition.

Will, C. M. (1996). The confrontation between general relativity and experiment: A 1995 update. In Hall, G. S. and Pulham, J. R., editors, *General Relativity*, number 46 in Scottish Graduate Series, pages 239–282, Bristol, UK. Institute of Physics Publishing.

Will, C. M. (1998). The confrontation between general relativity and experiment: A 1998 update. In Dixon, L. J., editor, *Gravity: From the Hubble Length to the Planck Length*. SLAC.

Will, C. M. (2006). The confrontation between general relativity and experiment. *Living Reviews in Relativity*, 9(3).

Will, C. M. (2010). Resource Letter PTG-1: Precision Tests of Gravity. *American Journal of Physics*.

Xanthopoulos, B. C. (1979). Multipole moments in general relativity. *Journal of Physics A*, 12:1025–1028.

Yunes, N., Buonanno, A., Hughes, S. A., Coleman Miller, M. and Pan, Y. (2010). Modeling Extreme Mass Ratio Inspirals within the Effective-One-Body Approach. *Physical Review Letters*, 104:091102.

Yunes, N. and Pretorius, F. (2009b). Dynamical Chern-Simons Modified Gravity: Spinning Black Holes in the Slow-Rotation Approximation. *Physical Review D*, 79:084043.

Yunes, N. and Pretorius, F. (2009a). Fundamental Theoretical Bias in Gravitational Wave Astrophysics and the Parameterized Post-Einsteinian Framework. *Physical Review D*, 80:122003.

Yunes, N. and Sopuerta, C. F. (2008). Perturbations of Schwarzschild Black Holes in Chern-Simons Modified Gravity. *Physical Review D*, 77:064007.

Yunes, N. and Sopuerta, C. F. (2010). Testing Effective Quantum Gravity with Gravitational Waves from Extreme-Mass-Ratio Inspirals. *Journal of Physics: Conference Series*, 228:012051.

*Selected abstracts*

*April to August 2010*

## Characterizing Spinning Black Hole Binaries in Eccentric Orbits with LISA

**Authors:** Key, Joey Shapiro; Cornish, Neil J.

**Eprint:** http://arxiv.org/abs/1006.3759

**Keywords:** astro-ph.CO; data analysis; gr-qc; massive binaries of black holes; parameter estimation

**Abstract:** The Laser Interferometer Space Antenna (LISA) is designed to detect gravitational wave signals from astrophysical sources, including those from coalescing binary systems of compact objects such as black holes. Colliding galaxies have central black holes that sink to the center of the merged galaxy and begin to orbit one another and emit gravitational waves. Some galaxy evolution models predict that the binary black hole system will enter the LISA band with significant orbital eccentricity, while other models suggest that the orbits will already have circularized. Using a full seventeen parameter waveform model that includes the effects of orbital eccentricity, spin precession and higher harmonics, we investigate how well the source parameters can be inferred from simulated LISA data. Defining the reference eccentricity as the value one year before merger, we find that for typical LISA sources, it will be possible to measure the eccentricity to an accuracy of parts in a thousand. The accuracy with which the eccentricity can be measured depends only very weakly on the eccentricity, making it possible to distinguish circular orbits from those with very small eccentricities. LISA measurements of the orbital eccentricity can provide strong constraints on theories of galaxy mergers in the early universe.

## The Galactic Centre star S2 as a dynamical probe for intermediate-mass black holes

**Authors:** Gualandris, Alessia; Gillessen, Stefan; Merritt, David

**Eprint:** http://arxiv.org/abs/1006.3563

**Keywords:** astro-ph.GA; astrophysics; IMRI; intermediate-mass black holes; stellar dynamics; supermassive black holes

**Abstract:** We study the short-term effects of an intermediate mass black hole (IBH) on the orbit of star S2 (S02), the shortest period star known to orbit the supermassive black hole (SBH) in the centre of the Milky Way. Near-infrared imaging and spectroscopic observations allow an accurate determination of the orbit of the star. Given S2's short orbital period and large eccentricity, general relativity (GR) needs to be taken into account, and its effects are potentially measurable with current technology. We show that perturbations due to an IBH in orbit around the SBH can produce a shift in the apoapsis of S2 that is as large or even larger than the GR shift. An IBH will also induce changes in the plane of S2's orbit at a level as large as one degree per period. We apply observational orbital fitting techniques to simulations of the S-cluster in the presence of an IBH and find that an IBH more massive than about 1000 solar masses at the distance of the S-stars will be detectable at the next periapse passage of S2, which will occur in 2018.

## Constraints on Black Hole Growth, Quasar Lifetimes, and Eddington Ratio Distributions from the SDSS Broad Line Quasar Black Hole Mass Function

**Authors:** Kelly, Brandon C.; Vestergaard, Marianne; Fan, Xiaohui; Hopkins, Philip; Hernquist, Lars; Siemiginowska, Aneta

**Abstract:** We present an estimate of the black hole mass function (BHMF) of broad line quasars (BLQSOs) that self-consistently corrects for incompleteness and the statistical uncertainty in the mass estimates, based on a sample of 9886 quasars at $1 <$ z 1 it is highly incomplete at $M_{BH} < 10^9 M_\odot$ and $L/L_{\rm Edd} 1$, where the BLQSO phase occurs at the end of a fueling event when black hole feedback unbinds the accreting gas, halting the accretion flow.

## Highly accurate and efficient self-force computations using time-domain methods: Error estimates, validation, and optimization

**Authors:** Thornburg, Jonathan

**Abstract:**

If a small "particle" of mass $\mu M$ (with $\mu \ll 1$) orbits a Schwarzschild or Kerr black hole of mass $M$, the particle is subject to an $\varnothing(\mu)$ radiation-reaction "self-force". Here I argue that it's valuable to compute this self-force highly accurately (relative error of $\lesssim 10^{-6}$) and efficiently, and I describe techniques for doing this and for obtaining and validating error estimates for the computation. I use an adaptive-mesh-refinement (AMR) time-domain numerical integration of the perturbation equations in the Barack-Ori mode-sum regularization formalism; this is efficient, yet allows easy generalization to arbitrary particle orbits. I focus on the model problem of a scalar particle in a circular geodesic orbit in Schwarzschild spacetime.

The mode-sum formalism gives the self-force as an infinite sum of regularized spherical-harmonic modes $\sum_{\ell=0}^{\infty} F_{\ell,\mathrm{reg}}$, with $F_{\ell,\mathrm{reg}}$ (and an "internal" error estimate) computed numerically for $\ell \lesssim 30$ and estimated for larger $\sim \ell$ by fitting an asymptotic "tail" series. Here I validate the internal error estimates for the individual $F_{\ell,\mathrm{reg}}$ using a large set of numerical self-force computations of widely-varying accuracies. I present numerical evidence that the actual numerical errors in $F_{\ell,\mathrm{reg}}$ for different $\sim \ell$ are at most weakly correlated, so the usual statistical error estimates are valid for computing the self-force. I show that the tail fit is numerically ill-conditioned, but this can be mostly alleviated by renormalizing the basis functions to have similar magnitudes.

Using AMR, fixed mesh refinement, and extended-precision floating-point arithmetic, I obtain the (contravariant) radial component of the self-force for a particle in a circular geodesic orbit of areal radius $r = 10M$ to within $1 \sim$ ppm relative error.

## Full-analytic frequency-domain 1pN-accurate gravitational wave forms from eccentric compact binaries

**Authors:** Tessmer, Manuel; Schaefer, Gerhard

**Abstract:** The article provides ready-to-use 1pN-accurate frequency-domain gravitational wave forms for eccentric nonspinning compact binaries of arbitrary mass ratio including the first post-Newtonian (1pN) point particle corrections to the far-zone gravitational wave amplitude, given in terms of tensor spherical harmonics.

The averaged equations for the decay of the eccentricity and growth of radial frequency due to radiation reaction are used to provide stationary phase approximations to the frequency-domain wave forms.

## Papaloizou-Pringle Instability of Magnetized Accretion Tori

**Authors:** Fu, Wen; Lai, Dong

**Abstract:** Hot accretion tori around a compact object are known to be susceptible to a global hydrodynamical instability, the so-called Papaloizou-Pringle (PP) instability, arising from the interaction of non-axisymmetric waves across the corotation radius, where the wave pattern speed matches the fluid rotation rate. However, accretion tori produced in various astrophysical situations (e.g., collapsars and neutron star binary mergers) are likely to be highly magnetized. We study the effect of magnetic fields on the PP instability in incompressible tori with various magnetic strengths and structures. In general, toroidal magnetic fields have significant effects on the PP instability: For thin tori (with the fractional width relative to the outer torus radius much less than unity), the instability is suppressed at large field strengths with the corresponding toroidal Alfven speed $v_{A\phi} \gtrsim 0.2r\Omega$ (where $\Omega$ is the flow rotation rate). For thicker tori (with the fractional width of order 0.4 or larger), which are hydrodynamically stable, the instability sets in for sufficiently strong magnetic fields (with $v_{A\phi} \gtrsim 0.2r\Omega$). Our results suggest that highly magnetized accretion tori may be subjected to global instability even when it is stable against the usual magneto-rotational instability.

## Accretion onto Intermediate Mass Black Holes Regulated by Radiative Feedback I. Spherical Symmetric Accretion

**Authors:** Park, KwangHo; Ricotti, Massimo

**Abstract:** We study the effect of radiative feedback on accretion onto intermediate mass black holes (IMBHs) using the hydrodynamical code ZEUS-MP with a radiative transfer algorithm. In this paper, the first of a series, we assume accretion from a uniformly dense gas with zero angular momentum. Our 1D and 2D simulations explore how X-ray and UV radiation emitted near the black hole regulates the gas supply from large scales. Both 1D and 2D simulations show similar accretion rate and period between peaks in accretion, meaning that the hydro-instabilities that develop in 2D simulations do not affect the mean flow properties. We present a suite of simulations exploring accretion across a large parameter space, including different radiative efficiencies and radiation spectra, black hole masses, density and temperature, $T_\infty$, of the neighboring gas. In agreement with previous studies we find regular oscillatory behavior of the accretion rate, with duty cycle $\sim 7\%$, mean accretion rate 3-6% $(T_\infty/10^4 K)^{2.5}$ of the Bondi rate and peak accretion $\sim 10$ times the mean. We derive parametric formulas for the period between bursts, the mean accretion rate and the peak luminosity of the bursts and thus provide a formulation of how feedback regulated accretion operates. The temperature profile of the hot ionized gas is crucial in determining the accretion rate, while the period of the bursts is proportional to the mean size of the Strömgren sphere. We also find that softer spectrum of radiation produces higher accretion rate. This study is a first step to model the growth of seed black holes in the early universe and to make a prediction of the number and the luminosity of ultra-luminous X-ray sources in galaxies produced by IMBHs accreting from the interstellar medium.

## Higher order moment models of dense stellar systems: Applications to the modeling of the stellar velocity distribution function

**Authors:** Schneider, Justus; Amaro-Seoane, Pau; Spurzem, Rainer

**Abstract:** Dense stellar systems such as globular clusters, galactic nuclei and nuclear star clusters are ideal loci to study stellar dynamics due to the very high densities reached, usually a million times higher than in the solar neighborhood; they are unique laboratories to study processes related to relaxation. There are a number of different techniques to model the global evolution of such a system. In statistical models we assume that relaxation is the result of a large number of two-body gravitational encounters with a net local effect. We present two moment models that are based on the collisional Boltzmann equation. By taking moments of the Boltzmann equation one obtains an infinite set of differential moment equations where the equation for the moment of order $n$ contains moments of order $n+1$. In our models we assume spherical symmetry but we do not require dynamical equilibrium. We truncate the infinite set of moment equations at order $n = 4$ for the first model and at order $n = 5$ for the second model. The collisional terms on the right-hand side of the moment equations account for two-body relaxation and are computed by means of the Rosenbluth potentials. We complete the set of moment equations with closure relations which constrain the degree of anisotropy of our model by expressing moments of order $n + 1$ by moments of order $n$. The accuracy of this approach relies on the number of moments included from the infinite series. Since both models include fourth order moments we can study mechanisms in more detail that increase or decrease the number of high velocity stars. The resulting model allows us to derive a velocity distribution function, with unprecedented accuracy, compared to previous moment models.

## Spectropolarimetric evidence for a kicked supermassive black hole in the Quasar E1821+643

**Authors:** Robinson, Andrew; Young, Stuart; Axon, David J.; Kharb, Preeti; Smith, James E.

**Abstract:** We report spectropolarimetric observations of the quasar E1821+643 (z=0.297), which suggest that it may be an example of gravitational recoil due to anisotropic emission of gravitational waves following the merger of a supermassive black hole (SMBH) binary. In total flux, the broad Balmer lines are redshifted by $\sim 1000$ km/s relative to the narrow lines and have highly red asymmetric profiles, whereas in polarized flux the broad $H_\alpha$ line exhibits a blueshift of similar magnitude and a strong blue asymmetry. We show that these observations are consistent with a scattering model in which the broad-line region has two components, moving with different bulk velocities away from the observer and towards a scattering region at rest in the host galaxy. If the high velocity system is identified as gas bound to the SMBH, this implies that the SMBH is itself moving with a velocity $\sim 2100$ km/s relative to the host galaxy. We discuss some implications of the recoil hypothesis and also briefly consider whether our observations can be explained in terms of scattering of broad-line emission originating from the active component of an SMBH binary, or from an outflowing wind.

## Self consistent model for the evolution of eccentric massive black hole binaries in stellar environments: implications for gravitational wave observations

**Authors:** Sesana, A.

**Eprint:** http://arxiv.org/abs/1006.0730

**Keywords:** astro-ph.CO; astrophysics; gr-qc; massive binaries of black holes; stellar dynamics

**Abstract:** We construct evolutionary tracks for massive black hole binaries (MBHBs) embedded in a surrounding distribution of stars. The dynamics of the binary is evolved by taking into account the erosion of the central stellar cusp bound to the massive black holes, the scattering of unbound stars feeding the binary loss cone, and the emission of gravitational waves (GWs). Stellar dynamics is treated in a hybrid fashion by coupling the results of numerical 3-body scattering experiments of bound and unbound stars to an analytical framework for the evolution of the stellar density distribution and for the efficiency of the binary loss cone refilling. Our main focus is on the behaviour of the binary eccentricity, in the attempt of addressing its importance in the merger process and its possible impact for GW detection with the planned Laser Interferometer Space Antenna (*LISA*), and ongoing and forthcoming pulsar timing array (PTA) campaigns. We produce a family of evolutionary tracks extensively sampling the relevant parameters of the system which are the binary mass, mass ratio and initial eccentricity, the slope of the stellar density distribution, its normalization and the efficiency of loss cone refilling. We find that, in general, stellar dynamics causes a dramatic increase of the MBHB eccentricity, especially for initially already mildly eccentric and/or unequal mass binaries. When applied to standard MBHB population models, our results predict eccentricities in the ranges $10^{-3} - 0.2$ and $0.03 - 0.3$ for sources detectable by *LISA* and PTA respectively. Such figures may have a significant impact on the signal modelling, on source detection, and on the development of parameter estimation algorithms.

## MYRIAD: A new N-body code for simulations of Star Clusters

**Authors:** Konstantinidis, Simos; Kokkotas, Kostas D.

**Eprint:** http://arxiv.org/abs/1006.3326

**Abstract:** We present a new C++ code for collisional N-body simulations of star clusters. The code uses the Hermite fourth-order scheme with block time steps, for advancing the particles in time, while the forces and neighboring particles are computed using the GRAPE-6 board. Special treatment is used for close encounters, binary and multiple sub-systems that either form dynamically or exist in the initial configuration. The structure of the code is modular and allows the appropriate treatment of more physical phenomena, such as stellar and binary evolution, stellar collisions and evolution of close black-hole binaries. Moreover, it can be easily modified so that the part of the code that uses GRAPE-6, could be replaced by another module that uses other accelerating-hardware like the Graphics Processing Units (GPUs). Appropriate choice of the free parameters give a good accuracy and speed for simulations of star clusters up to and beyond core collapse. Simulations of Plummer models consisting of equal-mass stars reached core collapse at $t \sim 17$ half-mass relaxation times, which compares very well with existing results, while the cumulative relative error in the energy remained below 0.001. Also, comparisons with published results of other codes for the time of core collapse for different initial conditions, show excellent agreement. Simulations of King models with an initial mass-function, similar to those found in the literature, reached core collapse at $t \sim 0.17$, which is slightly smaller than the expected result from previous works. Finally, the code accuracy becomes comparable and even better than the accuracy of existing codes, when a number of close binary systems is dynamically created in a simulation. This is due to the high accuracy of the method that is used for close binary and multiple sub-systems.

## Statistical constraints on binary black hole inspiral dynamics

**Authors:** Galley, Chad R.; Herrmann, Frank; Silberholz, John; Tiglio, Manuel; Guerberoff, Gustavo

**Abstract:** We perform a statistical analysis of the binary black hole problem in the post-Newtonian approximation by systematically sampling and evolving the parameter space of initial configurations for quasi-circular inspirals. Through a principal component analysis of spin and orbital angular momentum variables we systematically look for uncorrelated quantities and find three of them which are highly conserved in a statistical sense, both as functions of time and with respect to variations in initial spin orientations. We also look for and find the variables that account for the largest variations in the problem. We present binary black hole simulations of the full Einstein equations analyzing to what extent these results might carry over to the full theory in the inspiral and merger regimes. Among other applications these results should be useful both in semi-analytical and numerical building of templates of gravitational waves for gravitational wave detectors.

## The Lagrange Equilibrium Points $L_4$ and $L_5$ in a Black Hole Binary System

**Authors:** Schnittman, Jeremy D.

**Abstract:** We calculate the location and stability of the $L_4$ and $L_5$ Lagrange equilibrium points in the circular restricted three-body problem as the binary system evolves via gravitational radiation losses. Relative to the purely Newtonian case, we find that the $L_4$ equilibrium point moves towards the secondary mass and becomes slightly less stable, while the $L_5$ point moves away from the secondary and gains in stability. We discuss a number of astrophysical applications of these results, in particular as a mechanism for producing electromagnetic counterparts to gravitational-wave signals.

## The Massive Black Hole and Nuclear Star Cluster in the Center of the Milky Way

**Authors:** Genzel, Reinhard; Eisenhauer, Frank; Gillessen, Stefan

**Abstract:** The Galactic Center is an excellent laboratory for studying phenomena and physical occurring in many other galactic nuclei. The Center of our Milky

Way is by far the closest galactic nucleus, and observations with exquisite resolution and sensitivity cover 18 orders of magnitude in energy of electromagnetic radiation. Theoretical simulations have become increasingly more powerful in explaining these measurements. This review summarizes the recent progress in observational and theoretical work on the central parsec, with a strong emphasis on the current empirical evidence for a central massive black hole and on the properties of the surrounding dense star cluster. We present the current evidence, from the analysis of the orbits of more than two dozen stars and from the measurements of the size and motion of the central compact radio source, Sgr A*, that this radio source must be a massive black hole of about $4.4 \times 10^6 M_\odot$, beyond any reasonable doubt. We report what is known about the structure and evolution of the dense nuclear star cluster surrounding this black hole, including the astounding fact that stars have been forming in the vicinity of Sgr A* recently, apparently with a top-heavy stellar mass function. We discuss a dense concentration of fainter stars centered in the immediate vicinity of the massive black hole, three of which have orbital peri-bothroi of less than one light day. This 'S-star cluster' appears to consist mainly of young early-type stars, in contrast to the predicted properties of an equilibrium 'stellar cusp' around a black hole. This constitutes a remarkable and presently not fully understood 'paradox of youth'. We also summarize more briefly what is known about the emission properties of the accreting gas onto Sgr A* and how this emission is beginning to delineate the physical properties in the hot accretion zone around the event horizon.

## Resonant relaxation and the warp of the stellar disc in the Galactic centre

**Authors:** Kocsis, Bence; Tremaine, Scott

**Abstract:** Observations of the spatial distribution and kinematics of young stars in the Galactic centre can be interpreted as showing that the stars occupy one, or possibly two, discs of radii ∼ 0.05-0.5 pc. The most prominent ('clockwise') disc exhibits a strong warp: the normals to the mean orbital planes in the inner and outer third of the disc differ by ∼ 60 deg. Using an analytical model based on Laplace-Lagrange theory, we show that such warps arise naturally and inevitably through vector resonant relaxation between the disc and the surrounding old stellar cluster.

## Next to leading order spin-orbit effects in the motion of inspiralling compact binaries

**Authors:** Porto, Rafael A.

**Eprint:** http://arxiv.org/abs/1005.5730

**Keywords:** astro-ph.CO; gr-qc; hep-ph; hep-th; massive binaries of black holes; spin

**Abstract:** Using effective field theory (EFT) techniques we calculate the next-to-leading order (NLO) spin-orbit contributions to the gravitational potential of inspiralling compact binaries. We use the covariant spin supplementarity condition (SSC), and explicitly prove the equivalence with previous results by Faye et al. in arXiv:gr-qc/0605139. We also show that the direct application of the Newton-Wigner SSC at the level of the action leads to the correct dynamics using a canonical (Dirac) algebra. This paper then completes the calculation of the necessary spin dynamics within the EFT formalism that will be used in a separate paper to compute the spin contributions to the energy flux and phase evolution to NLO.

## Gravitational Waves, Sources, and Detectors

**Authors:** Schutz, Bernard F; Ricci, Franco

**Eprint:** http://arxiv.org/abs/1005.4735

**Keywords:** astro-ph.HE; gr-qc; notes of lectures

**Abstract:** Notes of lectures for graduate students that were given at Lake Como in 1999, covering the theory of linearized gravitational waves, their sources, and the prospects at the time for detecting gravitational waves. The lectures remain of interest for pedagogical reasons, and in particular because they contain a treatment of current-quadrupole gravitational radiation (in connection with the r-modes of neutron stars) that is not readily available in other sources.

## Gravitational signature of Schwarzschild black holes in dynamical Chern-Simons gravity

**Authors:** Molina, C.; Pani, Paolo; Cardoso, Vitor; Gualtieri, Leonardo

**Eprint:** http://arxiv.org/abs/1004.4007

Notes & News for GW science

**Abstract:** Dynamical Chern-Simons gravity is an extension of General Relativity in which the gravitational field is coupled to a scalar field through a parity-violating Chern-Simons term. In this framework, we study perturbations of spherically symmetric black hole spacetimes, assuming that the background scalar field vanishes. Our results suggest that these spacetimes are stable, and small perturbations die away as a ringdown. However, in contrast to standard General Relativity, the gravitational waveforms are also driven by the scalar field. Thus, the gravitational oscillation modes of black holes carry imprints of the coupling to the scalar field. This is a smoking gun for Chern-Simons theory and could be tested with gravitational-wave detectors, such as LIGO or LISA. For negative values of the coupling constant, ghosts are known to arise, and we explicitly verify their appearance numerically. Our results are validated using both time evolution and frequency domain methods.

## Relativistic models of magnetars: structure and deformations

**Authors:** Colaiuda, A.; Ferrari, V.; Gualtieri, L.; Pons, J. A.

**Abstract:** We find numerical solutions of the coupled system of Einstein-Maxwell's equations with a linear approach, in which the magnetic field acts as a perturbation of a spherical neutron star. In our study, magnetic fields having both poloidal and toroidal components are considered, and higher order multipoles are also included. We evaluate the deformations induced by different field configurations, paying special attention to those for which the star has a prolate shape. We also explore the dependence of the stellar deformation on the particular choice of the equation of state and on the mass of the star. Our results show that, for neutron stars with mass M = 1.4 $M_\odot$ and surface magnetic fields of the order of $10^{15}$ G, a quadrupole ellipticity of the order of $10^{-6} - 10^{-5}$ should be expected. Low mass neutron stars are in principle subject to larger deformations (quadrupole ellipticities up to $10^{-3}$ in the most extreme case). The effect of quadrupolar magnetic fields is comparable to that of dipolar components. A magnetic field permeating the whole star is normally needed to obtain negative quadrupole ellipticities, while fields confined to the crust typically produce positive quadrupole ellipticities.

## Stochastic backgrounds of gravitational waves from extragalactic sources

**Authors:** Schneider, Raffaella; Marassi, Stefania; Ferrari, Valeria

**Eprint:** http://arxiv.org/abs/1005.0977

**Keywords:** astro-ph.CO; back/foreground; cosmology; gr-qc

**Abstract:** Astrophysical sources emit gravitational waves in a large variety of processes occurred since the beginning of star and galaxy formation. These waves permeate our high redshift Universe, and form a background which is the result of the superposition of different components, each associated to a specific astrophysical process. Each component has different spectral properties and features that it is important to investigate in view of a possible, future detection. In this contribution, we will review recent theoretical predictions for backgrounds produced by extragalactic sources and discuss their detectability with current and future gravitational wave observatories.

## Supermassive Black Hole Formation at High Redshifts Through a Primordial Magnetic Field

**Authors:** Sethi, Shiv K.; Haiman, Zoltán; Pandey, Kanhaiya

**Eprint:** http://arxiv.org/abs/1005.2942

**Keywords:** astro-ph.CO; cosmology; supermassive black holes

**Abstract:** It has been proposed that primordial gas in early dark matter halos, with virial temperatures above $10^4$ K, can avoid fragmentation and undergo rapid collapse, possibly resulting in a supermassive black hole (SMBH). This requires the gas to avoid cooling and to remain at temperatures near $T=10^4$ K. We show that this condition can be satisfied in the presence of a sufficiently strong primordial magnetic field, which heats the collapsing gas via ambipolar diffusion. If the field has a strength above B = 3.6 (comoving) nG, the collapsing gas is kept warm ($T=10^4$K) until it reaches the critical density $n_{crit} = 10^3 cm^{-3}$ at which the roto-vibrational states of $H_2$ approach local thermodynamic equilibrium. $H_2$-cooling then remains inefficient, and the gas temperature stays near $10^4$K, even as it continues to collapse to higher densities. The critical magnetic field strength required to permanently suppress $H_2$-cooling is somewhat higher than upper limit of approx. 2 nG from the cosmic microwave background (CMB). However, it can be realized in the rare (2-3)-sigma regions of the spatially fluctuating B-field; these regions contain a sufficient number of halos to account for the z=6 quasar BHs.

## Autonomous perturbations of LISA orbits

**Authors:** Pucacco, Giuseppe; Bassan, Massimo; Visco, Massimo

**Abstract:** We investigate autonomous perturbations on the orbits of LISA, namely the effects produced by fields that can be expressed only in terms of the position, but not of time in the Hill frame. This first step in the study of the LISA orbits has been the subject of recent papers which implement analytical techniques based on a "post-epicyclic" approximation in the Hill frame to find optimal unperturbed orbits. The natural step forward is to analyze the perturbations to purely Keplerian orbits. In the present work a particular emphasis is put on the tidal field of the Earth assumed to be stationary in the Hill frame. An accurate interpretation of the global structure of the perturbed solution sheds light on possible implications on injection in orbit when the time base-line of the mission is longer than that assumed in previous papers. Other relevant classes of autonomous perturbations are those given by the corrections to the Solar field responsible for a slow precession and a global stationary field, associated to sources like the interplanetary dust or a local dark matter component. The inclusion of simple linear contributions in the expansion of these fields produces secular solutions that can be compared with the measurements and possibly used to evaluate some morphological property of the perturbing components.

## Supermassive black hole spin-flip during the inspiral

**Authors:** Gergely, Laszlo A.; Biermann, Peter L.; Caramete, Laurentiu I.

**Abstract:** During post-Newtonian evolution of a compact binary, a mass ratio different from one provides a second small parameter, which can lead to unexpected results. We present a statistics of supermassive black hole candidates, which enables us first to derive their mass distribution, then to establish a logarithmically even probability of the mass ratios at their encounter. In the mass ratio range $(1/30, 1/3)$ of supermassive black hole mergers representing 40% of all possible cases, the combined effect of spin-orbit precession and gravitational radiation leads to a spin-flip of the dominant spin during the inspiral phase of the merger. This provides a mechanism for explaining a large set of observations on X-shaped radio galaxies. In another 40%, with mass ratios (1/30,1/1000) a spin-flip never happens, while in the remaining 20% of mergers with mass ratios (1/3,1) it may occur during the plunge. We analyze the magnitude of the spin-flip angle occurring during the inspiral as function of the mass ratio and original relative orientation of the spin and orbital angular momentum. We also derive a formula for the final spin at the end of the inspiral in this mass ratio range.

## Experimental Demonstration of Time-Delay Interferometry for the Laser Interferometer Space Antenna

**Authors:** de Vine, Glenn; Ware, Brent; McKenzie, Kirk; Spero, Robert E.; Klipstein, William M.; Shaddock, Daniel A.

**Abstract:** We report on the first demonstration of time-delay interferometry (TDI) for LISA, the Laser Interferometer Space Antenna. TDI was implemented in a laboratory experiment designed to mimic the noise couplings that will occur in LISA. TDI suppressed laser frequency noise by approximately $10^9$ and clock phase noise by $6 \times 10^4$, recovering the intrinsic displacement noise floor of our laboratory test bed. This removal of laser frequency noise and clock phase noise in post-processing marks the first experimental validation of the LISA measurement scheme.

## A Displaced Supermassive Black Hole in M87

**Authors:** Batcheldor, D.; Robinson, A.; Axon, D. J.; Perlman, E. S.; Merritt, D.

**Abstract:** Isophotal analysis of M87, using data from the Advanced Camera for Surveys, reveals a projected displacement of 6.8 +/- 0.8 pc (~ 0.1 arcsec) between the nuclear point source (presumed to be the location of the supermassive black hole, SMBH) and the photo-center of the galaxy. The displacement is along a position angle of 307 +/- 17 degrees and is consistent with the jet axis. This suggests the active

SMBH in M87 does not currently reside at the galaxy center of mass, but is displaced in the counter-jet direction. Possible explanations for the displacement include orbital motion of an SMBH binary, gravitational perturbations due to massive objects (e.g., globular clusters), acceleration by an asymmetric or intrinsically one-sided jet, and gravitational recoil resulting from the coalescence of an SMBH binary. The displacement direction favors the latter two mechanisms. However, jet asymmetry is only viable, at the observed accretion rate, for a jet age of > 0.1 Gyr and if the galaxy restoring force is negligible. This could be the case in the low density core of M87. A moderate recoil ~ 1 Myr ago might explain the disturbed nature of the nuclear gas disk, could be aligned with the jet axis, and can produce the observed offset. Alternatively, the displacement could be due to residual oscillations resulting from a large recoil that occurred in the aftermath of a major merger any time in the last 10 Gyr.

## Spin effects in the phasing of gravitational waves from binaries on eccentric orbits

**Authors:** Klein, Antoine; Jetzer, Philippe

**Abstract:** We compute here the spin-orbit and spin-spin couplings needed for an accurate computation of the phasing of gravitational waves emitted by comparable-mass binaries on eccentric orbits at the second post-Newtonian (PN) order. We use a quasi-Keplerian parametrization of the orbit free of divergencies in the zero eccentricity limit. We find that spin-spin couplings induce a residual eccentricity for coalescing binaries at 2PN, of the order of $10^{-4}$-$10^{-3}$ for supermassive black hole binaries in the LISA band. Spin-orbit precession also induces a non-trivial pattern in the evolution of the eccentricity, which could help to reduce the errors on the determination of the eccentricity and spins in a gravitational wave measurement.

## Detection, Localization and Characterization of Gravitational Wave Bursts in a Pulsar Timing Array

**Authors:** Finn, Lee Samuel; Lommen, Andrea N.

**Abstract:** Efforts to detect gravitational waves by timing an array of pulsars have focused traditionally on stationary gravitational waves: e.g., stochastic or periodic signals. Gravitational wave bursts — signals whose duration is much shorter than the observation period — will also arise in the pulsar timing array waveband. Sources that give rise to detectable bursts include the formation or coalescence of supermassive black holes (SMBHs), the periapsis passage of compact objects in highly elliptic or unbound orbits about a SMBH, or cusps on cosmic strings. Here we describe how pulsar timing array data may be analyzed to detect and characterize these bursts. Our analysis addresses, in a mutually consistent manner, a hierarchy of three questions: *i) What are the odds that a dataset includes the signal from a gravitational wave burst?* ii) Assuming the presence of a burst, what is the direction to its source? and *iii) Assuming the burst propagation direction, what is the burst waveform's time dependence in each of its polarization states? Applying our analysis to synthetic data sets we find that we can* detect gravitational waves even when the radiation is too weak to either localize the source of infer the waveform, and *detect and* localize sources even when the radiation amplitude is too weak to permit the waveform to be determined. While the context of our discussion is gravitational wave detection via pulsar timing arrays, the analysis itself is directly applicable to gravitational wave detection using either ground or space-based detector data.

## Witnessing the Birth of a Quasar

**Authors:** Tanaka, Takamitsu; Haiman, Zoltan; Menou, Kristen

**Abstract:** The coalescence of a supermassive black hole binary (SMBHB) is thought to be accompanied by an electromagnetic (EM) afterglow, produced by the viscous infall of the surrounding circumbinary gas disk after the merger. It has been proposed that once the merger has been detected in gravitational waves (GWs) by LISA, follow-up EM searches for this afterglow can help identify the EM counterpart of the LISA source. Here we study whether the afterglows may be sufficiently bright and numerous to be detectable in EM surveys alone. The viscous afterglow, which lasts for years to decades for SMBHBs in LISA's sensitivity window, is characterized by rapid increases in both the bolometric luminosity and in the spectral hardness of the source. If quasar activity is triggered by the same major galaxy mergers that produce SMBHBs, then the afterglow could be interpreted as a signature of the birth of a quasar. Using an idealized model for the post-merger viscous spreading of the circumbinary disk and the resulting light curve, and using the observed luminosity function of quasars as a proxy for the SMBHB merger rate, we delineate the survey requirements for identifying such birthing quasars. If circumbinary disks have a high disk surface density and viscosity, an all-sky soft X-ray survey with a sensitivity of ~ 10%/yr. If >1% of the X-ray emission is reprocessed into optical frequencies, birthing quasars could also be identified in optical transient surveys such as the LSST. Distinguishing a birthing quasar from other variable sources may be facilitated by the monotonic hardening of its spectrum, but will likely remain challenging. This reinforces the notion that joint EM-plus-GW observations offer the best prospects for identifying the EM signatures of SMBHB mergers.

## Computing waveforms for spinning compact binaries in quasi-eccentric orbits

**Authors:** Cornish, Neil J.; Key, Joey Shapiro

**Abstract:** Several scenarios have been proposed in which the orbits of binary black holes enter the band of a gravitational wave detector with significant eccentricity. To avoid missing these signals or biasing the parameter estimation it is important that we consider waveform models that account for eccentricity. The ingredients needed to compute post-Newtonian (PN) waveforms produced by spinning black holes inspiralling on quasi-eccentric orbits have been available for almost two decades at 2 PN order, and this work has recently been extended to 2.5 PN order. However, the computational cost of directly implementing these waveforms is high, requiring many steps per orbit to evolve the system of coupled differential equations. Here we employ a separation of timescales and a generalized Keplarian parameterization of the orbits to produce efficient waveforms describing spinning black hole binaries with arbitrary spin orientations on quasi-eccentric orbits to 1.5 PN order. Our solution includes the spin contributions to the decay of the semi-major axis and eccentricity. We outline a scheme for extending our approach to higher post-Newtonian order.

# Reducing the weak lensing noise for the gravitational wave Hubble diagram using the non-Gaussianity of the magnification distribution

**Authors:** Hirata, Christopher M.; Holz, Daniel E.; Cutler, Curt

**Eprint:** http://arxiv.org/abs/1004.3988

**Keywords:** astro-ph.CO; cosmology

**Abstract:** Gravitational wave sources are a promising cosmological standard candle because their intrinsic luminosities are determined by fundamental physics (and are insensitive to dust extinction). They are, however, affected by weak lensing magnification due to the gravitational lensing from structures along the line of sight. This lensing is a source of uncertainty in the distance determination, even in the limit of perfect standard candle measurements. It is commonly believed that the uncertainty in the distance to an ensemble of gravitational wave sources is limited by the standard deviation of the lensing magnification distribution divided by the square root of the number of sources. Here we show that by exploiting the non-Gaussian nature of the lensing magnification distribution, we can improve this distance determination, typically by a factor of 2–3; we provide a fitting formula for the effective distance accuracy as a function of redshift for sources where the lensing noise dominates.

# On the transition from nuclear-cluster to black-hole dominated galaxy cores

**Authors:** Bekki, Kenji; Graham, Alister W.

**Eprint:** http://arxiv.org/abs/1004.3627

**Keywords:** astro-ph.CO; astrophysics; cosmology; massive binaries of black holes; supermassive black holes

**Abstract:** Giant elliptical galaxies, believed to be built from the merger of lesser galaxies, are known to house a massive black hole at their center rather than a compact star cluster. If low- and intermediate-mass galaxies do indeed partake in the hierarchical merger scenario, then one needs to explain why their dense nuclear star clusters are not preserved in merger events. A valuable clue may the recent revelation that nuclear star clusters and massive black holes frequently co-exist in intermediate mass bulges and elliptical galaxies. In an effort to understand the physical mechanism responsible for the disappearance of nuclear star clusters, we have numerically investigated the evolution of merging star clusters with seed black holes.

Using black holes that are 1-5% of their host nuclear cluster mass, we reveal how their binary coalescence during a merger dynamically heats the newly wed star cluster, expanding it, significantly lowering its central stellar density, and thus making it susceptible to tidal destruction during galaxy merging. Moreover, this mechanism provides a pathway to explain the observed reduction in the nucleus-to-galaxy stellar mass ratio as one proceeds from dwarf to giant elliptical galaxies.

## Cosmology with Standard Sirens: the Importance of the Shape of the Lensing Magnification Distribution

**Authors:** Shang, Cien; Haiman, Zoltan

**Abstract:** The gravitational waves (GWs) emitted by inspiraling binary black holes, expected to be detected by the Laser Interferometer Space Antenna (LISA), could be used to determine the luminosity distance to these sources with the unprecedented precision of $<\sim$ 1%. We study cosmological parameter constraints from such standard sirens, in the presence of gravitational lensing by large-scale structure. Lensing introduces magnification with a probability distribution function (PDF) whose shape is highly skewed and depends on cosmological parameters. We use Monte-Carlo simulations to generate mock samples of standard sirens, including a small intrinsic scatter, as well as the additional, larger scatter from lensing, in their inferred distances. We derive constraints on cosmological parameters, by simultaneously fitting the mean and the distribution of the residuals on the distance vs redshift ($d_L$ - z) Hubble diagram. We find that for standard sirens at redshift z ~ 1, the sensitivity to a single cosmological parameter, such as the matter density $\Omega_m$, or the dark energy equation of state w, is $\sim$ 50% − 80% tighter when the skewed lensing PDF is used, compared to the sensitivity derived from a Gaussian PDF with the same variance. When these two parameters are constrained simultaneously, the skewness yields a further enhanced improvement (by $\sim$ 120%), owing to the correlation between the parameters. The sensitivity to the amplitude of the matter power spectrum, $\sigma_8$ from the cosmological dependence of the PDF alone, however, is $\sim$ 20% worse than that from the Gaussian PDF. At higher redshifts, the PDF resembles a Gaussian more closely, and the effects of the skewness become less prominent. These results highlight the importance of obtaining an accurate and reliable PDF of the lensing convergence, in order to realize the full potential of standard sirens as cosmological probes.

## Momentum-driven winds and positive AGN feedback

**Authors:** Silk, Joe; Nusser, Adi

**Abstract:** Force balance considerations put a limit on the rate of AGN radiation momentum output, $L/c$, capable of driving galactic superwinds. We show that this condition is insufficient: black holes obeying the observed $M_\bullet - \sigma$ relation cannot supply enough energy in radiation which can drive the gas out by pressure alone. The shortfall is by up to an order of magnitude in most, but not all, cases. We propose that outflow-triggering of star formation by enhancing the intercloud medium turbulent pressure and squeezing clouds can supply the necessary boost, and suggest possible tests of this hypothesis. We further point out that the time-scales for Bondi accretion and for orbital decay of merging clumps by dynamical friction in the nuclear disk around a central black hole both follow a similar scaling with mass, favoring the most massive black holes, but the latter process is up to two orders of magnitude more rapid at $z \gtrsim 10$. The combination of accretion and coalescence results in earlier formation of more massive black holes, and, in particular, can account for the masses of the black holes inferred to power AGN at $z \sim 6$.

## Inflow-Outflow Model with Conduction and Self-Consistent Feeding for Sgr A*

**Authors:** Shcherbakov, Roman V.; Baganoff, Frederick K.

**Abstract:** We propose a two-temperature radial inflow-outflow model near Sgr A* with self-consistent feeding and conduction. Stellar winds from individual stars are considered to find the rates of mass injection and energy injection. These source terms help to partially eliminate the boundary conditions on the inflow. Electron thermal conduction is crucial for inhibiting the accretion. Energy diffuses out from several gravitational radii, unbinding more gas at several arcseconds and limiting the accretion rate to <1% of Bondi rate. We successfully fit the X-Ray surface brightness profile found from the extensive Chandra observations and reveal the X-Ray point source in the center. The super-resolution technique allows us to infer the presence and estimate the unabsorbed luminosity $L \approx 4 \cdot 10^{32} \mathrm{ergs}^{-1}$ of the point source. The employed relativistic heat capacity and direct heating of electrons naturally lead to low electron temperature $T_e \approx 4 \cdot 10^{10}$ K near the black hole. Within the same model we fit 86 GHz optically thick emission and obtain the order of magnitude agreement of Faraday rotation measure, thus achieving a single accretion model suitable at all radii.

## Parsec-Scale Localization of the Quasar SDSS J1536+0441A, a Candidate Binary Black Hole System

**Authors:** Wrobel, J. M.; Laor, A.

**Abstract:** The radio-quiet quasar SDSS J1536+0441A shows two broad-line emission systems, recently interpreted as a binary black hole (BBH) system with a subparsec separation; as a double-peaked emitter; or as both types of systems. The NRAO VLBA was used to search for 8.4 GHz emission from SDSS J1536+0441A, focusing on the optical localization region for the broad-line emission, of area 5400 mas$^2$ (0.15 kpc$^2$). One source was detected, with a diameter of less than 1.63 mas (8.5 pc) and a brightness temperature $T_b > 1.2x10^7$ K. New NRAO VLA photometry at 22.5 GHz, and earlier photometry at 8.5 GHz, gives a rising spectral slope of $\alpha = 0.35+/-0.08$. The slope implies an optically thick synchrotron source, with a radius of about 0.04 pc, and thus $T_b \sim 5x10^{10}$ K. The implied radio-sphere at rest frame 31.2 GHz has a radius of 800 gravitational radii, just below the size of the broad line region in this object. Observations at higher frequencies can probe whether or not the radio-sphere is as compact as expected from the coronal framework for the radio emission of radio-quiet quasars.

## A correlation between central supermassive black holes and the globular cluster systems of early-type galaxies

**Authors:** Burkert, Andreas; Tremaine, Scott

**Abstract:** Elliptical, lenticular, and early-type spiral galaxies show a remarkably tight power-law correlation between the mass $M_{BH}$ of their central supermassive black hole (SMBH) and the number $N_{GC}$ of globular clusters: $M_{BH} =$

$m$ times $N_{\mathrm{GC}}^{1.11+/-0.04}$ with $m = 1.3 \times 10^5$ solar masses. Thus, to a good approximation the SMBH mass is the same as the total mass of the globular clusters. Based on a limited sample of 13 galaxies, this relation appears to be a better predictor of SMBH mass (rms scatter 0.2 dex) than the $M_{\mathrm{BH}} - \sigma$ relation between SMBH mass and velocity dispersion sigma. The small scatter reflects the fact that galaxies with high globular cluster specific frequency $S_N$ tend to harbor SMBHs that are more massive than expected from the $M_{\mathrm{BH}} - \sigma$ relation. A possible explanation is that both large black-hole masses and large globular cluster populations are associated with recent major mergers.

# Conservative self-force correction to the innermost stable circular orbit: comparison with multiple post-Newtonian-based methods

**Authors:** Favata, Marc

**Abstract:** [abridged] Barack and Sago have recently computed the shift of the innermost stable circular orbit (ISCO) due to the conservative self-force that arises from the finite-mass of an orbiting test-particle. This is one of the first concrete results of the self-force program, and provides an exact point of comparison with approximate post-Newtonian (PN) computations of the ISCO. Here this exact ISCO shift is compared with nearly all known PN-based methods. These include both "non-resummed" and "resummed" approaches (the latter reproduce the test-particle limit by construction). The best agreement with the exact result is found from effective-one-body (EOB) calculations that are fit to numerical relativity simulations. However, if one considers uncalibrated methods based only on the currently-known 3PN-order conservative dynamics, the best agreement is found from the gauge-invariant ISCO condition of Blanchet and Iyer (2003). This method reproduces the exact test-particle limit without any resummation. A comparison of PN methods with the equal-mass ISCO is also performed. The results of this study suggest that the EOB approach—while exactly incorporating the conservative test-particle dynamics—does not (in the absence of calibration) incorporate conservative self-force effects more accurately than standard PN methods. I also consider how the conservative self-force ISCO shift, combined with numerical relativity computations of the ISCO, can be used to constrain our knowledge of (1) the EOB effective metric, (2) phenomenological inspiral-merger-ringdown templates, and (3) 4PN and 5PN order terms in the PN orbital energy. These constraints could help in constructing better gravitational-wave templates. Lastly, I suggest a new method to calibrate unknown PN-terms in inspiral templates using "low-cost" numerical-relativity calculations.

## The Globular Cluster/Central Black Hole Connection in Galaxies

**Authors:** Harris, Gretchen L. H.; Harris, William E.

**Abstract:** We explore the relation between the total globular cluster population in a galaxy ($N_{GC}$) and the the mass of its central black hole ($M_{BH}$). Using a sample of 33 galaxies, twice as large as the original sample discussed by Burkert & Tremaine (2010), we find that $N_{GC}$ for elliptical and spiral galaxies increases in almost precisely direct proportion to $M_{BH}$. The S0-type galaxies by contrast do not follow a clear trend, showing large scatter in $M_{BH}$ at a given $N_{GC}$. After accounting for observational measurement uncertainty, we find that the mean relation defined by the E and S galaxies must also have an intrinsic or "cosmic" scatter of +-0.2 in either $\log N_{GC}$ or $\log M_{BH}$. The residuals from this correlation show no trend with globular cluster specific frequency. We suggest that these two types of galaxy subsystems (central black hole and globular cluster system) may be closely correlated because they both originated at high redshift during the main epoch of hierarchical merging, and both require extremely high-density conditions for formation. Lastly, we note that roughly 10% of the galaxies in our sample (one E, one S, and two S0) deviate strongly from the main trend, all in the sense that their $M_{BH}$ is at least 10x smaller than would be predicted by the mean relation.

## A Local Baseline of the Black Hole Mass Scaling Relations for Active Galaxies. I. Methodology and Results of Pilot Study

**Authors:** Bennert, Vardha Nicola; Auger, Matthew W.; Treu, Tommaso; Woo, Jong-Hak; Malkan, Matthew A.

**Abstract:** We present high-quality Keck/LRIS longslit spectroscopy of a pilot sample of 25 local active galaxies selected from the SDSS ($0.02 < z$10$^7 M_\odot$) to study the relations between black hole mass (MBH) and host-galaxy properties. We determine stellar kinematics of the host galaxy with an unprecedented level of spatial resolution, deriving stellar-velocity dispersion profiles and rotation curves from three spectral regions (including CaH&K, MgIb triplet, and CaII triplet). In addition, we perform surface photometry on SDSS images, using a newly developed code for joint multi-band analysis. BH masses are estimated from the width of the Hbeta emission line and the host-galaxy free 5100A AGN luminosity. Combining results from spectroscopy and imaging allows us to study four MBH scaling relations: MBH-sigma, MBH-L(sph), MBH-M(sph,*), MBH-M(sph,dyn). We find the following results. First, stellar-velocity dispersions determined from aperture spectra (e.g. SDSS fiber spectra or unresolved data from distant galaxies) can be biased, depending on aperture size, AGN contamination, and host-galaxy morphology. However, such a bias cannot explain the offset seen in the MBH-sigma relation at higher redshifts. Second, while the CaT region is the cleanest to determine stellar-velocity dispersions, both the MgIb region, corrected for FeII emission, and the CaHK region, although often swamped by the AGN powerlaw continuum and emission lines, can give results accurate to within a few percent. Third, the MBH scaling relations of our pilot sample agree in slope and scatter with those of other local active and inactive galaxies. In the next papers of the series we will quantify the scaling relations, exploiting the full sample of ~ 100 objects.

## X-ray study of HLX1: intermediate-mass black hole or foreground neutron star?

**Authors:** Soria, Roberto; Zampieri, Luca; Zane, Silvia; Wu, Kinwah

**Abstract:** We re-assess the XMM-Newton and Swift observations of HLX1, to examine the evidence for its identification as an intermediate-mass black hole. We show that the X-ray spectral and timing properties are equally consistent with an intermediate-mass black hole in a high state, or with a foreground neutron star with a luminosity of about a few times $10^{32}$ erg/s ~ $10^{-6} L_{Edd}$, located at a distance of about 1.5 to 3 kpc. Contrary to previously published results, we find that the X-ray spectral change between the two XMM-Newton observations of 2004 and 2008 (going from power-law dominated to thermal dominated) is not associated with a change in the X-ray luminosity. The thermal component becomes more dominant (and hotter) during the 2009 outburst seen by Swift, but in a way that is consistent with either scenario.

## Croatian Black Hole School 2010 lecture notes on IMBHs in GCs

**Authors:** Pasquato, Mario

**Eprint:** http://arxiv.org/abs/1008.4477

**Keywords:** astro-ph.GA; globular clusters; gr-qc; intermediate-mass black holes; notes of lectures; stellar dynamics

**Abstract:** Black holes are fascinating objects. As a class of solutions to the Einstein equations they have been studied a great deal, yielding a wealth of theoretical results. But do they really exist? What do astronomers really mean when they claim to have observational evidence of their existence? To answer these questions, I will focus on a particular range of black-hole masses, approximately from 100 to 10000 solar masses. Black holes of this size are named Intermediate Mass Black Holes (IMBHs) and their existence is still heavily disputed, so they will be perfect for illustrating the observational challenges faced by a black hole hunter

## Is a Circular Orbit Possible According to General Relativity?

**Authors:** Hioe, F. T.; Kuebel, David

**Eprint:** http://arxiv.org/abs/1008.3553

**Keywords:** EMRI; general relativity; gr-qc

**Abstract:** The following results are presented concerning the possibility of a circular orbit for a particle in a gravitational field according to general relativity in the Schwarzschild metric: (1) A stable circular orbit is not possible. The most "circular" orbit must have a small but nonzero eccentricity and is thus elliptical and precesses. (2) An unstable circular orbit at a distance between 1.5 and 2.25 Schwarzschild radii from a black hole is possible. (3) There is no so-called innermost stable circular orbit.

## A Radio Census of Binary Supermassive Black Holes

**Authors:** Burke-Spolaor, Sarah

**Abstract:** Using archival VLBI data for 3114 radio-luminous active galactic nuclei, we searched for binary supermassive black holes using a radio spectral index mapping technique which targets spatially resolved, double radio-emitting nuclei. Only one source was detected as a double nucleus. This result is compared with a cosmological merger rate model and interpreted in terms of (1) implications for postmerger timescales for centralisation of the two black holes, (2) implications for the possibility of "stalled" systems, and (3) the relationship of radio activity in nuclei to mergers. Our analysis suggests that the binary evolution of paired supermassive black holes (both of masses $geqslant 10^8 M_\odot$) spends less than 500 Myr in progression from the merging of galactic stellar cores to within the purported stalling radius for supermassive black hole pairs. The data show no evidence for an excess of stalled binary systems at small separations. We see circumstantial evidence that the relative state of radio emission between paired supermassive black holes is correlated within orbital separations of 2.5 kpc.

## Intermediate-mass-ratio black hole binaries: intertwining numerical and perturbative techniques

**Authors:** Lousto, Carlos O.; Nakano, Hiroyuki; Zlochower, Yosef; Campanelli, Manuela

**Abstract:** We describe in detail full numerical and perturbative techniques to compute the gravitational radiation from intermediate mass ratio (IMR) black-holebinary (BHB) inspirals and mergers. We perform a series of full numerical simulations of nonspinning black holes with mass ratios q=1/10 and q=1/15 from different initial separations and for different finite difference resolutions. The highest resolution runs reach phase accuracies with errors <0.05 radians when the gravitational wave frequency is 0.2/M. In order to perform those full numerical runs, we adapted the gauge of the moving punctures approach with a variable damping term for the shift. We also derive an extrapolation (to infinite radius) formula for the waveform extracted at finite radius. For the perturbative evolutions we use the full numerical tracks, transformed into the Schwarzschild gauge, in the source terms of the Regge-Wheller-Zerilli Schwarzschild perturbations formalism. We then extend this perturbative formalism to take into account small intrinsic spins of the large black hole, and validate it by computing the quasinormal mode (QNM) frequencies, where we find good agreement for spins a/M<0.3. Including the final spins improves the overlap functions when comparing full numerical and perturbative waveforms, reaching 99.5% for the leading (l,m)=(2,2) and (3,3) modes, and 98.3% for the nonleading (2,1) mode in the q=1/10 case, which includes 8 orbits before merger. For the q=1/15 case, we obtain overlaps near 99.7% for all three modes. We discuss the modeling of the full inspiral and merger based on a combined matching of Post-Newtonian, Full Numerical, and Geodesic trajectories.

## The growth of supermassive black holes fed by accretion disks

**Authors:** Armijo, M. A. Montesinos; Pacheco, J. A. de Freitas

**Abstract:**

Supermassive black holes are probably present in the centre of the majority of the galaxies. There is a consensus that these exotic objects are formed by the growth of seeds either by accreting mass from a circumnuclear disk and/or by coalescences during merger episodes.

The mass fraction of the disk captured by the central object and the related timescale are still open questions, as well as how these quantities depend on parameters like the initial mass of the disk or the seed or on the angular momentum transport mechanism. This paper is addressed to these particular aspects of the accretion disk evolution and of the growth of seeds.

The time-dependent hydrodynamic equations were solved numerically for an axisymmetric disk in which the gravitational potential includes contributions both from the central object and from the disk itself. The numerical code is based on a Eulerian formalism, using a finite difference method of second-order, according to the Van Leer upwind algorithm on a staggered mesh.

The present simulations indicate that seeds capture about a half of the initial disk mass, a result weakly dependent on model parameters. The timescales required for accreting 50% of the disk mass are in the range 130-540 Myr, depending on the adopted parameters. These timescales permit to explain the presence of bright quasars at z ~ 6.5. Moreover, at the end of the disk evolution, a "torus-like" geometry develops, offering a natural explanation for the presence of these structures in the central regions of AGNs, representing an additional support to the unified model.

## A Tidal Disruption Flare in Abell 1689 from an Archival X-ray Survey of Galaxy Clusters

**Authors:** Maksym, Peter; Ulmer, Melville P.; Eracleous, Michael

**Eprint:** http://arxiv.org/abs/1008.4140

**Keywords:** astro-ph.HE; astrophysics; EM counterparts; observations; supermassive black holes

**Abstract:** Theory suggests that a star making a close passage by a supermassive black hole at the center of a galaxy can under most circumstances be expected to emit a giant flare of radiation as it is disrupted and a portion of the resulting stream of shock-heated stellar debris falls back onto the black hole itself. We examine the first results of an ongoing archival survey of galaxy clusters using *Chandra* and *XMM-Newton*-selected data, and report a likely tidal disruption flare from SDSS J131122.15-012345.6 in Abell 1689. The flare is observed to vary by a factor of $\gtrsim$ 30 over at least 2 years, to have maximum $L_X(0.3 - 3.0$ keV$) \gtrsim 5 \times 10^{42}$ erg s$^{-1}$ and to emit as a blackbody with $kT \sim 0.12$ keV. From the galaxy population as determined by existing studies of the cluster, we estimate a tidal disruption rate of $1.2 \times 10^{-4}$ galaxy$^{-1}$ year$^{-1}$ if we assume a contribution to the observable rate from galaxies whose range of luminosities corresponds to a central black hole mass ($M_\bullet$) between $10^6$ and $10^8 M_\odot$.

## Spectroscopic Signatures of the Tidal Disruption of Stars by Massive Black Holes

**Authors:** Strubbe, Linda E.; Quataert, Eliot

**Eprint:** http://arxiv.org/abs/1008.4131

**Keywords:** astro-ph.CO; astrophysics; EM counterparts; supermassive black holes

**Abstract:** During the tidal disruption of a star by a massive black hole (BH) of mass MBH *lesssim* $10^7 M_\odot$, stellar debris falls back to the BH at a rate well above the Eddington rate. A fraction of this gas is subsequently blown away from the BH, producing an optically bright flare of radiation. We predict the spectra and spectral evolution of tidal disruption events, focusing on the photoionized gas outside this outflow's photosphere. The spectrum will show absorption lines that are strongly blueshifted relative to the host galaxy, very broad (0.01-0.1c), and strongest at UV wavelengths (e.g., C IV, Ly alpha, O VI), lasting ~ 1 month for a $10^6 M_\odot$ BH. Meanwhile, supernovae in galactic nuclei are a significant source of confusion in optical surveys for tidal disruption events: we estimate that nuclear Type Ia supernovae are two orders of magnitude more common than tidal disruption events at z ~ 0.1 for ground-based surveys. Nuclear Type II supernovae occur at a comparable rate but can be excluded by pre-selecting red galaxies. Supernova contamination can be reduced to a manageable level using high-resolution follow-up imaging with adaptive optics or the Hubble Space Telescope. Our predictions should help optical transient surveys capitalize on their potential for discovering tidal disruption events.

## A Redshift for the Intermediate Mass Black Hole Candidate HLX-1: Confirmation of its Association with the Galaxy ESO 243-49

**Authors:** Wiersema, Klaas; Farrell, Sean A.; Webb, Natalie A.; Servillat, Mathieu; Maccarone, Thomas J.; Barret, Didier; Godet, Olivier

**Abstract:** In this Letter we report a spectroscopic confirmation of the association of HLX-1, the brightest ultra-luminous X-ray source, with the galaxy ESO 243-49. At the host galaxy distance of 95 Mpc, the maximum observed 0.2 - 10 keV luminosity is 1.2E42 erg/s. This luminosity is ~ 400 times above the Eddington limit for a 20 Msun black hole, and has been interpreted as implying an accreting intermediate mass black hole with a mass in excess of 500 Msun (assuming the luminosity is a factor of 10 above the Eddington value). However, a number of other ultra-luminous X-ray sources have been later identified as background active galaxies or foreground sources. It has recently been claimed that HLX-1 could be a quiescent neutron star X-ray binary at a Galactic distance of only 2.5 kpc, so a definitive association with the host galaxy is crucial in order to confirm the nature of the object. Here we report the detection of the Halpha emission line for the recently identified optical counterpart at a redshift consistent with that of ESO 243-49. This finding definitively places HLX-1 inside ESO 243-49, confirming the extreme maximum luminosity and strengthening the case for it containing an accreting intermediate mass black hole of more than 500 Msun.

## Gravitational recoils of supermassive black holes in hydrodynamical simulations of gas rich galaxies

**Authors:** Sijacki, Debora; Springel, Volker; Haehnelt, Martin

**Abstract:** We study the evolution of gravitationally recoiled supermassive black holes (BHs) in massive gas-rich galaxies by means of high-resolution hydrodynamical simulations. We find that the presence of a massive gaseous disc allows recoiled BHs to return to the centre on a much shorter timescale than for purely stellar discs. Also, BH accretion and feedback can strongly modify the orbit of recoiled BHs and hence their return timescale, besides affecting the distribution of gas and stars in the galactic centre. However, the dynamical interaction of kicked BHs with the surrounding medium is in general complex and can facilitate both a fast return to the centre as well as a significant delay. The Bondi-Hoyle-Lyttleton accretion rates of the recoiling BHs in our simulated galaxies are favourably high for the detection of off-centred AGN if kicked within gas-rich discs – up to a few per cent of the Eddington accretion rate – and are highly variable on timescales of a few $10^7$ yrs. In major merger simulations of gas-rich galaxies, we find that gravitational recoils increase the scatter in the BH mass – host galaxy relationships compared to simulations without kicks, with the BH mass being more sensitive to recoil kicks than the bulge mass. A generic result of our numerical models is that the clumpy massive discs suggested by recent high-redshift observations, as well as the remnants of gas-rich mergers, exhibit a gravitational potential that falls steeply in the central regions, due to the dissipative concentration of baryons. As a result, supermassive BHs should only rarely be able to escape from massive galaxies at high redshifts, which is the epoch where the bulk of BH recoils is expected to occur.[Abridged]

## Gravitational Radiations from a Spinning Compact Object circling a Supermassive Kerr Black Hole

**Authors:** Han, Wen-Biao

**Abstract:** The gravitational waves and energy radiations from a spinning compact object with stellar mass in a circular orbit in the equatorial plane of a supermassive

Kerr black hole are investigated in this paper. The effect how the spin acts on energy and angular moment fluxes is discussed in detail. The calculation results indicate that the spin of small body should be considered in waveform-template production for the upcoming gravitational wave detections. It is clear that when the direction of spin axes is as same as the orbitally angular momentum ("positive" spin), spin can decrease the energy fluxes which radiate to infinity. For anti-direction spin ("negative"), the energy fluxes to infinity be enlarged. And the relation between fluxes (both infinity and horizon) and spin looks like a quadratic function. From frequency shift due to spin, we estimate the wave-phase accumulation during inspiralling process of the particle. We find that the time of particle inspiral into the black hole is longer for "positive" spin and shorter for "negative" comparing with non-spinning particle. Especially, for extreme spin value, the energy radiation near the horizon of the extreme Kerr black hole is much more than the non-spinning one. And consequently, the maximum binging energy of the extreme spinning particle is much bigger than the non-spinning particle.

## Length requirements for numerical-relativity waveforms

**Authors:** Hannam, Mark; Husa, Sascha; Ohme, Frank; Ajith, P.

**Abstract:** One way to produce complete inspiral-merger-ringdown gravitational waveforms from black-hole-binary systems is to connect post-Newtonian (PN) and numerical-relativity (NR) results to create "hybrid" waveforms. Hybrid waveforms are central to the construction of some phenomenological models for GW search templates, and for tests of GW search pipelines. The dominant error source in hybrid waveforms arises from the PN contribution, and can be reduced by increasing the number of NR GW cycles that are included in the hybrid. Hybrid waveforms are considered sufficiently accurate for GW detection if their mismatch error is below 3% (i.e., a fitting factor about 0.97). We address the question of the length requirements of NR waveforms such that the final hybrid waveforms meet this requirement, considering nonspinning binaries with $q = M_2/M_1 \in [1, 4]$ and equal-mass binaries with $\chi = S_i/M_i^2 \in [-0.5, 0.5]$. We conclude that for the cases we study simulations must contain between three (in the equal-mass nonspinning case) and ten (the $\chi = 0.5$ case) orbits before merger, but there is also evidence that these are the regions of parameter space for which the least number of cycles will be needed.

## Variability and stability in optical blazar jets: the case of OJ287

**Authors:** Villforth, C.; Nilsson, K.; Heidt, J.; Pursimo, T.

**Abstract:** OJ287 is a BL Lac object at redshift z=0.306 that has shown double-peaked bursts at regular intervals of ~ 12 yr during the last ~ 40 yr. Due to this behavior, it has been suggested that OJ287 might host a close supermassive binary black hole. We present optical photopolarimetric monitoring data from 2005-2009, during which the latest double-peaked outburst occurred. We find a stable component in the optical jet: the optical polarization core. The optical polarization indicates that the magnetic field is oriented parallel to the jet. Using historical optical polarization data, we trace the evolution of the optical polarization core and find that it has showed a swing in the Stokes plane indicating a reorientation of the jet magnetic field. We also find that changes in the optical jet magnetic field seem tightly related to the double-peaked bursts. We use our findings as a new constraint on possible binary black hole models. Combining all available observations, we find that none of the proposed binary black bole models is able to fully explain the observations. We suggest a new approach to understanding OJ287 that is based on the assumption that changes in the jet magnetic field drive the regular outbursts.

## Nuclear Star Clusters from Clustered Star Formation

**Authors:** Agarwal, Meghann; Milosavljevic, Milos

**Abstract:** Photometrically distinct nuclear star clusters (NSCs) are common in late-type-disk and spheroidal galaxies. The formation of NSCs is inevitable in the context of normal star formation in which a majority of stars form in clusters. A young, mass-losing cluster embedded in an isolated star-forming galaxy remains gravitationally bound over a period determined by its initial mass and the galactic tidal field. The cluster migrates radially toward the center of the galaxy and becomes integrated in the NSC if it reaches the center. The rate at which the NSC grows by accreting young clusters can be estimated from empirical cluster formation rates and dissolution times. We model cluster migration and dissolution and find that the NSCs in late-type disks and in spheroidals could have assembled from migrating clusters. The resulting stellar nucleus contains a small fraction of the stellar mass of the galaxy; this fraction is sensitive to the high-mass truncation of the initial cluster mass function (ICMF). The resulting NSC masses are consistent with the observed values, but generically, the final NSCs are surrounded by a spatially more extended excess over the inward-extrapolated exponential (or Sersic) law of the outer galaxy. We suggest that the excess can be related to the pseudobulge phenomenon in disks, though not all of the pseudobulge mass assembles this way. Comparison with observed NSC masses can be used to constrain the truncation mass scale of the ICMF and the fraction of clusters suffering prompt dissolution. We infer truncation mass scales of $\sim 10^5 M_\odot$ without (with 90%) prompt dissolution.

## The Orbital Structure of Triaxial Galaxies with Figure Rotation

**Authors:** Deibel, Alex T.; Valluri, Monica; Merritt, David

**Abstract:** We survey the properties of all orbit families in the rotating frame of a family of realistic triaxial potentials with central supermassive black holes. In such galaxies, most regular box orbits (vital for maintaining triaxiality) are associated with resonances which occupy two dimensional surfaces in configuration space. For slow figure rotation all orbit families are largely stable. At intermediate pattern speeds a significant fraction of the resonant box orbits as well as inner long-axis tubes are destabilized by the "envelope doubling" that arises from the coriolis forces, and are driven into the destabilizing center. Thus for pattern rotation periods .2 Gyr < Tp < 5 Gyr, the two orbit families that are most important for maintaining triaxiality are highly chaotic. As pattern speed increases there is also a sharp decrease in the overall fraction of prograde short-axis tubes and a corresponding increase in the retrograde variety. At the highest pattern speeds (close to that of triaxial bars) box-like orbits undergo a sudden transition to a new family of stable retrograde loop-like orbits, which resemble orbits in 3-D bars, and circulate about the short axis. Our analysis implies that stable triaxial galaxies and dark matter halos (with central cusps and supermassive black holes) could either be slowly rotating or very rapidly rotating like fast bars, but equilibrium triaxial elliptical galaxies and dark matter halos with intermediate pattern speeds are unlikely to exist due to the high levels of stochasticity in both the box and inner long-axis tube orbit families.

# Widespread presence of shallow cusps in the surface-brightness profile of globular clusters

**Authors:** Vesperini, Enrico; Trenti, Michele

**Abstract:** Surface brightness profiles of globular clusters with shallow central cusps (Sigma $\sim$ $R^v$ with $-0.3 <\sim$ $v \sim$ $-0.3$ in the pre-core-collapse and core-collapse phases. Post-core-collapse clusters without an IMBH transition to steeper cusps, $-0.7 <\sim$ $v <\sim$ $-0.4$, only if the primordial binary fraction is very small, $f_{bin}$ - 0.3 even when $\lesssim -0.4$. Overall our analysis shows that a shallow cusp is not an unequivocal signature of a central IMBH and casts serious doubts on the usefulness of measuring v in the context of the hunt for IMBHs in globular clusters.

# A new globular cluster black hole in NGC 4472

**Authors:** Maccarone, Thomas J.; Kundu, Arunav; Zepf, Stephen E.; Rhode, Katherine L.

**Abstract:** We discuss CXOU$\sim$ 1229410+075744, a new black hole candidate in a globular cluster in the elliptical galaxy NGC$\sim$ 4472. By comparing two Chandra observations of the galaxy, we find a source that varies by at least a factor of 4, and has a peak luminosity of at least $2 \times 10^{39}$ ergs/sec. As such, the source varies by significantly more than the Eddington luminosity for a single neutron star, and is a strong candidate for being a globular cluster black hole. The source's X-ray spectrum also evolves in a manner consistent with what would be expected from a single accreting stellar mass black hole. We consider the properties of the host cluster of this source and the six other strong black hole X-ray binary candidates, and find that there is suggestive evidence that black hole X-ray binary formation is favored in bright and metal rich clusters, just as is the case for bright X-ray sources in general.

## Dynamical damping terms for symmetry-seeking shift conditions

**Authors:** Alic, Daniela; Rezzolla, Luciano; Hinder, Ian; Mösta, Philipp

**Abstract:** Suitable gauge conditions are fundamental for stable and accurate numerical-relativity simulations of inspiralling compact binaries. A number of well-studied conditions have been developed over the last decade for both the lapse and the shift and these have been successfully used both in vacuum and non-vacuum spacetimes when simulating binaries with comparable masses. At the same time, recent evidence has emerged that the standard "Gamma-driver" shift condition requires a careful and non-trivial tuning of its parameters to ensure long-term stable evolutions of unequal-mass binaries. We present a novel gauge condition in which the damping constant is promoted to be a dynamical variable and the solution of an evolution equation. We show that this choice removes the need for special tuning and provides a shift damping term which is free of instabilities in our simulations and dynamically adapts to the individual positions and masses of the binary black-hole system. Our gauge condition also reduces the variations in the coordinate size of the apparent horizon of the larger black hole and could therefore be useful when simulating binaries with very small mass ratios.

## Observational characteristics of accretion onto black holes

**Authors:** Done, Chris

**Abstract:** These notes resulted from a series of lectures at the IAC winter school. They are designed to help students, especially those just starting in subject, to get hold of the fundamental tools used to study accretion powered sources. As such, the references give a place to start reading, rather than representing a complete survey of work done in the field. I outline Compton scattering and blackbody radiation as the two predominant radiation mechanisms for accreting black holes, producing the hard X-ray tail and disc spectral components, respectively. The interaction of this radiation with matter can result in photo-electric absorption and/or reflection.

While the basic processes can be found in any textbook, here I focus on how these can be used as a toolkit to interpret the spectra and variability of black hole binaries (hereafter BHB) and Active Galactic Nuclei (AGN). I also discuss how to use these to physically interpret real data using the publicly available XSPEC spectral fitting package (Arnaud et al 1996), and how this has led to current models (and controversies) of the accretion flow in both BHB and AGN.

## Rapid Flux Variability of Sgr A*

**Authors:** Yusef-Zadeh, F.; Wardle, M.; Miller-Jones, J.; Roberts, D.; Porquet, D.; Grosso, N.

**Abstract:** Sgr A* exhibits flares in radio, millimeter and submm wavelengths with durations of $\sim 1$ hour. Using a structure function analysis, we investigate the variability of Sgr A* on time scales ranging from a few seconds to several hours, and find evidence for sub-minute time scale variability at radio wavelengths. These measurements suggest a strong case for continuous variability from sub-minute to hourly time scales. This short time scale variability constrains the size of the emitting region to be less than 0.1 AU. Assuming that the minute time scale fluctuations of the emission at 7 mm arise through the expansion of regions of optically thick synchrotron-emitting plasma, this suggests the presence of explosive, energetic expansion events at speeds close to $c$. The required rate of mass processing and energy loss of this component are estimated to be $\gtrsim 6 \times 10^{-10} M_\odot$ yr$^{-1}$ and $400 L_\odot$ respectively. The inferred scale length corresponding to one-minute light travel time is comparable to the time averaged spatially resolved 0.1AU scale observed at 1.3mm emission of Sgr A*. This steady component from Sgr A* is then interpreted mainly as an ensemble average of numerous weak and overlapping flares that are detected on short time scales. The nature of such short time scale variable emission or quiescent variability is not understood but could result from fluctuations in the accretion flow of Sgr A* that feed the base of an outflow or jet.

## Study of multi black hole and ring singularity apparent horizons

**Authors:** Jaramillo, Gabriela; Lousto, Carlos O.

**Keywords:** astro-ph.CO; astro-ph.GA; astro-ph.HE; gr-qc; massive binaries of black holes; numerical relativity

**Abstract:** We study critical black hole separations for the formation of a common apparent horizon in systems of $N$ - black holes in a time symmetric configuration. We study in detail the aligned equal mass cases for $N = 2, 3, 4, 5$, and relate them to the unequal mass binary black hole case. We then study the apparent horizon of the time symmetric initial geometry of a ring singularity of different radii. The apparent horizon is used as indicative of the location of the event horizon in an effort to predict a critical ring radius that would generate an event horizon of toroidal topology. We found that a good estimate for this ring critical radius is $20/(3\pi)M$. We briefly discuss the connection of this two cases through a discrete black hole 'necklace' configuration.

## Recoiling Massive Black Holes in Gas-Rich Galaxy Mergers

**Authors:** Guedes, Javiera; Madau, Piero; Mayer, Lucio; Callegari, Simone

**Abstract:** The asymmetric emission of gravitational waves produced during the coalescence of a massive black hole (MBH) binary imparts a velocity "kick" to the system that can displace the hole from the center of its host. Here we study the trajectories and observability of MBHs recoiling in three (one major, two minor) gas-rich galaxy merger remnants that were previously simulated at high resolution, and in which the pairing of the MBHs had been shown to be successful. We run new simulations of MBHs recoiling in the major merger remnant with Mach numbers in the range 1<M<6 km/s, and use simulation data to construct a semi-analytical model for the orbital evolution of MBHs in gas-rich systems. We show that: 1) in major merger remnants the energy deposited by the moving hole into the rotationally supported, turbulent medium makes a negligible contribution to the thermodynamics of the gas. This contribution is more significant in minor merger remnants, potentially allowing for electromagnetic signatures in this case; 2) in major mergers, the drag from high-density gas allows even MBHs with kick velocities of 1200 km/s to remain within 1 kpc from the host's center; 3) kinematically offset nuclei can be observed for timescales of a few Myr in major merger remnants in the case of recoil velocities in the range 700-1000 km/s; 4) in minor mergers remnants the effect of gas drag is weaker, and MBHs with recoil speeds in the range 300-600 km/s will wander through the host halo and may be detectable as spatially-offset active nuclei.

## The two states of Sgr A* in the near-infrared: bright episodic flares on top of low-level continuous variability

**Authors:** Dodds-Eden, K.; Gillessen, S.; Fritz, T. K.; Eisenhauer, F.; Trippe, S.; Genzel, R.; Ott, T.; Bartko, H.; Pfuhl, O.; Bower, G.; Goldwurm, A.; Porquet, D.; Trap, G.; Yusef-Zadeh, F.

**Abstract:** In this paper we examine properties of the variable source Sgr A* in the near-infrared (NIR) using a very extensive Ks-band data set from NACO/VLT observations taken 2004 to 2009. We investigate the variability of Sgr A* with two different photometric methods and analyze its flux distribution. We find Sgr A* is continuously variable (meaning the source is always 'on' and varying) in the near-infrared, and there also appears to be some medium-term variability on timescales of weeks to months. The flux distribution can be described by a lognormal distribution at low intrinsic fluxes (less than about 5 mJy, dereddened with $A_{Ks}$ = 2.5). The lognormal distribution has a median flux of ~ 1.6 mJy, but above 5 mJy the flux distribution is significantly flatter (high flux events are more common) than expected for the extrapolation of the lognormal distribution to high fluxes. We make a general identification of the low level emission above 5 mJy as flaring emission and of the low level emission as the quiescent state. We also report here the brightest Ks-band flare ever observed (from August 5th, 2008) which reached an intrinsic Ks-band flux of 27.5 mJy ($m_{Ks}$ = 13.5). This flare was a factor 27 increase over the median flux of Sgr A*, close to double the brightness of the star S2 in the Ks-band, and 40% brighter than the next brightest flare ever observed from Sgr A*.

## Massive Black Hole Binary Systems in Hierarchical Scenario of Structure Formation

**Authors:** Pereira, Eduardo S.; Miranda, Oswaldo D.

**Abstract:** Recently, it has increased the observational evidence that, in most galaxies there are massive black holes (MBH). On the other hand, the hierarchical scenario of structure formation describe which objects like galaxies and galaxy clusters are formatted by mergers of small objects. In this context, we can suppose that mergers of galaxies leads to the formation of MBH binary systems. It is expected that the merger of two MBH produces a gravitational waves signal detectable by the Laser Interferometer Space Antenna (LISA). In this work, we use the Press-Schechter formalism and its extention to take into account the analytical form for the merger rate of haloes that contains massive black holes. Also, we describe a way to determine the number of binary systems of MBH.

# Precession effect of the gravitational self-force in a Schwarzschild spacetime and the effective one-body formalism

**Authors:** Barack, Leor; Damour, Thibault; Sago, Norichika

**Abstract:** Using a recently presented numerical code for calculating the Lorenz-gauge gravitational self-force (GSF), we compute the $O(m)$ conservative correction to the precession rate of the small-eccentricity orbits of a particle of mass $m$ moving around a Schwarzschild black hole of mass $M \gg m$. Specifically, we study the gauge-invariant function $\rho(x)$, where $\rho$ is defined as the $O(m)$ part of the dimensionless ratio $(\hat{\Omega}_r/\hat{\Omega}_\varphi)^2$ between the squares of the radial and azimuthal frequencies of the orbit, and where $x = [Gc^{-3}(M + m)\hat{\Omega}_\varphi]^{2/3}$ is a gauge-invariant measure of the dimensionless gravitational potential (mass over radius) associated with the mean circular orbit. Our GSF computation of the function $\rho(x)$ in the interval $0 < x \leq 1/6$ determines, for the first time, the *strong-field behavior* of a combination of two of the basic functions entering the Effective One Body (EOB) description of the conservative dynamics of binary systems. We show that our results agree well in the weak-field regime (small $x$) with the 3rd post-Newtonian (PN) expansion of the EOB results, and that this agreement is improved when taking into account the analytic values of some of the logarithmic-running terms occurring at higher PN orders. Furthermore, we demonstrate that GSF data give access to higher-order PN terms of $\rho(x)$ and can be used to set useful new constraints on the values of yet-undetermined EOB parameters. Most significantly, we observe that an *excellent global representation* of $\rho(x)$ can be obtained using a simple two-point Pade approximant which combines 3PN knowledge at $x = 0$ with GSF information at a single strong-field point (say, $x = 1/6$).

## Event Horizon Deformations in Extreme Mass-Ratio Black Hole Mergers

**Authors:** Hamerly, Ryan; Chen, Yanbei

**Abstract:** We study the geometry of the event horizon of a spacetime in which a small compact object plunges into a large Schwarzschild black hole. We first use the Regge-Wheeler and Zerilli formalisms to calculate the metric perturbations induced by this small compact object, then find the new event horizon by propagating null geodesics near the unperturbed horizon. A caustic is shown to exist before the merger. Focusing on the geometry near the caustic, we show that it is determined predominantly by large-l perturbations, which in turn have simple asymptotic forms near the point at which the particle plunges into the horizon. It is therefore possible to obtain an analytic characterization of the geometry that is independent of the details of the plunge. We further show that among the leading-order horizon area increase, half arises from generators that enter the horizon through the caustic, and the rest arises from area increase near the caustic, induced by the gravitational field of the compact object.

## Testing GR with Galactic-centre Stars

**Authors:** Angelil, Raymond; Saha, Prasenjit

**Abstract:** The Galactic Centre S-stars orbiting the central supermassive black hole reach velocities of a few percent of the speed of light. The GR-induced perturbations to the redshift enter the dynamics via two distinct channels. The post-Newtonian regime perturbs the orbit from the Keplerian (Zucker et al., 2006, Kannan & Saha 2009), and the photons from the Minkowski (Angelil & Saha 2010). The inclusion of gravitational time dilation at order $v^2$ marks the first departure of the redshift from the line-of-sight velocities. The leading-order Schwarzschild terms curve space, and enter at order $v^3$. The classical Keplerian phenomenology dominates the total redshift. Spectral measurements of sufficient resolution will allow for the detection of these post-Newtonian effects. We estimate the spectral resolution required to detect each of these effects by fitting the redshift curve via the five keplerian elements plus black hole mass to mock data. We play with an exaggerated S2 orbit - one with a semi-major axis a fraction of that of the real S2. This amplifies the relativistic effects, and allows clear visual distinctions between the relativistic terms. We argue that spectral data of S2 with a dispersion of about 10km/s would allow for a clear detection of gravitational redshift, and about 1 km/s would suffice for leading-order space curvature detection.

## Constraining the Accretion Flow in Sgr A* by General Relativistic Dynamical and Polarized Radiative Modeling

**Authors:** Shcherbakov, Roman V.; Penna, Robert F.; McKinney, Jonathan C.

**Abstract:** The constraints on Sgr A* black hole (BH) and accretion flow parameters are found by fitting polarized sub-mm observations. The observations from 29 papers are averaged into a quasi-quiescent set. We run three-dimensional general relativistic magnetohydrodynamical (3D GRMHD) simulations for dimensionless spins $a = 0, 0.5, 0.7, 0.9, 0.98$ till $20000M$, construct an averaged dynamical model, perform GR polarized radiative transfer, and explore the parameter space of spin $a$, inclination angle $\theta$, position angle (PA), accretion rate $\dot{M}$, and electron temperature $T_e$ at $6M$ radius. The best-fitting model for spin $a = 0.9$ gives $\chi^2 = 0.99$ with $\theta = 59°$, $\dot{M} = 1.3 \cdot 10^{-8} M_\odot \text{year}^{-1}$, $T_e = 3.2 \cdot 10^{10}$ K at $6M$, the best-fitting model for spin $a = 0.5$ gives $\chi^2 = 0.84$ with $\theta = 70°$, $\dot{M} = 7.0 \cdot 10^{-8} M_\odot \text{year}^{-1}$, and $T_p/T_e = 22$ at $6M$ with $T_e = 3.50 \cdot 10^{10}$ K. We identify the physical phenomena leading to the matched linear polarization (LP), circular polarization (CP), and electric vector position angle (EVPA). Our statistical analysis reveals the most probable spin is $a = 0.9$. The spin $a = 0.5$ solutions are 10 times less probable despite giving lower minimum $\chi^2$ and spin $a = 0$ is excluded as having probability $P(a) < 1\%$. Polarized data allows us to tightly constrain some quantities. Inclination angle, electron temperature, and position angle have ranges $\theta = 59° \pm 9°$, $T_e = 3.4^{+1.2}_{-0.9} \times 10^{10}$ K, and PA = $96° \pm 30°$ with 90% confidence. The total range of accretion rate is large, but assuming spin $a = 0.9$ we get $\dot{M}(0.9) = 13^{+4}_{-3} \times 10^{-9} M_\odot \text{year}^{-1}$ interval with 90% confidence. The emission region sizes at 230 GHz of the best-fitting models are found to be marginally consistent with the observed by VLBI technique.

## Key questions about Galactic Center dynamics

**Authors:** Alexander, Tal

**Abstract:** I discuss four key questions about Galactic Center dynamics, their implications for understanding both the environment of the Galactic MBH and galactic nuclei in general, and the progress made in addressing them. The questions are (1) Is the stellar system around the MBH relaxed? (2) Is there a "dark cusp" around the MBH? (3) What is the origin of the stellar disk(s)?, and (4) What is the origin of the S-stars?

## Broad emission lines for negatively spinning black holes

**Authors:** Dauser, T.; Wilms, J.; Reynolds, C. S.; Brenneman, L. W.

**Abstract:** We present an extended scheme for the calculation of the profiles of emission lines from accretion discs around rotating black holes. The scheme includes discs with angular momenta which are parallel and antiparallel with respect to the black hole's angular momentum, as both configurations are assumed to be stable (King et al., 2005). We discuss line shapes for such discs and present a code for modelling observational data with this scheme in X-ray data analysis programs. Based on a Green's function approach, an arbitrary radius dependence of the disc emissivity and arbitrary limb darkening laws can be easily taken into account, while the amount of precomputed data is significantly reduced with respect to other available models.

## Fast Fisher Matrices and Lazy Likelihoods

**Authors:** Cornish, Neil J.

**Abstract:** Theoretical studies in gravitational wave astronomy often require the calculation of Fisher Information Matrices and Likelihood functions, which in a direct approach entail the costly step of computing gravitational waveforms. Here I describe an alternative technique that sidesteps the need to compute full waveforms, resulting in significant computational savings. I describe how related techniques can be used to speed up Bayesian inference applied to real gravitational wave data.

## VLT Kinematics for omega Centauri: Further Support for a Central Black Hole

**Authors:** Noyola, Eva; Gebhardt, Karl; Kissler-Patig, Markus; Lutzgendorf, Nora; Jalali, Behrang; de Zeeuw, P. Tim; Baumgardt, Holger

**Abstract:** The Galactic globular cluster omega Centauri is a prime candidate for hosting an intermediate mass black hole. Recent measurements lead to contradictory conclusions on this issue. We use VLT-FLAMES to obtain new integrated spectra for the central region of omega Centauri. We combine these data with existing measurements of the radial velocity dispersion profile taking into account a new derived center from kinematics and two different centers from the literature. The data support previous measurements performed for a smaller field of view and show a discrepancy with the results from a large proper motion data set. We see a rise in the radial velocity dispersion in the central region to 22.8+-1.2 km/s, which provides a strong sign for a central black hole. Isotropic dynamical models for omega Centauri imply black hole masses ranging from 3.0 to $5.2 \times 10^4$ solar masses depending on the center. The best-fitted mass is $4.7 + -1.0 \times 10^4$ solar masses.

## Black-hole binaries with non-precessing spins

**Authors:** Hannam, Mark; Husa, Sascha; Ohme, Frank; Mueller, Doreen; Brueggmann, Bernd

**Abstract:** We present gravitational waveforms for the last orbits and merger of black-hole-binary (BBH) systems along two branches of the BBH parameter space: equal-mass binaries with equal non-precessing spins, and nonspinning unequal-mass binaries. The waveforms are calculated from numerical solutions of Einstein's equations for black-hole binaries that complete between six and ten orbits before merger. Along the equal-mass spinning branch, the spin parameter of each BH is $\chi_i = S_i/M_i^2 \in [-0.85, 0.85]$, and along the unequal-mass branch the mass ratio is $q = M_2/M_1 \in [1, 4]$. We discuss the construction of low-eccentricity puncture initial data for these cases, the properties of the final merged BH, and compare the last 8-10 GW cycles up to $M\omega = 0.1$ with the phase and amplitude predicted by standard post-Newtonian (PN) approximants. As in previous studies, we find that the phase from the 3.5PN TaylorT4 approximant is most accurate for nonspinning binaries. For equal-mass spinning binaries the 3.5PN TaylorT1 approximant (including spin terms up to only 2.5PN order) gives the most robust performance, but it is possible to treat TaylorT4 in such a way that it gives the best accuracy for spins $\chi_i > -0.75$. When high-order amplitude corrections are included, the PN amplitude of the ($\ell = 2, m = \pm 2$) modes is larger than the NR amplitude by between 2-4%.

# On the falloff of radiated energy in black hole spacetimes

**Authors:** Burko, Lior M.; Hughes, Scott A.

**Abstract:** The goal of much research in relativity is to understand gravitational waves generated by a strong-field dynamical spacetime. Quantities of particular interest for many calculations are the Weyl scalar $\psi_4$, which is simply related to the flux of gravitational waves far from the source, and the flux of energy carried to distant observers, $\dot{E}$. Conservation laws guarantee that, in asympotically flat spacetimes, $\psi_4 \propto 1/r$ and $\dot{E} \propto 1/r^2$ as $r \to \infty$. Most calculations extract these quantities at some finite extraction radius. An understanding of finite radius corrections to $\psi_4$ and $\dot{E}$ allows us to more accurately infer their asymptotic values from a computation. In this paper, we show that, if the final state of the system is a black hole, then the leading correction to $\psi_4$ is $O(1/r^3)$, and that to the energy flux is $O(1/r^4)$ — not $O(1/r^2)$ and $O(1/r^3)$ as one might naively guess. Our argument only relies on the behavior of the curvature scalars for black hole spacetimes. Using black hole perturbation theory, we calculate the corrections to the leading falloff, showing that it is quite easy to correct for finite extraction radius effects.

## The Origin and Detection of High-Redshift Supermassive Black Holes

**Authors:** Haiman, Zoltán

**Eprint:** http://arxiv.org/abs/1007.4741

**Keywords:** astro-ph.CO; cosmology; supermassive black holes

**Abstract:** Supermassive black holes (SMBHs) are common in local galactic nuclei, and SMBHs as massive as several billion solar masses already exist at redshift z=6. These earliest SMBHs may arise by the combination of Eddington-limited growth and mergers of stellar-mass seed BHs left behind by the first generation of metal-free stars, or by the rapid direct collapse of gas in rare special environments where the gas can avoid fragmenting into stars. In this contribution, I review these two competing scenarios. I also briefly mention some more exotic ideas and how the different models may be distinguished in the future by LISA and other instruments.

## Accretion Disks in Active Galactic Nuclei: Gas Supply Driven by Star Formation

**Authors:** Wang, J. -M.; Yan, C. -S.; Gao, H. -Q.; Hu, C.; Li, Y. -R.; Zhang, S.

**Eprint:** http://arxiv.org/abs/1007.4060

**Keywords:** accretion discs; astro-ph.CO; astrophysics; EM counterparts

**Abstract:** Self-gravitating accretion disks collapse to star-forming(SF) regions extending to the inner edge of the dusty torus in active galactic nuclei (AGNs). A full set of equations including feedback of star formation is given to describe the dynamics of the regions. We explore the role of supernovae explosion (SNexp), acting to excite turbulent viscosity, in the transportation of angular momentum in the regions within 1pc scale. We find that accretion disks with typical rates in AGNs can be driven by SNexp in the regions and metals are produced spontaneously. The present model predicts a metallicity–luminosity relationship consistent with that observed in AGNs. As relics of SF regions, a ring (or belt) consisting of old stars remains for every episode of supermassive black hole activity. We suggest that multiple stellar rings with random directions interact and form a nuclear star cluster after episodes driven by star formation.

# The Carter Constant for Inclined Orbits About a Massive Kerr Black Hole: I. circular orbits

**Authors:** Komorowski, P. G.; Valluri, S. R.; Houde, M.

**Abstract:** In an extreme binary black hole system, an orbit will increase its angle of inclination ($\iota$) as it evolves in Kerr spacetime. We focus our attention on the behaviour of the Carter constant ($Q$) for near-polar orbits; and develop an analysis that is independent of and complements radiation reaction models. For a Schwarzschild black hole, the polar orbits represent the abutment between the prograde and retrograde orbits at which $Q$ is at its maximum value for given values of latus rectum ($\tilde{l}$) and eccentricity ($e$). The introduction of spin ($\tilde{S} = |\mathbf{J}|/M^2$) to the massive black hole causes this boundary, or abutment, to be moved towards greater orbital inclination; thus it no longer cleanly separates prograde and retrograde orbits.

To characterise the abutment of a Kerr black hole (KBH), we first investigated the last stable orbit (LSO) of a test-particle about a KBH, and then extended this work to general orbits. To develop a better understanding of the evolution of $Q$ we developed analytical formulae for $Q$ in terms of $\tilde{l}$, $e$, and $\tilde{S}$ to describe elliptical orbits at the abutment, polar orbits, and last stable orbits (LSO). By knowing the analytical form of $\partial Q/\partial \tilde{l}$ at the abutment, we were able to test a 2PN flux equation for Q. We also used these formulae to numerically calculate the $\partial \iota/\partial \tilde{l}$ of hypothetical circular orbits that evolve along the abutment. From these values we have determined that $\partial \iota/\partial \tilde{l} = -\left(122.7\tilde{S} - 36\tilde{S}^3\right)\tilde{l}^{-11/2} - \left(63/2\,\tilde{S} + 35/4\,\tilde{S}^3\right)\tilde{l}^{-9/2} - 15/2\,\tilde{S}\tilde{l}^{-7/2} - 9/2\,\tilde{S}\tilde{l}^{-5/2}$. By taking the limit of this equation for $\tilde{l} \to \infty$, and comparing it with the published result for the weak field radiation-reaction, we found the upper limit on $\left|\partial \iota/\partial \tilde{l}\right|$ for the full range of $\tilde{l}$ up to the LSO. Although we know the value of $\partial Q/\partial \tilde{l}$ at the abutment, we find that the second and higher derivatives of $Q$ with respect to $\tilde{l}$ exert an influence on $\partial \iota/\partial \tilde{l}$. Thus the abutment becomes an important analytical and numerical laboratory for studying the evolution of $Q$ and $\iota$ in Kerr spacetime and for testing current and future radiation back-reaction models for near-polar retrograde orbits.

## An expanded $M_{bh} - \sigma$ diagram, and a new calibration of active galactic nuclei masses

**Authors:** Graham, Alister W.; Onken, Christopher A.; Athanassoula, E.; Combes, Francoise

**Abstract:** We present an updated and improved $M_{bh}$-$\sigma$ diagram containing 64 galaxies for which $M_{bh}$ measurements (not just upper limits) are available. Due to new and increased black hole masses at the high-mass end, and a better representation of barred galaxies at the low-mass end, the "classical" (all morphological type) $M_{bh}$-$\sigma$ relation for predicting black hole masses is $\log(M_{bh}/M_\odot) = (8.13 \pm 0.05) + (5.13 \pm 0.34) \log[\sigma/200 \, \mathrm{km \, s^{-1}}]$, with an r.m.s. scatter of 0.43 dex. Modifying the regression analysis to correct for a hitherto over-looked sample bias in which black holes with masses $< 10^6 M_\odot$ are not (yet) detectable, the relation steepens further to give $\log(M_{bh}/M_\odot) = (8.15 \pm 0.06) + (5.95 \pm 0.44) \log[\sigma/200 \, \mathrm{km \, s^{-1}}]$. We have also updated the "barless" and "elliptical-only" $M_{bh}$-$\sigma$ relations introduced by Graham and Hu in 2008 due to the offset nature of barred galaxies. These relations have a total scatter as low as 0.34 dex and currently define the upper left envelope of points in the $M_{bh}$-$\sigma$ diagram. These relations also have a slope consistent with the value 5, in agreement with the prediction by Silk & Rees based on feedback from massive black holes in bulges built by monolithic-collapse.

Using updated virial products and velocity dispersions from 28 active galactic nuclei, we determine that the optimal scaling factor $f$ — which brings their virial products in line with the 64 directly measured black hole masses — is $2.8^{+0.7}_{-0.5}$. This is roughly half the value reported by Onken et al. and Woo et al., and consequently halves the mass estimates of most high-redshift quasars. Given that barred galaxies are, on average, located ~0.5 dex below the "barless" and "elliptical-only" $M_{bh}$-$\sigma$ relations, we have explored the results after separating the samples into barred and non-barred galaxies, and developed a preliminary corrective term to the velocity dispersion based on bar dynamics. In addition, given the recently recognised co-existence of massive black holes and nuclear star clusters, we present the first ever $(M_{bh} + M_{nc})$-$\sigma$ diagram and begin to explore how galaxies shift from their former location in the $M_{bh}$-$\sigma$ diagram.

## The peculiar optical spectrum of 4C+22.25: Imprint of a massive black hole binary?

**Authors:** Decarli, Roberto; Dotti, Massimo; Montuori, Carmen; Liimets, Tiina; Ederoclite, Alessandro

**Abstract:** We report the discovery of peculiar features in the optical spectrum of 4C+22.25, a flat spectrum radio quasar at z=0.4183 observed in the SDSS and in a dedicated spectroscopic follow-up from the Nordic Optical Telescope. The Hbeta and Halpha lines show broad profiles (FWHM~ 12,000 km/s), faint fluxes and extreme offsets (Delta v=8,700+/-1,300 km/s) with respect to the narrow emission lines. These features show no significant variation in a time lag of ~ 3.1 yr (rest frame). We rule out possible interpretations based on the superposition of two sources or on recoiling black holes, and we discuss the virtues and limitations of a massive black hole binary scenario.

# Precise Black Hole Masses From Megamaser Disks: Black Hole-Bulge Relations at Low Mass

**Authors:** Greene, J. E.; Peng, C. Y.; Kim, M.; Kuo, C. Y.; Braatz, J. A.; Impellizzeri, C. M. V.; Condon, J. J.; Lo, K. Y.; Henkel, C.; Reid, M. J.

**Abstract:**

The black hole (BH)-bulge correlations have greatly influenced the last decade of effort to understand galaxy evolution. Current knowledge of these correlations is limited predominantly to high BH masses ($M_{BH} \gtrsim 10^8 M_\odot$) that can be measured using direct stellar, gas, and maser kinematics. These objects, however, do not represent the demographics of more typical $L < L^*$ galaxies. This study transcends prior limitations to probe BHs that are an order of magnitude lower in mass, using BH mass measurements derived from the dynamics of $H_2O$ megamasers in circumnuclear disks. The masers trace the Keplerian rotation of circumnuclear molecular disks starting at radii of a few tenths of a pc from the central BH. Modeling of the rotation curves, presented by Kuo et al. (2010), yields BH masses with exquisite precision. We present stellar velocity dispersion measurements for a sample of nine megamaser disk galaxies based on long-slit observations using the B&C spectrograph on the Dupont telescope and the DIS spectrograph on the 3.5 m telescope at Apache Point. We also perform bulge-to-disk decomposition of a subset of five of these galaxies with SDSS imaging. The maser galaxies as a group fall below the relation defined by elliptical galaxies. We show, now with very precise BH mass measurements, that the low-scatter power-law relation between $M_{BH}$ and $\sigma$ seen in elliptical galaxies is not universal. The elliptical galaxy $M_{BH} - \sigma$ relation cannot be used to derive the BH mass function at low mass or the zeropoint for active BH masses. The processes (perhaps BH self-regulation or minor merging) that operate at higher mass have not effectively established a $M_{BH} - \sigma$ relation in this low-mass regime.

## High-Accuracy Comparison between the Post-Newtonian and Self-Force Dynamics of Black-Hole Binaries

**Authors:** Blanchet, Luc; Detweiler, Steven; Tiec, Alexandre Le; Whiting, Bernard F.

**Abstract:** The relativistic motion of a compact binary system moving in circular orbit is investigated using the post-Newtonian (PN) approximation and the perturbative self-force (SF) formalism. A particular gauge-invariant observable quantity is computed as a function of the binary's orbital frequency. The conservative effect induced by the gravitational SF is obtained numerically with high precision, and compared to the PN prediction developed to high order. The PN calculation involves the computation of the 3PN regularized metric at the location of the particle. Its divergent self-field is regularized by means of dimensional regularization. The poles proportional to 1/(d-3) which occur within dimensional regularization at the 3PN order disappear from the final gauge-invariant result. The leading 4PN and next-to-leading 5PN conservative logarithmic contributions originating from gravitational-wave tails are also obtained. Making use of these exact PN results, some previously unknown PN coefficients are measured up to the very high 7PN order by fitting to the numerical self-force data. Using just the 2PN and new logarithmic terms, the value of the 3PN coefficient is also confirmed numerically with very high precision. The consistency of this cross-cultural comparison provides a crucial test of the very different regularization methods used in both SF and PN formalisms, and illustrates the complementarity of these approximation schemes when modelling compact binary systems.

## Linear Stability Analysis and the Speed of Gravitational Waves in Dynamical Chern-Simons Modified Gravity

**Authors:** Garfinkle, David; Pretorius, Frans; Yunes, Nicolas

**Abstract:** We perform a linear stability analysis of dynamical Chern-Simons modified gravity in the geometric optics approximation and find that it is linearly stable on the backgrounds considered. Our analysis also reveals that gravitational waves in the modified theory travel at the speed of light in Minkowski spacetime. However, on a Schwarzschild background the characteristic speed of propagation along a given direction splits into two modes, one subluminal and one superluminal. The width of the splitting depends on the azimuthal components of the propagation vector, is linearly proportional to the mass of the black hole, and decreases with the third inverse power of the distance from the black hole. Radial propagation is unaffected, implying that as probed by gravitational waves the location of the event horizon of the spacetime is unaltered. The analysis further reveals that when a high frequency, pure gravitational wave is scattered from a black hole, a scalar wave of comparable amplitude is excited, and vice-versa.

## A hybrid method for understanding black-hole mergers: head-on case

**Authors:** Nichols, David A.; Chen, Yanbei

**Abstract:** Black-hole-binary coalescence is often divided into three stages: inspiral, merger and ringdown. The post-Newtonian (PN) approximation treats the inspiral phase, black-hole perturbation (BHP) theory describes the ringdown, and the nonlinear dynamics of spacetime characterize the merger. In this paper, we introduce a hybrid method that incorporates elements of PN and BHP theories, and we apply it to the head-on collision of black holes with transverse, anti-parallel spins. We compare our approximation technique with a full numerical-relativity simulation, and we find good agreement between the gravitational waveforms and the radiated energy and momentum. Our results suggest that PN and BHP theories may suffice to explain the main features of outgoing gravitational radiation for head-on mergers. This would further imply that linear perturbations to exact black-hole solutions can capture the nonlinear aspects of head-on binary-black-hole mergers accessible to observers far from the collision.

# Towards relativistic orbit fitting of Galactic center stars and pulsars

**Authors:** Angelil, Raymond; Saha, Prasenjit; Merritt, David

**Abstract:** The S stars orbiting the Galactic center black hole reach speeds of up to a few percent the speed of light during pericenter passage. This makes, for example, S2 at pericenter much more relativistic than known binary pulsars, and opens up new possibilities for testing general relativity. This paper develops a technique for fitting nearly-Keplerian orbits with perturbations from Schwarzschild curvature, frame dragging, and spin-induced torque, to redshift measurements distributed along the orbit but concentrated around pericenter. Both orbital and light-path effects are taken into account. It turns out that absolute calibration of rest-frame frequency is not required. Hence, if pulsars on orbits similar to the S stars are discovered, the technique described here can be applied without change, allowing the much greater accuracies of pulsar timing to be taken advantage of. For example, pulse timing of 3 microsec over one hour amounts to an effective redshift precision of 30 cm/s, enough to measure frame dragging and the quadrupole moment from an S2-like orbit, provided problems like the Newtonian "foreground" due to other masses can be overcome. On the other hand, if stars with orbital periods of order a month are discovered, the same could be accomplished with stellar spectroscopy from the E-ELT at the level of 1 km/s.

---

### *Intention and purpose of GW Notes*

*A succinct explanation*

---

The electronic publishing service **arXiv** is a dynamic, well-respected source of news of recent work and is updated daily. But, perhaps due to the large volume of new work submitted, it is probable that a member of our community might easily overlook relevant material. This new e-journal and its blog, **The LISA Brownbag (http://www.lisa-science.org/brownbag)**, both produced by the AEI, propose to offer scientist of the Gravitational Wave community the opportunity to more easily follow advances in the three areas mentioned: Astrophysics, General Relativity and Data Analysis. We hope to achieve this by selecting the most significant e-prints and list them in abstract form with a link to the full paper in both a single e-journal (GW Newsletter) and a blog (The LISA Brownbag). Of course, *this also implies that the paper will have its impact increased, since it will reach a broader public*, so that we encourage you to not forget submitting your own work

In addition to the abstracts, in each PDF issue of GW Notes, we will offer you a previously unpublished article written by a senior researcher in one of these three domains, which addresses the interests of all readers.

Thus the aim of The LISA Brownbag and GW Notes is twofold:

- Whenever you see an interesting paper on GWs science and LISA, you can submit the **arXiv** number to our **submission page (http://brownbag.lisascience.org)**. This is straightforward: No registration is required (although recommended) to simply type in the number in the entry field of the page, indicate some keywords and that's it

- We will publish a new full article in each issue, if available. This "feature article" will be from the fields of Astrophysics, General Relativity or the Data Analysis of gravitational waves and LISA. We will prepare a more detailed guide for authors, but for now would like to simply remind submitters that they are writing for colleagues in closely related but not identical fields, and that cross-fertilization and collaboration is an important goal of our concept

Subscribers get the issue distributed in PDF form. Additionally, they will be able to submit special announcements, such as meetings, workshops and jobs openings, to the list of registered people. For this, please register at the **registration page (http://lists.aei.mpg.de/cgi-bin/mailman/listinfo/lisa_brownbag)** by filling in your e-mail address and choosing a password.

> ### *The Astro-GR meetings*
>
> *Past, present and future*

Sixty two scientists attended the **Astro-GR@AEI** meeting, which took place September 18-22 2006 at the **Max-Planck Institut für Gravitationsphysik (Albert Einstein-Institut)** in Golm, Germany. The meeting was the brainchild of an AEI postdoc, who had the vision of bringing together Astrophysicists and experts in General Relativity and gravitational-wave Data Analysis to discuss sources for **LISA**, the planned Laser Interferometer Space Antenna. More specifically, the main topics were EMRIs and IMRIs (Extreme and Intermediate Mass-Ratio Inspiral events), i.e. captures of stellar-mass compact objects by supermassive black holes and coalescence of intermediate-mass black holes with supermassive black holes.

The general consensus was that the meeting was both interesting and quite stimulating. It was generally agreed that someone should step up and host a second round of this meeting. Monica Colpi kindly did so and this led to **Astro-GR@Como**, which was very similar in its informal format, though with a focus on all sources, meant to trigger new ideas, as a kind of brainstroming meeting.

Also, in the same year, in the two first weeks of September, we had another workshop in the Astro-GR series with a new "flavour", namely, the **Two Weeks At The AEI (2W@AEI)**, in which the interaction between the attendees be even higher than what was reached in the previous meetings. To this end, we reduced the number of talks, allowing participants more opportunity to collaborate. Moreover, participants got office facilities and we combined the regular talks with the so-called "powerpointless" seminar, which will were totally informal and open-ended, on a blackboard. The next one was held in Barcelona in 2009 at the beginning of September, **Astro-GR@BCN** and next 2010 it will be the turn of Paris, at the APC.

**LISA Astro-GR@Paris (Paris, APC Monday 13th to Friday September 17th 2010)**

If you are interested in hosting in the future an Astro-GR meeting, please contact us. We are open to new formats, as long as the *Five Golden Rules* are respected.

A proper Astro-GR meeting **MUST** closely follow the *Five Golden Rules*:

I. Bring together Astrophysicists, Cosmologists, Relativists and Data Analysts

II. Motivate new collaborations and projects

III. Be run in the style of Aspen, ITP, Newton Institute and Modest meetings, with plenty of time for discussions

IIII. Grant access to the slides in a cross-platform format, such as PDF and, within reason, to the recorded movies of the talks in a free format which everybody can play like **Theora**, for those who could not attend, following the good principles of **Open Access**

IHI. Keep It Simple and... Spontaneous